\documentclass[aps,prc,superscriptaddress,showpacs,amssymb,amsmath,amsfonts,twocolumn,floatfix]{revtex4-1}
\usepackage{graphicx,amsmath,color,epstopdf}

\begin{document}

\newcommand{\piplusRxn}{\gamma \ p \rightarrow n \ \pi^+}
\newcommand{\xRxn}{\gamma \ p\rightarrow \pi^+X}
\newcommand{\xRxnp}{\gamma \ p\rightarrow p X}
\newcommand{\ppipiRxn}{\gamma \ p\rightarrow p \ \pi^+ \ \pi^-}
\newcommand{\ppipiyRxn}{\gamma \ p\rightarrow p \ \pi^+ \ \pi^- \ Y}
\newcommand{\Eg}{E_{\gamma}}
\newcommand{\cosThetaCm}{\cos\theta_{\rm c.m.}}
\newcommand{\cosThetaCmP}{\cos\theta^p_{\rm c.m.}}
\newcommand{\cosThetaCmPi}{\cos\theta^{\pi}_{\rm c.m.}}
\newcommand{\ppizero}{\gamma \ p \rightarrow p \ \pi^0}
\newcommand{\piNRxn}{\gamma \ p \rightarrow n \ \pi^+}
\newcommand{\ffunc}{f^{i,j}_a(\varphi)}

\hyphenation {po-la-ri-za-tion}

\title{Beam asymmetry $\Sigma$ for $\pi^+$ and $\pi^0$ photoproduction on the 
proton for photon energies from 1.102 to 1.862~GeV}

\newcommand*{\ANL}{Argonne National Laboratory, Argonne, Illinois 60439}
\newcommand*{\ANLindex}{1}
\affiliation{\ANL}
\newcommand*{\ASU}{Arizona State University, Tempe, Arizona 85287-1504}
\newcommand*{\ASUindex}{2}
\affiliation{\ASU}
\newcommand*{\CSUDH}{California State University, Dominguez Hills, Carson, CA 90747}
\newcommand*{\CSUDHindex}{3}
\affiliation{\CSUDH}
\newcommand*{\CMU}{Carnegie Mellon University, Pittsburgh, Pennsylvania 15213}
\newcommand*{\CMUindex}{4}
\affiliation{\CMU}
\newcommand*{\CUA}{Catholic University of America, Washington, D.C. 20064}
\newcommand*{\CUAindex}{5}
\affiliation{\CUA}
\newcommand*{\SACLAY}{CEA, Centre de Saclay, Irfu/Service de Physique Nucl\'eaire, 91191 Gif-sur-Yvette, France}
\newcommand*{\SACLAYindex}{6}
\affiliation{\SACLAY}
\newcommand*{\CNU}{Christopher Newport University, Newport News, Virginia 23606}
\newcommand*{\CNUindex}{7}
\affiliation{\CNU}
\newcommand*{\UCONN}{University of Connecticut, Storrs, Connecticut 06269}
\newcommand*{\UCONNindex}{8}
\affiliation{\UCONN}
\newcommand*{\EDINBURGH}{Edinburgh University, Edinburgh EH9 3JZ, United Kingdom}
\newcommand*{\EDINBURGHindex}{9}
\affiliation{\EDINBURGH}
\newcommand*{\FU}{Fairfield University, Fairfield CT 06824}
\newcommand*{\FUindex}{10}
\affiliation{\FU}
\newcommand*{\FIU}{Florida International University, Miami, Florida 33199}
\newcommand*{\FIUindex}{11}
\affiliation{\FIU}
\newcommand*{\FSU}{Florida State University, Tallahassee, Florida 32306}
\newcommand*{\FSUindex}{12}
\affiliation{\FSU}
\newcommand*{\Genova}{Universit$\grave{a}$ di Genova, 16146 Genova, Italy}
\newcommand*{\Genovaindex}{13}
\affiliation{\Genova}
\newcommand*{\GWUI}{The George Washington University, Washington, DC 20052}
\newcommand*{\GWUIindex}{14}
\affiliation{\GWUI}
\newcommand*{\ISU}{Idaho State University, Pocatello, Idaho 83209}
\newcommand*{\ISUindex}{15}
\affiliation{\ISU}
\newcommand*{\INFNFE}{INFN, Sezione di Ferrara, 44100 Ferrara, Italy}
\newcommand*{\INFNFEindex}{16}
\affiliation{\INFNFE}
\newcommand*{\INFNFR}{INFN, Laboratori Nazionali di Frascati, 00044 Frascati, Italy}
\newcommand*{\INFNFRindex}{17}
\affiliation{\INFNFR}
\newcommand*{\INFNGE}{INFN, Sezione di Genova, 16146 Genova, Italy}
\newcommand*{\INFNGEindex}{18}
\affiliation{\INFNGE}
\newcommand*{\INFNRO}{INFN, Sezione di Roma Tor Vergata, 00133 Rome, Italy}
\newcommand*{\INFNROindex}{19}
\affiliation{\INFNRO}
\newcommand*{\ORSAY}{Institut de Physique Nucl\'eaire ORSAY, Orsay, France}
\newcommand*{\ORSAYindex}{20}
\affiliation{\ORSAY}
\newcommand*{\ITEP}{Institute of Theoretical and Experimental Physics, Moscow, 117259, Russia}
\newcommand*{\ITEPindex}{21}
\affiliation{\ITEP}
\newcommand*{\JMU}{James Madison University, Harrisonburg, Virginia 22807}
\newcommand*{\JMUindex}{22}
\affiliation{\JMU}
\newcommand*{\KNU}{Kyungpook National University, Daegu 702-701, Republic of Korea}
\newcommand*{\KNUindex}{23}
\affiliation{\KNU}
\newcommand*{\LPSC}{LPSC, Universite Joseph Fourier, CNRS/IN2P3, INPG, Grenoble, France}
\newcommand*{\LPSCindex}{24}
\affiliation{\LPSC}
\newcommand*{\UNH}{University of New Hampshire, Durham, New Hampshire 03824-3568}
\newcommand*{\UNHindex}{25}
\affiliation{\UNH}
\newcommand*{\NSU}{Norfolk State University, Norfolk, Virginia 23504}
\newcommand*{\NSUindex}{26}
\affiliation{\NSU}
\newcommand*{\OHIOU}{Ohio University, Athens, Ohio  45701}
\newcommand*{\OHIOUindex}{27}
\affiliation{\OHIOU}
\newcommand*{\ODU}{Old Dominion University, Norfolk, Virginia 23529}
\newcommand*{\ODUindex}{28}
\affiliation{\ODU}

\newcommand*{\PNPI}{Petersburg Nuclear Physics Institute, 188300 Gatchina, Russia}
\newcommand*{\PNPIindex}{29}
\affiliation{\PNPI}

\newcommand*{\RPI}{Rensselaer Polytechnic Institute, Troy, New York 12180-3590}
\newcommand*{\RPIindex}{30}
\affiliation{\RPI}
\newcommand*{\URICH}{University of Richmond, Richmond, Virginia 23173}
\newcommand*{\URICHindex}{31}
\affiliation{\URICH}
\newcommand*{\ROMAII}{Universita' di Roma Tor Vergata, 00133 Rome Italy}
\newcommand*{\ROMAIIindex}{32}
\affiliation{\ROMAII}
\newcommand*{\MSU}{Skobeltsyn Nuclear Physics Institute, 119899 Moscow, Russia}
\newcommand*{\MSUindex}{33}
\affiliation{\MSU}
\newcommand*{\SCAROLINA}{University of South Carolina, Columbia, South Carolina 29208}
\newcommand*{\SCAROLINAindex}{34}
\affiliation{\SCAROLINA}
\newcommand*{\JLAB}{Thomas Jefferson National Accelerator Facility, Newport News, Virginia 23606}
\newcommand*{\JLABindex}{35}
\affiliation{\JLAB}
\newcommand*{\UTFSM}{Universidad T\'{e}cnica Federico Santa Mar\'{i}a, Casilla 110-V Valpara\'{i}so, Chile}
\newcommand*{\UTFSMindex}{36}
\affiliation{\UTFSM}
\newcommand*{\GLASGOW}{University of Glasgow, Glasgow G12 8QQ, United Kingdom}
\newcommand*{\GLASGOWindex}{37}
\affiliation{\GLASGOW}
\newcommand*{\VIRGINIA}{University of Virginia, Charlottesville, Virginia 22901}
\newcommand*{\VIRGINIAindex}{38}
\affiliation{\VIRGINIA}
\newcommand*{\WM}{College of William and Mary, Williamsburg, Virginia 23187-8795}
\newcommand*{\WMindex}{39}
\affiliation{\WM}
\newcommand*{\YEREVAN}{Yerevan Physics Institute, 375036 Yerevan, Armenia}
\newcommand*{\YEREVANindex}{40}
\affiliation{\YEREVAN}

\newcommand*{\NOWMSU}{Skobeltsyn Nuclear Physics Institute, 119899 Moscow, Russia}
\newcommand*{\NOWORSAY}{Institut de Physique Nucl\'eaire ORSAY, Orsay, France}
\newcommand*{\NOWINFNGE}{INFN, Sezione di Genova, 16146 Genova, Italy}
\newcommand*{\NOWROMAII}{Universita' di Roma Tor Vergata, 00133 Rome Italy}


\newcommand*{\NOWCUA}{Catholic University of America, Washington, D.C. 20064}
\newcommand*{\NOWJLAB}{Thomas Jefferson National Accelerator Facility, Newport News, Virginia 23606, USA}

\author {M.~Dugger}
\affiliation{\ASU}
\author {B.G.~Ritchie}
\affiliation{\ASU}
\author {P.~Collins}
\altaffiliation[Current address:]{\NOWCUA}
\affiliation{\ASU}
\author {E.~Pasyuk}
\altaffiliation[Current address:]{\NOWJLAB}
\affiliation{\ASU}
\author {W.J.~Briscoe}
\affiliation{\GWUI}
\author {I.I.~Strakovsky}
\affiliation{\GWUI}
\author {R.L.~Workman}
\affiliation{\GWUI}
\author {Y.~Azimov}
\affiliation{\PNPI}

\author {K.P. ~Adhikari} 
\affiliation{\ODU}
\author {D.~Adikaram} 
\affiliation{\ODU}
% ...

\collaboration{The CLAS Collaboration}
\noaffiliation



\begin{abstract}
Beam asymmetries for the reactions $\gamma p \rightarrow p \pi^0$ and
$\gamma p \rightarrow n \pi^+$ have been measured with the
CEBAF Large Acceptance Spectrometer (CLAS) and a tagged,
linearly polarized photon beam
with energies from 1.102 to 1.862 GeV. A Fourier moment technique
for extracting beam asymmetries from experimental data is described.
The results reported here 
possess greater precision and finer energy resolution than previous
measurements. 
Our data for both pion reactions appear to favor the SAID and Bonn-Gatchina 
scattering analyses
over the older Mainz MAID predictions. 
After incorporating the present set of beam asymmetries into 
the world database, exploratory fits made with the SAID analysis indicate that
the largest changes from previous fits are for properties of the 
$\Delta(1700)3/2^-$ and $\Delta(1905)5/2^+$ states.
\end{abstract}


\maketitle


\section{Introduction}
\label{sec:Intro}

The properties of the resonances for the
non-strange baryons have been determined almost entirely
from the results of pion-nucleon scattering analyses~\cite{PDG}. 
Other reactions have mainly served to fix branching ratios and  
photo-couplings. With the refinement of multi-channel fits and
the availability of high-precision photoproduction data for
both single- and double-meson production, identifications of 
some new states have
emerged mainly due to evidence from reactions not involving
single-pion-nucleon initial or final states~\cite{PDG}. However,
beyond elastic pion-nucleon scattering, single-pion 
photoproduction remains the most studied source of resonance 
information.

Much of the effort aimed at providing complete or 
nearly-complete information for meson-nucleon photoproduction reactions 
has been directed to measuring double-polarization observables. 
However,  often overlooked is that the data coverage for several single-polarization 
observables, also vital   
in determining the properties of the nucleon resonance spectrum, still remains incomplete. 
More complete datasets for those single-polarization observables can also
offer important constraints on analyses of the photoproduction reaction. 

In this work, using linearly-polarized photons and an unpolarized target, 
we provide a large set of beam asymmetry $\Sigma$ measurements from 1.102 to 1.862~GeV 
in laboratory photon energy, corresponding to a 
center-of-mass energy $W$ range of 1.7 to 2.1~GeV.
As will be seen, this dataset greatly constrains
multipole analyses above the second-resonance 
region in part simply due to the size of the dataset provided:
with these new $\Sigma$ asymmetry data from the CEBAF 
Large Acceptance Spectrometer (CLAS) at Jefferson Lab in Hall~B, 
the number of measurements in the world 
database for the processes $\gamma p\to\pi^0p$ and $\gamma p\to\pi^+n$ 
between 1100 and 1900~MeV for $E_\gamma$ is more than doubled.
We will show here that there are unexpectedly large deviations between
these data and some of the most extensive multipole analyses 
covering the resonance region. We have included these $\Sigma$ 
data in a new partial wave analysis and will compare that analysis with competing 
predictions in this paper. 
 
The paper is organized in the following manner: We give a brief
background of the experimental conditions for this study in
Sec.~\ref{sec:runPeriod}.  An overview of the methods used
to extract the beam asymmetry results reported here is given 
in Sec.~\ref{sec:pid} through Sec.~\ref{sec:RelNorm}, and the 
uncertainty estimates for the 
$\Sigma$ data obtained are given in Sec.~\ref{sec:Errs}.  The 
resulting data are summarized and described in 
Sec.~\ref{sec:results}. They are then are compared to various 
predictions and a new analysis presented here in Sec.~\ref{sec:fit}, where we also 
compare multipoles obtained with and without including the 
present data set. Conclusions are presented in Sec.~\ref{sec:conc}. 
\section{The running period}
\label{sec:runPeriod}

The beam asymmetries for the $\piNRxn$ and $\ppizero$ reactions 
described in this paper were
part of a set of
experiments running at the same time with the same experimental
configuration (cryogenic hydrogen target, bremsstrahlung photon tagger \cite{tag}, 
and CLAS \cite{CLAS}) called the ``{\tt{g8b}}''
run period. 
The ``{\tt{g8a}}'' and ``{\tt{g8b}}'' run periods were the first Jefferson Lab 
experiments to use 
the coherent bremsstahlung technique
to produce polarized photons.

During the {\tt{g8b}} running period, 
a bremsstrahlung photon beam with enhanced linear polarization was incident on a 
40-cm-long
liquid hydrogen target placed 20 cm upstream from the center of CLAS.
The enhancement of  linear polarization was accomplished through the coherent 
bremsstrahlung process by having 
the CEBAF electron beam, with an energy of 4.55 GeV, incident  on a 50-$\mu$m-thick
diamond radiator. The photon polarization plane (defined as the plane containing 
the electric-field vector) and
the coherent edge energy of the enhanced polarization photon spectrum 
were controlled by adjusting the orientation of the diamond radiator
using a remotely-controlled goniometer.
The degrees of photon beam polarization are estimated
via a bremsstrahlung calculation using knowledge of
the goniometer orientation and the degree of collimation \cite{CBSA}.
Data with an unpolarized photon beam also were taken periodically using a graphite radiator 
(``amorphous runs'').
For all data runs, the CLAS magnetic field was set to 50\% of its maximum nominal field,
with positive particles bending outward away from the axis determined
by the incident photon beam.
The event trigger required the coincidence of a post-bremsstrahlung electron
passing through the focal plane of the photon tagger and at least one
charged particle detected in CLAS.

The {\tt{g8b}} run period was divided into intervals with different
coherent edge energies, nominally set to 1.3, 1.5, 1.7,
1.9, and 2.1 GeV. In addition to the differing coherent edge energies 
(all measured at the same electron beam energy of 4.55 GeV),
the data were
further grouped into runs where the polarization plane
was parallel to the floor (denoted as PARA) or
perpendicular to the floor (denoted as PERP) or where the beam
was unpolarized (amorphous). For the entire 1.9 GeV data set, the
polarization plane was flipped between PARA and PERP automatically (``auto-flip'').
Some of the
1.7 GeV data set was taken with auto-flip while for some runs, the
polarization plane of the 1.7~GeV data was manually controlled (``manual''). For the 1.3 and 
1.5~GeV data sets, all data used were of the manual type. (The 2.1 GeV data set
was not utilized in the analysis.)

\section{Particle identification; kinematic variables}
\label{sec:pid}

\begin{figure}
\includegraphics[scale=0.35]{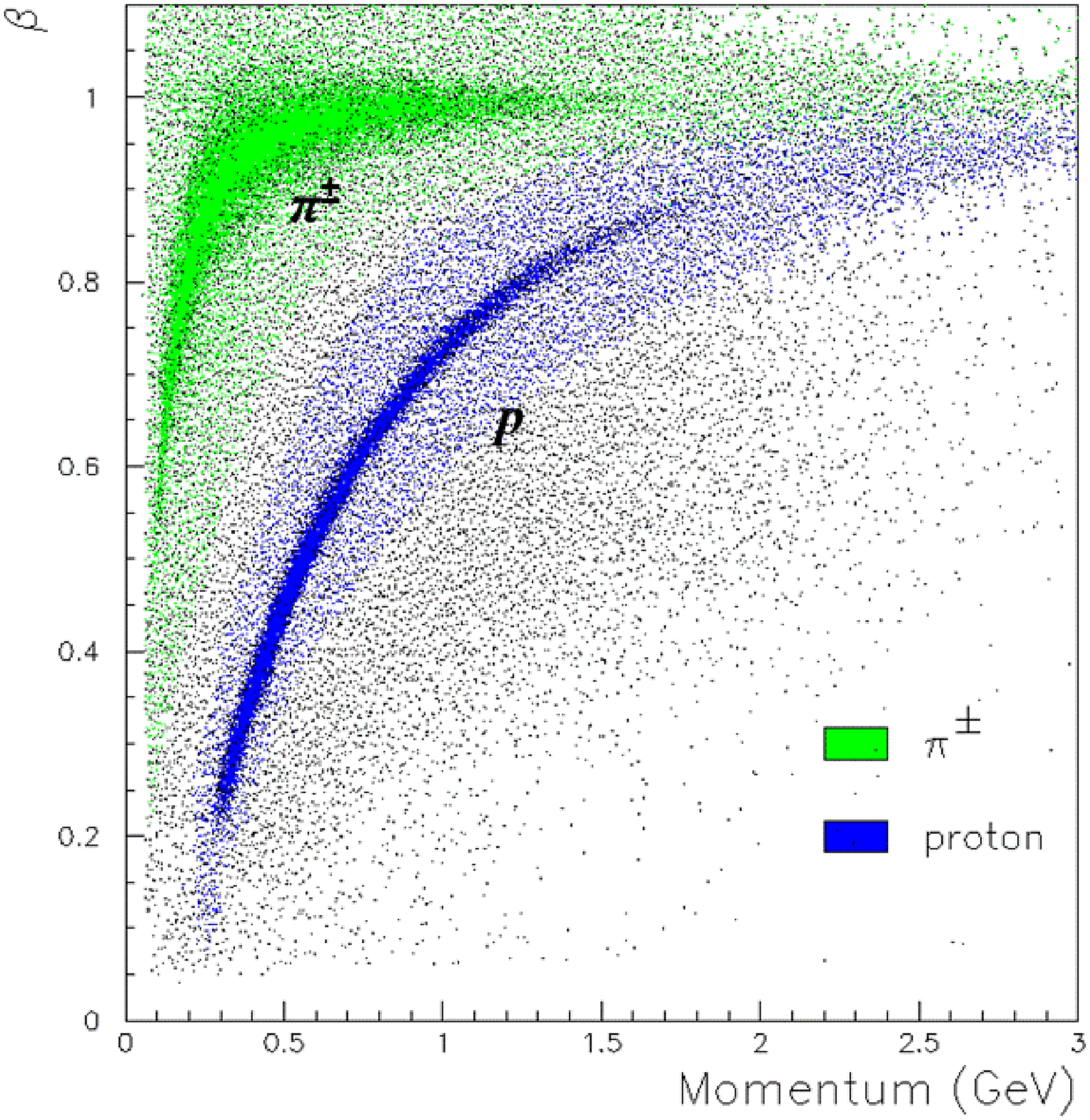}
\includegraphics[scale=0.35]{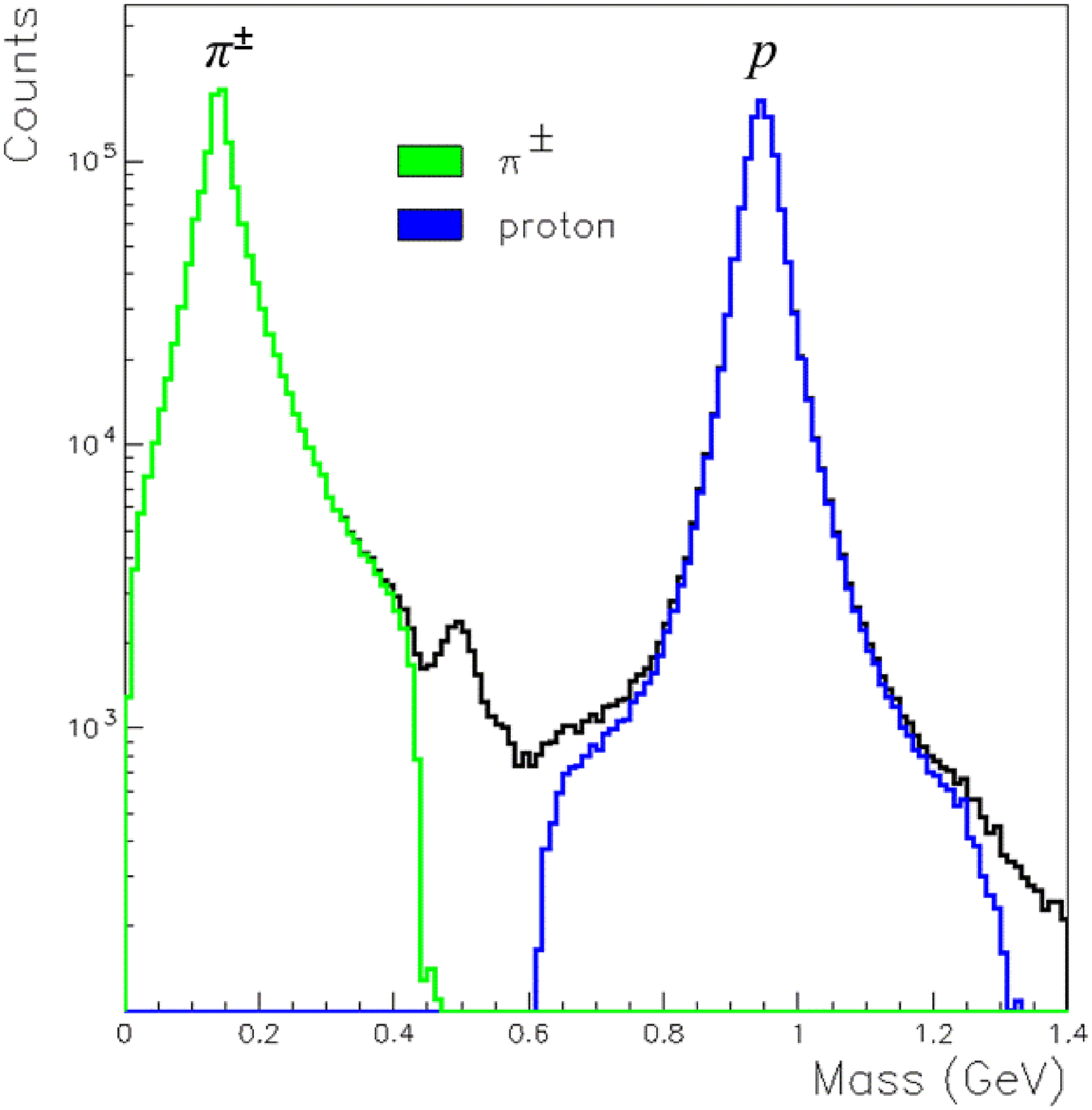}
\caption{(Color online) Top panel: Velocity $\beta$ versus momentum $p$ for 
        all charged particles passing through CLAS in this experiment.
        Bottom panel: Mass of charged particles passing through CLAS in 
        this experiment as determined by $\beta$ and $p$.
        In each figure, pions and protons identified through the {\tt{GPID}} 
        algorithm are colored green and blue, respectively.\label{fig:pid}}
\end{figure}

For  experiments using the bremsstrahlung photon beam,
the CLAS target region is surrounded by a scintillator array known as 
the ``start counter'',
which is used to establish the vertex time for the event \cite{startCounter}.
Particles then pass through drift chambers, which provide tracking information that yields
momentum and angular information on charged particles passing through CLAS \cite{DC}.
Particles then pass through the time-of-flight scintillator array \cite{TOF}, 
which, using the vertex time for the event, 
measures the time taken for the particle to pass from the start counter through the 
drift chambers. This time information
determines the velocity of the charged particles passing through CLAS and, 
when coupled with the momentum information provided by the drift chambers,
provides for determination of the mass and charge of the particle.


\begin{figure}
\includegraphics[scale=0.35]{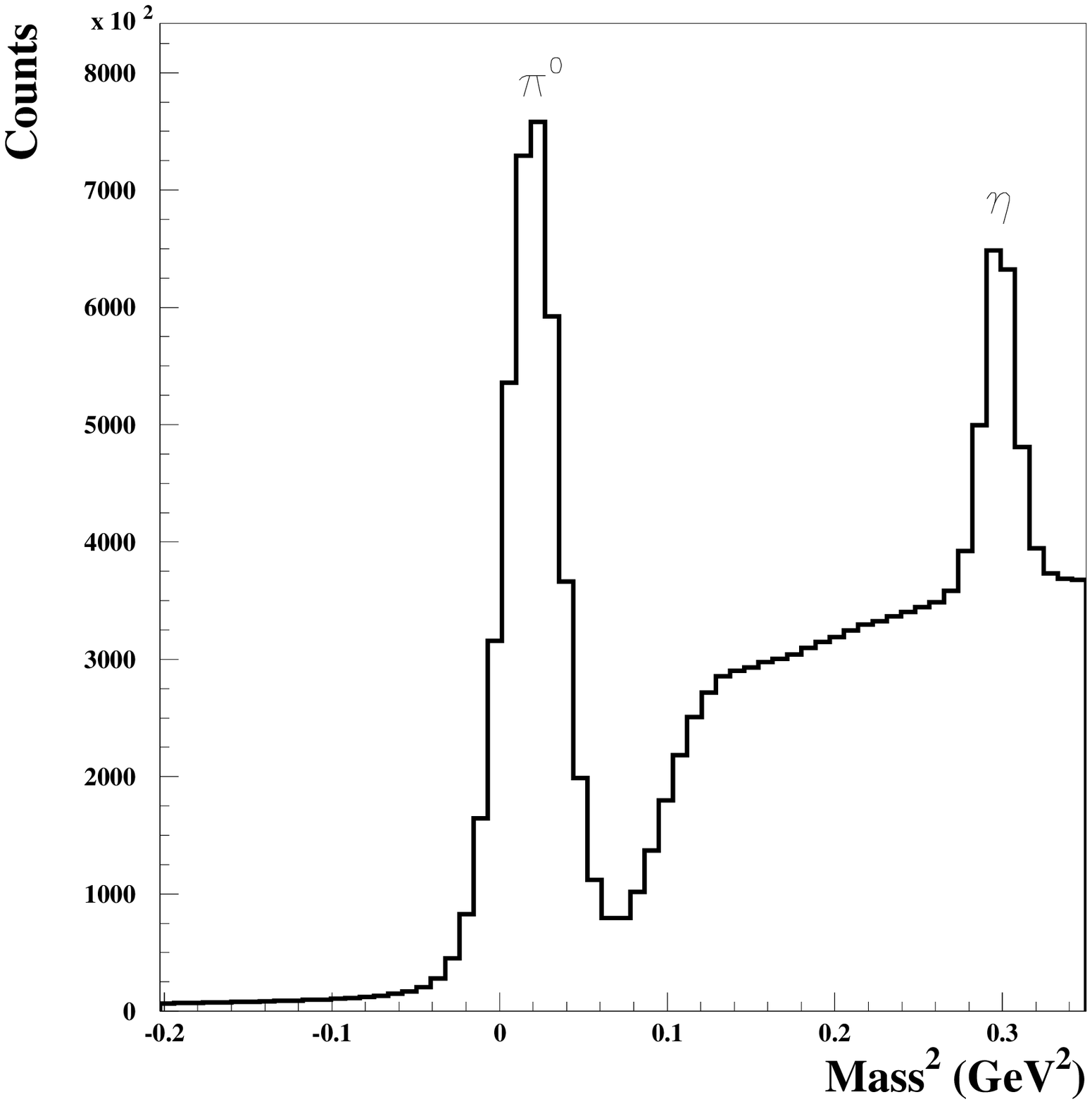}
\includegraphics[scale=0.35]{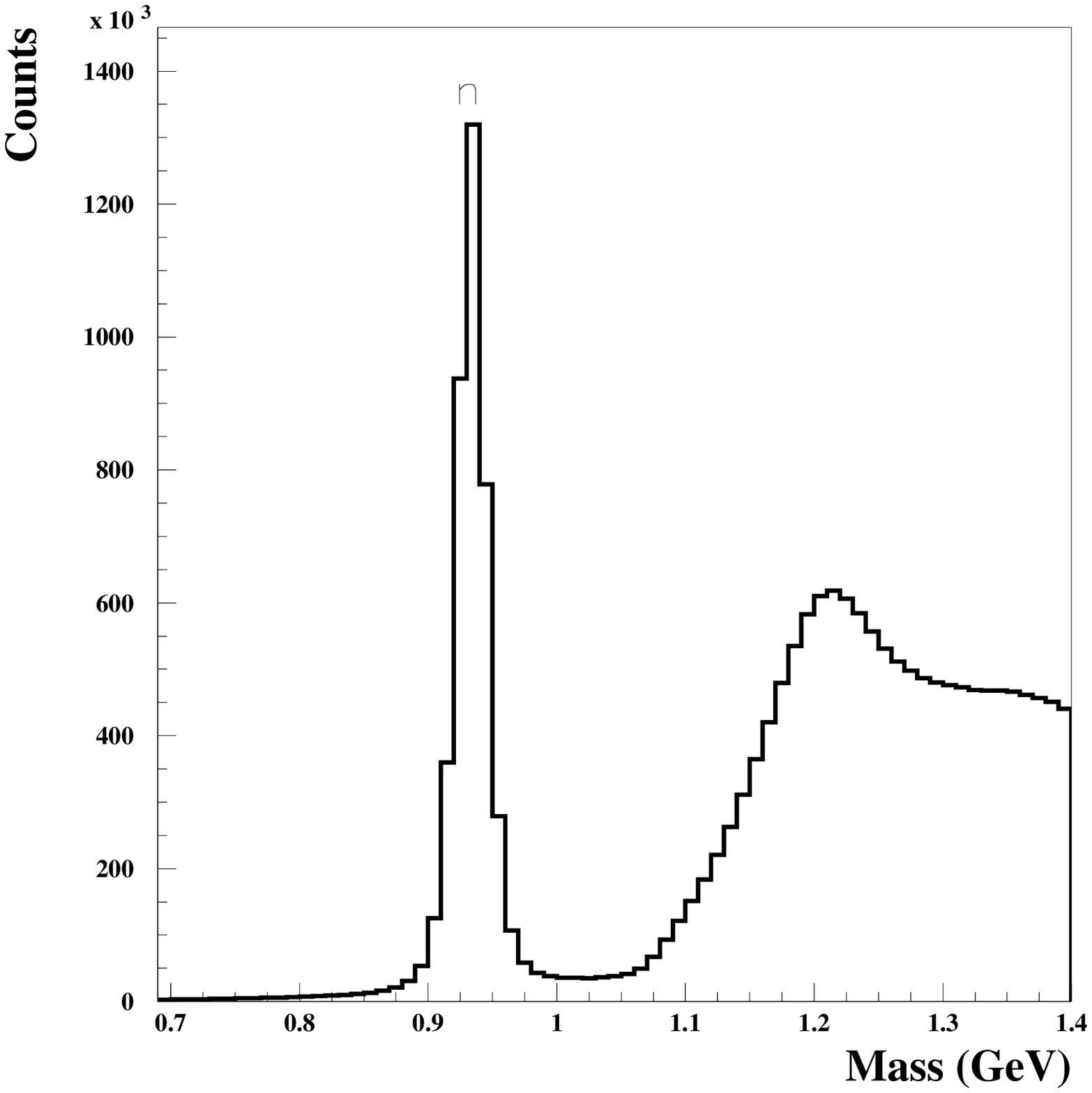}
\caption{Top panel:
        Spectrum of squared missing mass $M_X^2$ from the reaction 
        $\gamma \ p \rightarrow p X$.
        Bottom panel: Missing mass $M_X$ from the reaction $\gamma \ p 
        \rightarrow \pi^+ X$. The $\pi^0$, $\eta$, and neutron peaks 
        are indicated on the plots. \label{fig:missmass}}
\end{figure}


Using the information obtained from the start counter, drift chambers, and TOF array for each 
particle scattered into CLAS,
particle identification was performed with the {\tt{GPID}} algorithm 
(described in \cite{gpid}).
Plots showing $\beta$ versus $p$ and the mass distribution of the charged particles 
detected in CLAS,
as determined by the {\tt{GPID}} algorithm,
are given in Fig.~\ref{fig:pid}.  As discussed in Ref.\ \cite{gpid},
the {\tt{GPID}} method uses the CLAS-measured momentum
of the particle whose identity is to be determined,
and calculates theoretical values of $\beta$
for the particle to be any one of all possible identities.
Each one of the possible identities is tested by comparing the 
``theoretical'' value of $\beta$ for a given particle type 
(using the reconstructed momentum information from CLAS) to 
the ``measured'' value of $\beta$ (as determined from time-of-flight information).
The particle is assigned the identity that provides the closest expected value
of $\beta$ to the empirically measured value of $\beta$.
The identification for protons and pions 
is illustrated 
in Figure~\ref{fig:pid}.


\begin{figure}
\includegraphics[scale=0.4]{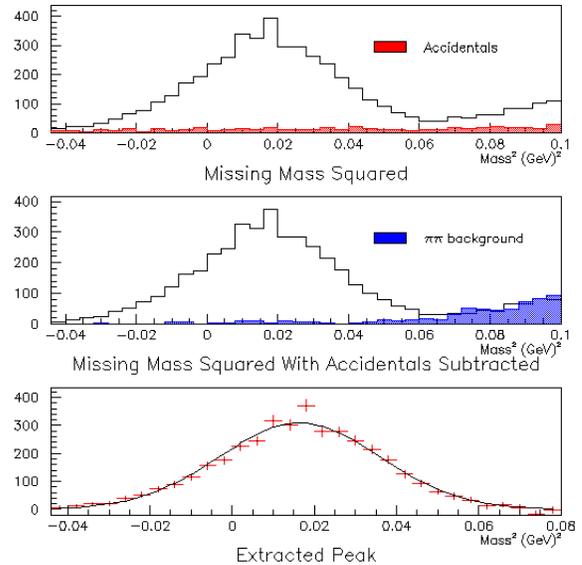}
\caption{(Color online) Yield extraction examples from previously published 
        {\tt{g1c}} $\pi^0$ differential cross sections \cite{ASUpi0}. Shown 
        are data for the $\pi^0$ yield extraction for $\Eg = 1.425$ GeV and 
        $\cosThetaCmP$ = -0.45.  The top panel is the missing mass yield 
        for this bin, with the accidental contribution displayed in red. The
        accidental contribution is seen to be small and linear. The middle 
        panel shows the missing mass distribution with the accidental 
        contributions subtracted; the blue region indicates the 2-$\pi$ 
        contribution determined. The bottom panel shows the extracted 
        $\pi^0$ yield after contributions for accidentals and two-pion 
        photoproduction have been subtracted from the missing mass 
        distribution. \label{fig:g1cfits}}
\end{figure}


\begin{figure}
\includegraphics[scale=0.4]{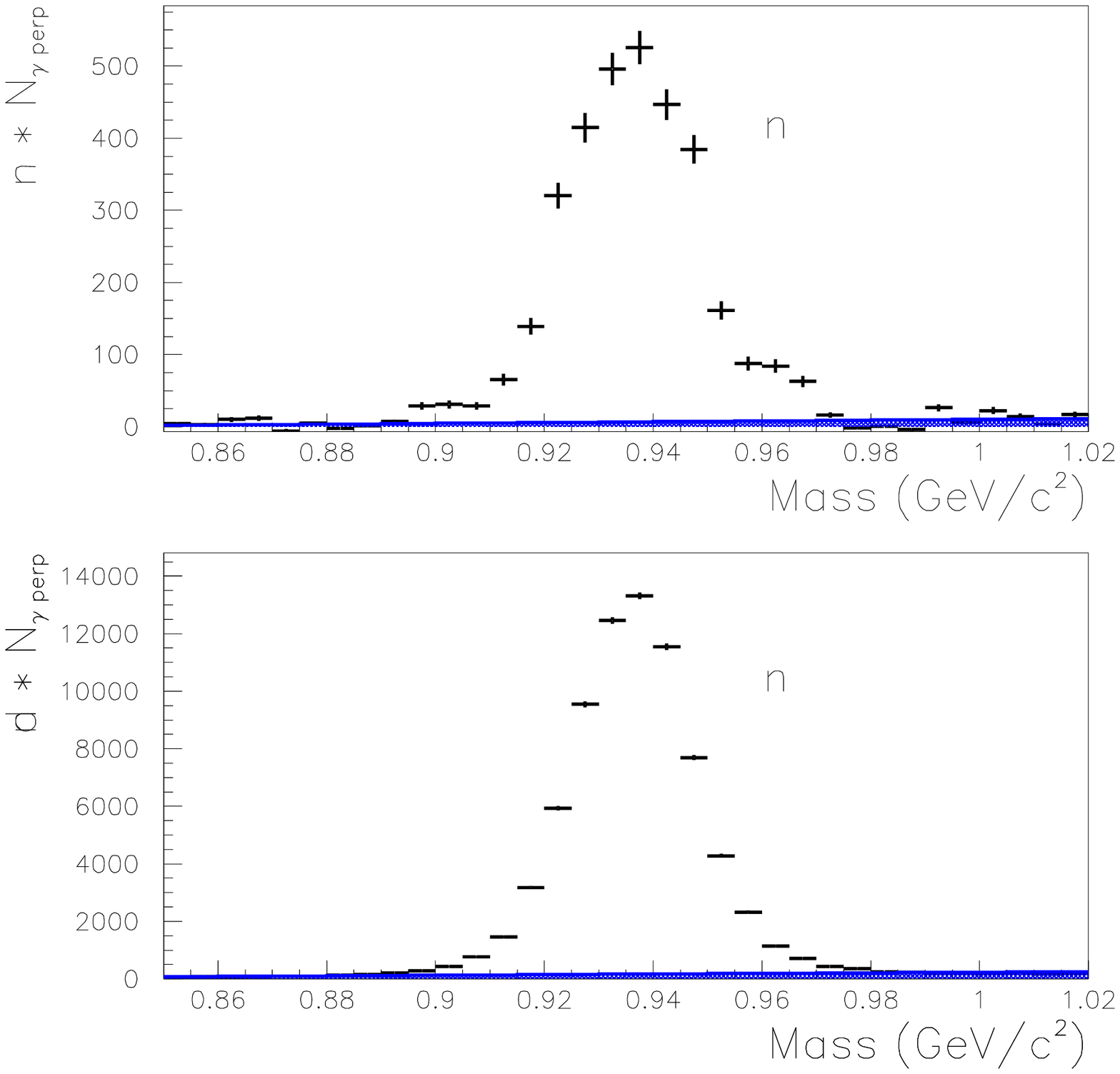}
\includegraphics[scale=0.4]{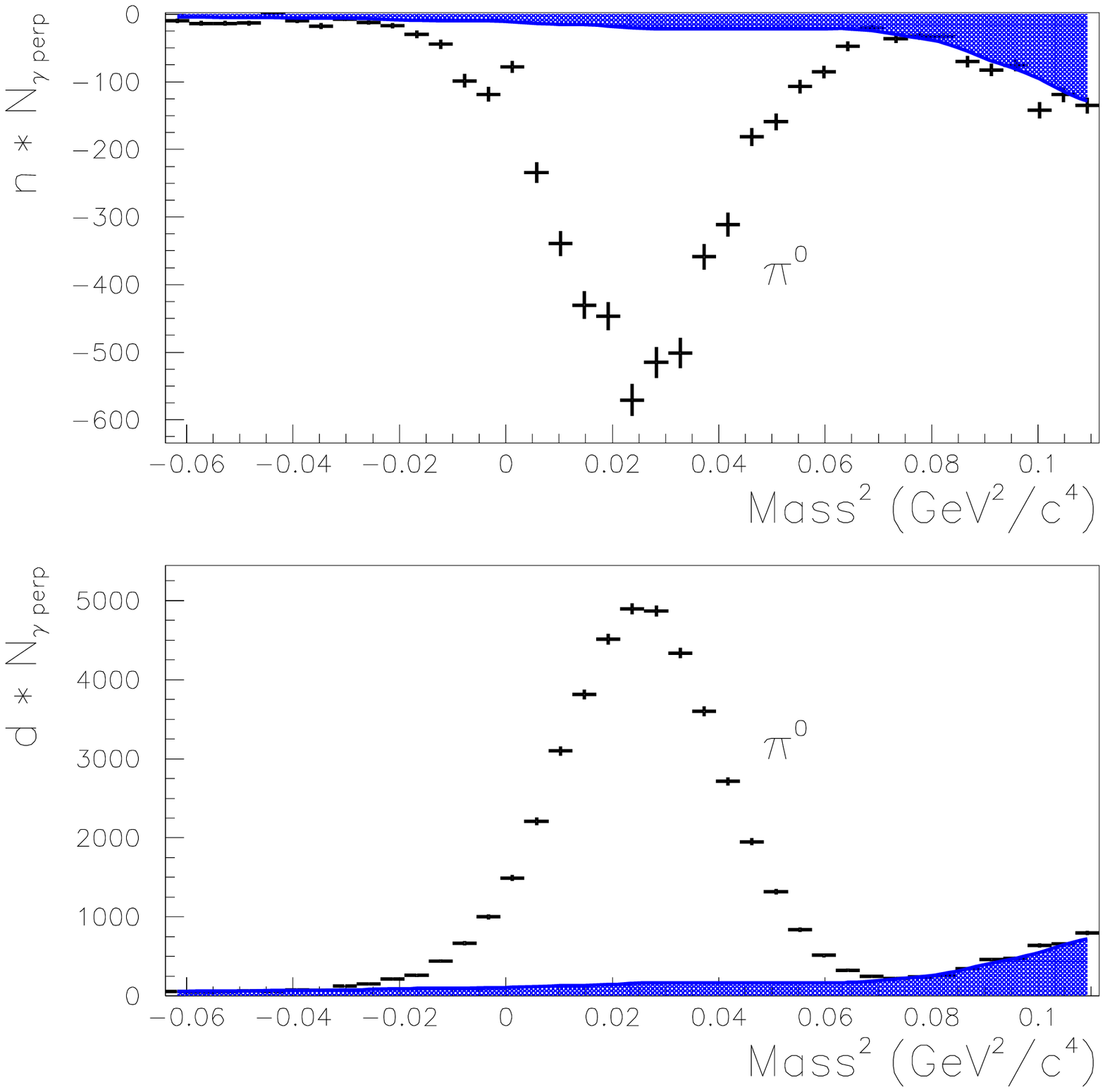}
\caption{(Color online) Extraction of the quantities $n$ and $d$ (defined 
        in Eqn. \ref{eqn:nd}) for neutron (top two panels) and $\pi^0$ 
        (bottom two panels) for
        the kinematic bin $\theta^{\pi}_{c.m.} = 123^\circ$ and
        $E_{\gamma} = 1229$ MeV.
        Top two panels: Reaction $\gamma p \rightarrow  \pi^+ n$. The
        $x$-axis is missing mass $M_X$ from
        the reaction $\gamma p \rightarrow \pi^+ X$.
        Bottom two panels: Reaction $\gamma p \rightarrow p \pi^0$.
        The $x$-axis is squared missing mass $M_X^2$ from
        the reaction $\gamma p \rightarrow p X$.
        The blue shaded region in each plot represents contributions from 
        background.\label{fig:g8bfits}}
\end{figure}


\section{Missing mass reconstruction for $\mathbf{\pi}$ $\mathbf{N}$ final states}

The kinematic quantities determined from the time-of-flight and drift chamber systems
yield good momentum definition for the proton and $\pi^+$.  
The energy and momentum determined for each particle by CLAS were
corrected for energy lost by that particle in passing through the material 
in both the target cell and the start counter in order to reconstruct
the momentum at the vertex where the photoproduction reaction occurred
using the standard CLAS algorithm for those corrections, {\tt{ELOSS}} \cite{eloss}.
In addition to the energy loss correction, 
a CLAS momentum correction was used. The CLAS momentum correction 
optimized the momentum determination through kinematic fitting.

The scattering angle and momentum information for each particle was used to construct a 
missing mass $M_X$ based on the
assumption that the reaction observed was $\gamma \ p \rightarrow \pi^+ X$ or $\gamma \ 
p \rightarrow p X$, where $X$ is the other body in the
two-body final state using the relation
\begin{eqnarray}
\nonumber & M_X \!\! = \!\!   
& \sqrt{m^2_{\pi^+} \!\! +  m^2_p +  2\Eg m_p \!\! -2 E_{\pi^+}\!\!\left(m_p \!\! 
+\Eg \right)\!\!+\!\!2\Eg p_{z  \pi^+}}
\end{eqnarray}
for the $\gamma \ p \rightarrow \pi^+ X$ reaction, and
\begin{equation}
 M_X \!\!  =  \!\!   \sqrt{2 m^2_p + 2\Eg m_p -2 E_p \left( m_p + \Eg \right) + 2 \Eg p_{zp}} 
\nonumber
\end{equation}
when the reaction is $\gamma \ p \rightarrow p X$, where $M_X$ is the mass of
the missing particle, $\Eg$ is the incident photon energy,
$m$ denotes mass, $p$ is the momentum, $p_z$ denotes the $z$-component of the momentum,
and subscripts define the particle type.

Based on these assumptions, 
the missing mass spectrum for data in the
full spectrometer acceptance for all photon energies within the 1.3 GeV
coherent edge setting is shown in Fig.~\ref{fig:missmass}.
The neutron and $\pi^0$ peaks are clearly seen.

\section{Fourier moment technique for extracting beam asymmetry $\Sigma$}
\label{sec:momentMethod}


\begin{figure}
\includegraphics[scale=0.4]{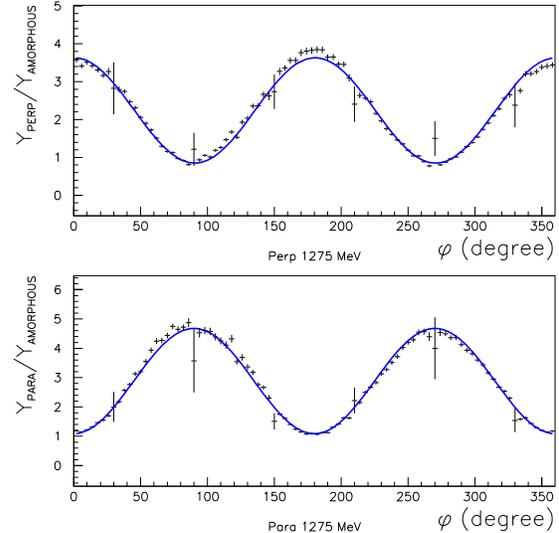}
\caption{Rough $\pi^0$ normalized yields for forward center-of-mass angles.
        Top panel shows the normalized yields for runs with the PERP polarization
        orientation, while the bottom shows the normalized yields for runs
        with the PARA orientation.} \label{fig:phi1275}
\end{figure}


\begin{figure}
\includegraphics[scale=0.4]{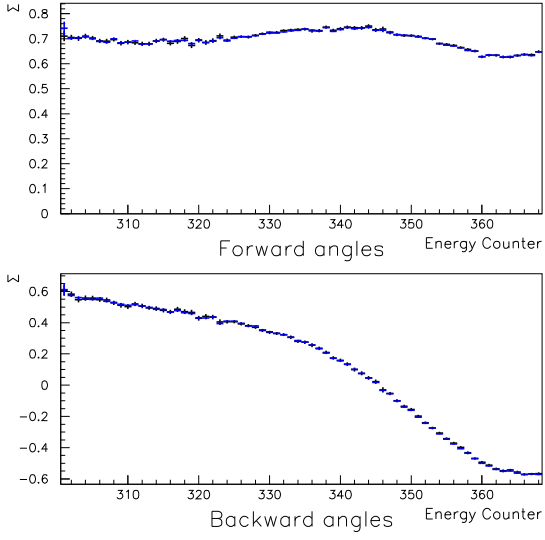}
\caption{(Color online) Rough $\pi^0$ beam asymmetries 
        $\Sigma$ for forward (top panel), 
        and backward (bottom panel) center-of-mass angles determined for 
        polarized photons associated with the indicated photon tagger 
        energy counter. Blue points represent values of $\Sigma$ determined 
        by the Fourier moment method, while black points represent values 
        of $\Sigma$ determined by the $\phi$-bin method averaged over 
        polarization orientations.} \label{fig:sigRough}
\end{figure}


Traditionally, beam asymmetries have been extracted by breaking the azimuthal acceptance of the 
spectrometer into a very large number of bins, extracting the meson yields for those bins, and then fitting 
that distribution of yields with a linear-plus-cosine expression to determine $\Sigma$. 
As a more efficient procedure, 
the beam asymmetries for this experiment were extracted using a Fourier moment analysis of the polar and 
azimuthal scattering angle distributions of the particles detected in CLAS. An overview of the technique
used to extract the beam asymmetries is presented here.

\subsection{Definition of observables}

Meson photoproduction differential cross sections may be written as
\begin{equation}
  \frac{d^2{\sigma}}{d\Omega {dE}_{\gamma}},  \nonumber
\label{eq:dxsec}
\end{equation}
\noindent
where ${d}E_{\gamma}$ is the infinitesimal incident photon energy bin width
and $d\Omega$ is the infinitesimal solid angle element
in which the photoproduced meson is detected.
(All quantities are center-of-mass quantities unless otherwise indicated.)
Practically, however, the cross sections are measured in terms of finite
kinematic bins in photon energy and scattering angle.
Thus, what is measured is more accurately written
\begin{equation}
\Delta\sigma^{i,j,k}  =
{\int_{E_{i-1}}^{E_{i}}}
{\int_{\theta_{j-1}}^{\theta_{j}}}
{\int_{\varphi_{k-1}}^{\varphi_{k}}}
\frac{d^{2}{\sigma}}{d\cos(\theta){dE}_{\gamma}}
{dE}_{\gamma} \sin(\theta) d\theta d\varphi ,
\label{eq:Dxsec}
\end{equation}
\noindent
where the indices $i,j,$ and $k$ denote the individual bin boundaries
for incident photon energy $\Eg$,
scattering polar angle $\theta$,
and azimuthal scattering angle $\varphi$, respectively.

Experimentally, $\Delta\sigma^{i,j,k}$ in Eqn (\ref{eq:Dxsec}) is approximated by the relation
\begin{equation}
  \Delta\sigma^{i,j,k}  \approx
\frac{Y^{i,j,k}}{N_\gamma^i \rho L \epsilon^{i,j,k}} ,
\label{eq:expxsec}
\end{equation}
\noindent
where $Y^{i,j,k}$ is the meson yield in kinematic bin $i,j,k$,
$N_\gamma^i$ is the incident number of photons for bin $i$,
$\rho$ is the target density,
$L$ is the target length,
and $\epsilon^{i,j,k}$ is the detector efficiency for kinematic bin $i,j,k$.


\begin{figure}
\includegraphics[scale=0.4]{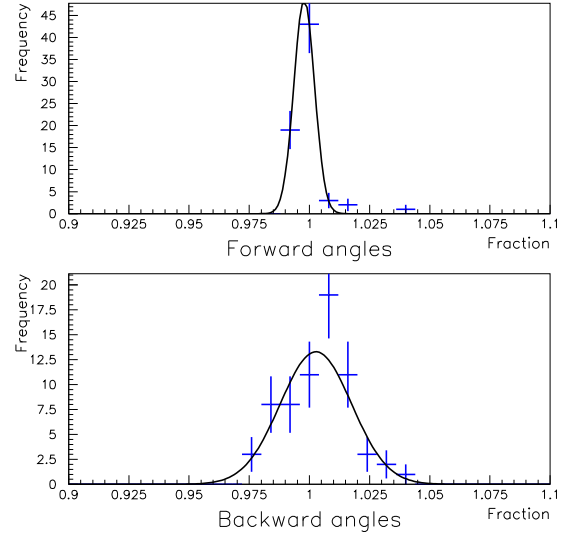}
\caption{Frequency distribution for the rough $\pi^0$
        data shown in Fig. \ref{fig:sigRough} of the ratios of 
        $\Sigma$ from the Fourier moment method divided by $\Sigma$ 
        obtained from the $\phi$-bin method for each photon tagger 
        energy counter.  
        Top panel: Forward center-of-mass angles. 
        Bottom panel: Backward center-of-mass angles. 
        A Gaussian fit to each of these distributions is also shown. }
        \label{fig:sigFrac}
\end{figure}


As defined above, the photon beam polarization orientations used for the running period had either
the electric field vector parallel to the Hall B floor
(with the degree of polarization denoted by $P_\|$)
or perpendicular to the floor (with the corresponding degree of polarization $P_\bot$).
The differential cross sections for the various incident photon beam polarizations
are labeled in the following fashion:

(a) for perpendicular beam polarization,
\begin{equation}
  \frac{d{\sigma}_\bot}{d\Omega} =  \frac{d{\sigma}_a}{d\Omega} [ 1 + P_\bot \Sigma \cos(2 \varphi) ]
\label{eq:dperpxsec}
\end{equation}

(b) for parallel beam polarization,
\begin{equation}
  \frac{d{\sigma}_\|}{d\Omega} =  \frac{d{\sigma}_a}{d\Omega} [ 1 - P_\| \Sigma \cos(2 \varphi) ] .
\label{eq:dparaxsec}
\end{equation}
\noindent
The unpolarized differential cross sections for a given reaction
extracted from the amorphous carbon
radiator is 
\begin{equation}
  \frac{d{\sigma}_a}{d\Omega} =  \frac{1}{2} \left( 
\frac{d{\sigma}_{\bot}}{d\Omega} + \frac{d{\sigma}_{\|}}{d\Omega}
\right) .
\label{eq:amoxsec}
\end{equation}

\subsection{Azimuthal moments for determining $\Sigma$}

For this analysis, two additional  $\varphi$-dependent quantities are defined:

\begin{equation}
f^{i,j}(\varphi)  =  \rho L
{\int_{E_{i-1}}^{E_{i}}}
{\int_{\theta_{j-1}}^{\theta_{j}}}
\epsilon(E_\gamma,\theta,\varphi)
\frac{d^{2}{\sigma}}{d\Omega{dE}_{\gamma}}
{dE}_{\gamma} \sin\theta d\theta \nonumber
\end{equation}
\noindent
and
\begin{equation}
  \tilde{Y}^{i,j}(\varphi) =  \frac{Y^{i,j} (\varphi )}{N_\gamma^i},
\label{eq:Ydef}
\end{equation}
\noindent
where the former defines the normalized yield density with respect to azimuthal angle, and
the latter is simply the normalized yield for a given $\varphi$ for bin $i,j$. 
These normalized yields may be further labeled by the photon beam polarization as 
$\tilde{Y}^{i,j}_a $, $\tilde{Y}^{i,j}_\bot$, and
$\tilde{Y}^{i,j}_\| $, which would be the yield for the amorphous target, the yield 
with the perpendicularly polarized beam, and the yield with 
the parallel polarized beam, respectively.  

Using the appropriate
definitions given in
Eqn~(\ref{eq:dperpxsec}), (\ref{eq:dparaxsec}), and (\ref{eq:amoxsec}),
the three normalized yields in  Eqs.~(\ref{eq:Ydef}) may be written as  
\begin{equation}
  \tilde{Y}^{i,j}_a =  {\int_{0}^{2\pi}} f^{i,j}_a(\varphi) d\varphi
\label{eq:Yfa}
\end{equation}
\begin{equation}
  \tilde{Y}^{i,j}_\bot =  {\int_{0}^{2\pi}} f^{i,j}_a(\varphi)[ 1 + P_\bot^i \Sigma^{i,j} \cos(2 \varphi) ] d\varphi
\label{eq:Yfperp}
\end{equation}
\begin{equation}
  \tilde{Y}^{i,j}_\| =  {\int_{0}^{2\pi}} f^{i,j}_a(\varphi)[ 1 - P_\|^i \Sigma^{i,j} \cos(2 \varphi) ] d\varphi  .
\label{eq:Yfpara}
\end{equation}

With these definitions, {\emph{all}} yields are now expressed
in terms of various integrals involving the normalized yield density \protect{$\ffunc$},
which is the normalized yield density for the {\emph{amorphous}} carbon radiator.
This function includes all physics effects
modulated by the experimental acceptance $\epsilon^{i,j}$.
The quantity $\ffunc$ is then expanded in a Fourier series as
\begin{equation}
f^{i,j}_a(\varphi)  =  a_0 + \sum_{n=1}^{\infty} [ a_n \cos(n\varphi) + b_n \sin(n\varphi) ]  ,
\label{eq:fseries}
\end{equation}
\noindent
where each term of the series represents the $n^{\rm{th}}$ Fourier moment of $f^{i,j}_a$.

As usual, one can construct, event by event, a missing mass histogram for
the reaction $\gamma p \rightarrow p X$ or $\gamma p \rightarrow \pi^+ X$.
In the approach used here,
moment-$n$ histograms
are constructed by taking each event in the $\gamma p \rightarrow p X$ or $\gamma p 
\rightarrow \pi^+ X$ missing mass histogram and
weighting each event by the value of $\cos(n\varphi)$ corresponding to that event
for the various yields in (\ref{eq:Yfa})-(\ref{eq:Yfpara}).

Of particular importance are the moment-2 histograms
\begin{eqnarray}
\tilde{Y}^{i,j}_{\bot 2} & \equiv & {\int_{0}^{2\pi}} f^{i,j}_\bot (\varphi) \cos(2 \varphi) d\varphi \nonumber \\
           & =      & {\int_{0}^{2\pi}} f^{i,j}_a(\varphi)  \cos(2 \varphi) d\varphi  \nonumber \\
& &+ P_\bot \Sigma^{i,j} {\int_{0}^{2\pi}} f^{i,j}_a (\varphi)  \cos^2(2 \varphi) d\varphi \label{eq:hperp2} \\
\nonumber
\end{eqnarray}
\noindent
and
\begin{eqnarray}
\tilde{Y}^{i,j}_{\| 2} & \equiv & {\int_{0}^{2\pi}} f^{i,j}_\| (\varphi) \cos(2 \varphi) d\varphi \nonumber \\
         &  =     & {\int_{0}^{2\pi}} f^{i,j}_a(\varphi)  \cos(2 \varphi) d\varphi \nonumber \\
& & - P_\| \Sigma^{i,j} {\int_{0}^{2\pi}} f^{i,j}_a(\varphi)  \cos^2(2 \varphi) d\varphi . \label{eq:hpara2}
 \\
\nonumber
\end{eqnarray}

Subtracting Eqn~(\ref{eq:hpara2}) from (\ref{eq:hperp2}) yields
\begin{eqnarray}
\tilde{Y}^{i,j}_\Sigma &\equiv &\tilde{Y}^{i,j}_{\bot 2} - \tilde{Y}^{i,j}_{\| 2} \label{eq:hsigmadef} \\
         & = &(P_\bot^i + P_\|^i) \Sigma^{i,j} {\int_{0}^{2\pi}} f^{i,j}_a (\varphi)  \cos^2(2 \varphi) d\varphi . \nonumber \\
\nonumber
\end{eqnarray}
Using the double-angle relationship for the cosine of an angle, and keeping
the Fourier series definition of $\ffunc$ \ from Eqn~(\ref{eq:fseries}) in mind,
this can be rewritten as
\begin{eqnarray}
\tilde{Y}^{i,j}_\Sigma & = & 
\pi (P_\bot^i + P_\|^i) \Sigma^{i,j} \left( a_0 + \frac {a_4}{2} \right) . \label{eq:hsigma} \\ \nonumber
\end{eqnarray}

The polarization varies continuously during the course of a
typical data run owing to fluctuations in the
relative alignment of the incident electron beam and the diamond.
Thus, the polarization must be determined
continuously during a data run
so that a photon-flux-weighted equivalent
value of polarization for each run can be determined.
The values of $P_\|$ and $P_\bot$ used in these equations
are assumed to be these photon-flux-weighted values.

With these photon-flux-weighted equivalent values for the
polarization $P_\|$ and $P_\bot$
and the histogram defined by Eqn~(\ref{eq:hsigma}),
one only needs the Fourier coefficients $a_0$ and $a_4$
for $\ffunc$ \ to determine $\tilde{Y}^{i,j}_\Sigma$.


\begin{figure*}[ht]
\includegraphics[height=1\textwidth, angle=90]{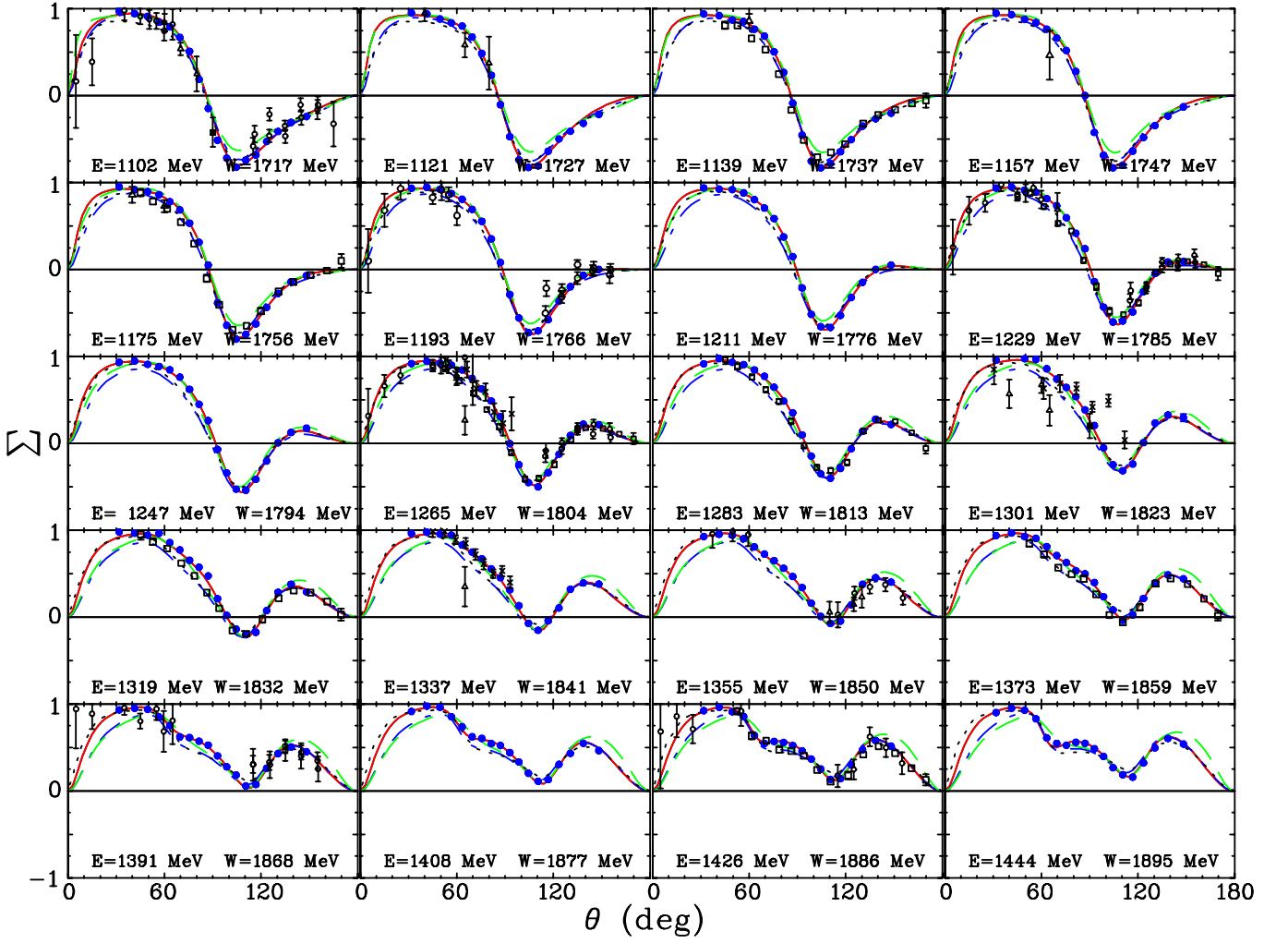}
\vspace*{0.2cm}
\caption{(Color online) Beam asymmetry $\Sigma$ for $\vec{\gamma}
        p\to\pi^0p$ at $E_\gamma = 1102 - 1444$~MeV versus pion
        center-of-mass production angle.  
        Photon energy is indicated by $E$, while the center-of-mass
        total energy is indicated by $W$.
        Solid (dash-dotted) lines correspond to the SAID DU13 
        (CM12~\protect\cite{cm12}) solution. 
        Dashed (short-dashed) lines give the MAID07~\protect\cite{Maid07} 
        (BG2011-02 BnGa~\protect\cite{BnGa}) predictions.  
        Experimental data are from the current (filled circles),
        Bonn \cite{CBELSA1,CBELSA2} (open circles),
        Yerevan \cite{Yerevan1,Yerevan2,Yerevan3,Yerevan4,Yerevan5,Yerevan6} 
                        (open triangle),
        GRAAL \cite{GRAAL1} (open square),
        CEA \cite{CEA} (filled square),
        DNPL \cite{DNPL1,DNPL2} (cross),
        and LEPS \cite{LEPS} (asterisk).
        Plotted uncertainties are statistical. 
        The plotted points from previously published experimental 
        data~\protect\cite{SAID} 
        are those data points within 3~MeV of the photon energy indicated 
        on each panel. \label{fig:g1}}
\end{figure*}


\begin{figure*}[th]
\includegraphics[height=1\textwidth, angle=90]{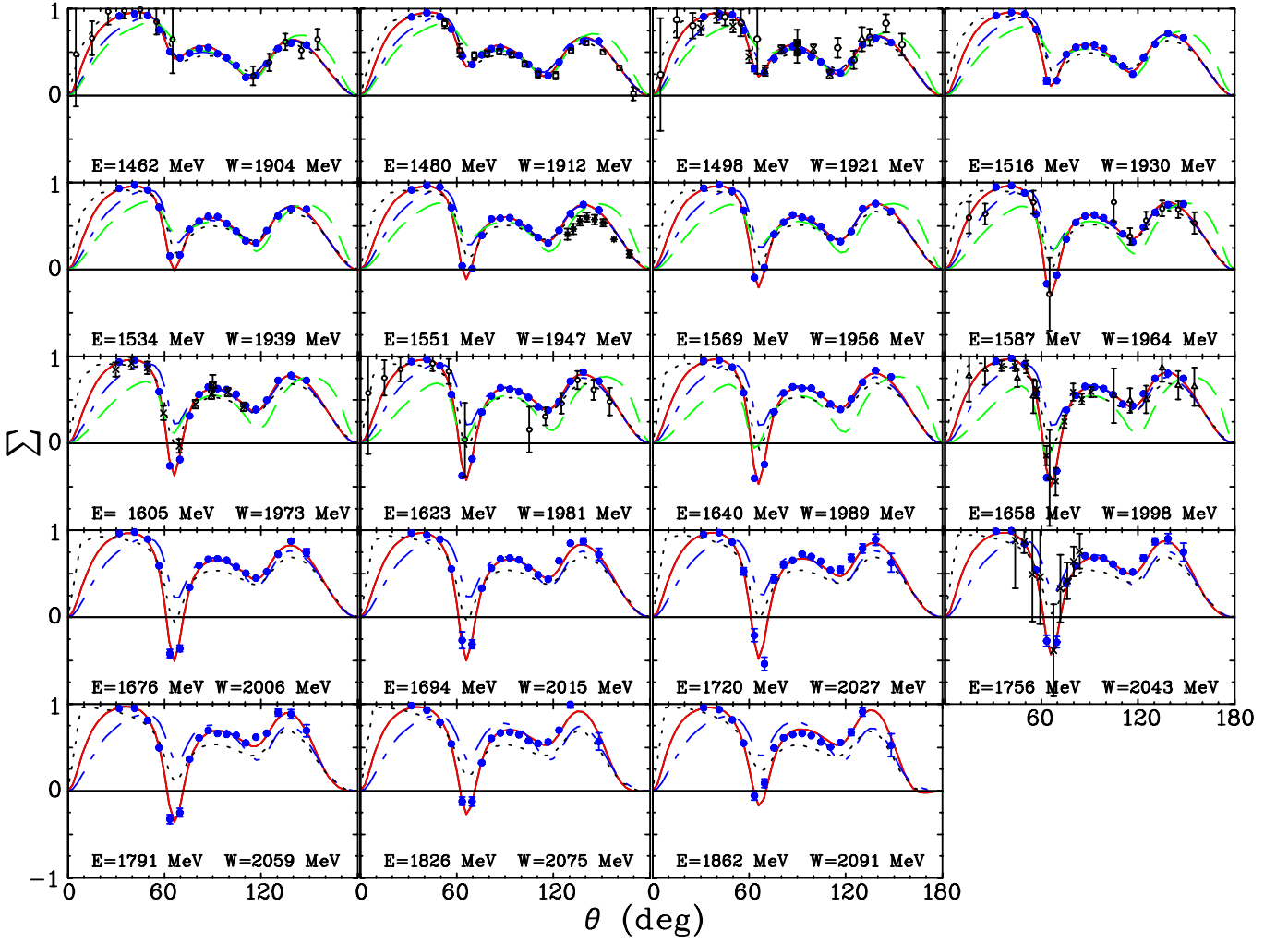}
\caption{(Color online) Beam asymmetry $\Sigma$ for $\vec{\gamma}
        p\to\pi^0p$ at $E_\gamma = 1462 - 1862$~MeV versus pion
        center-of-mass production angle.  The photon energy is
        shown as $E$. Notation as in Fig.~\protect\ref{fig:g1}.
        \label{fig:g2}}
\end{figure*}


\begin{figure*}[th]
\includegraphics[height=0.85\textwidth, angle=90]{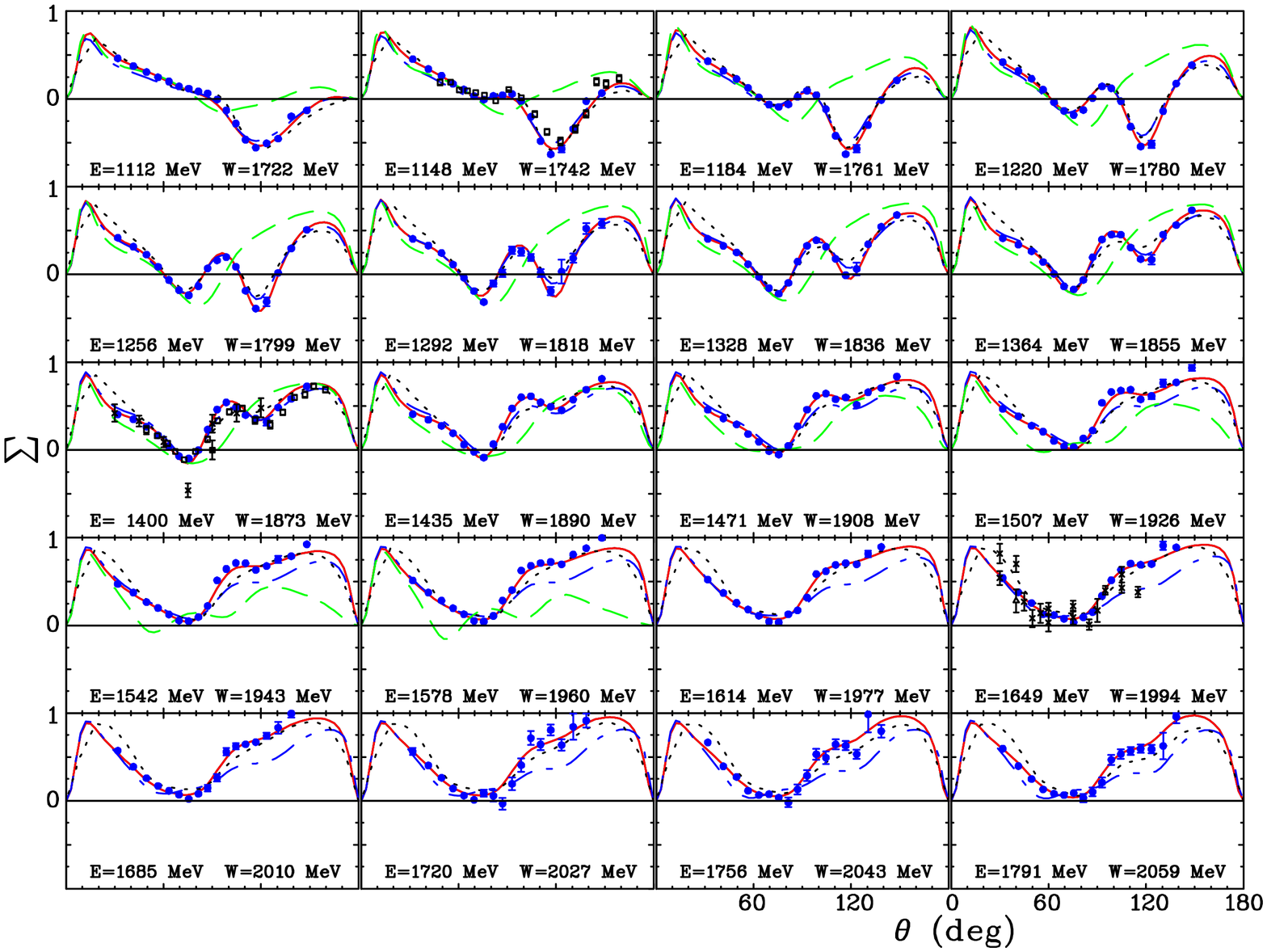}\hfill\\
\includegraphics[height=0.28\textwidth, angle=90]{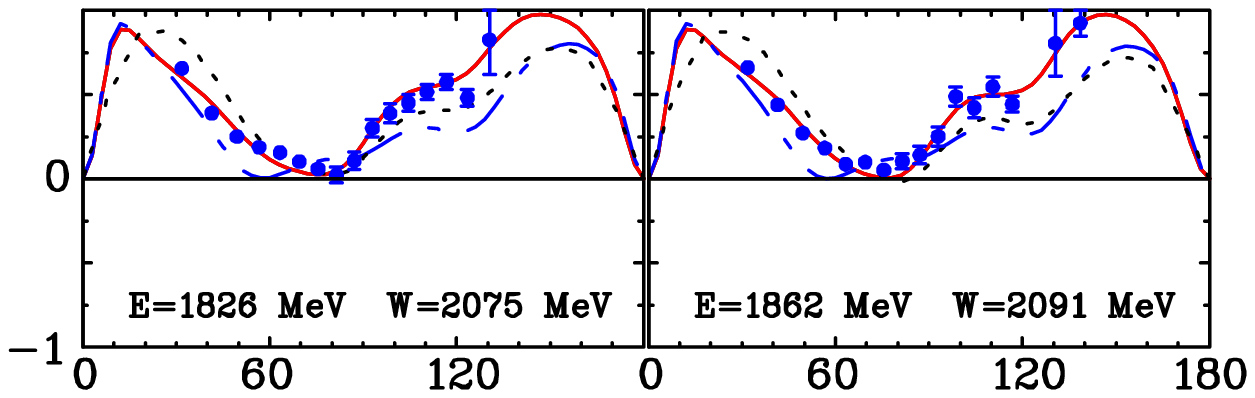}
\vspace{0.3in}
\caption{
(Color online) Beam asymmetry $\Sigma$ for $\vec{\gamma} 
        p\to\pi^+n$ at $E_\gamma = 1112 - 1862$~MeV versus pion
        center-of-mass production angle. The photon energy is
        shown as $E$. 
        Solid (dash-dotted) lines correspond to the SAID DU13 
        (CM12~\protect\cite{cm12}) solution. 
        Dashed (short-dashed) lines give the MAID07~\protect\cite{Maid07} 
        (BG2011-02 BnGa~\protect\cite{BnGa}) predictions.  
        Experimental data are from the current (filled circles),
        GRAAL \cite{GRAAL2} (open square),
        Yerevan \cite{Yerevan7} (open triangle),
        CEA \cite{CEA} (filled square),
        and DNPL \cite{DNPL2} (cross).
        Plotted uncertainties are
        statistical. The plotted points from previously published
        experimental data~\protect\cite{SAID} are those data
        points within 3~MeV of the photon energy indicated on
        each panel. \label{fig:g3}}
\end{figure*}


\begin{figure*}[th]
\includegraphics[height=1\textwidth, angle=90]{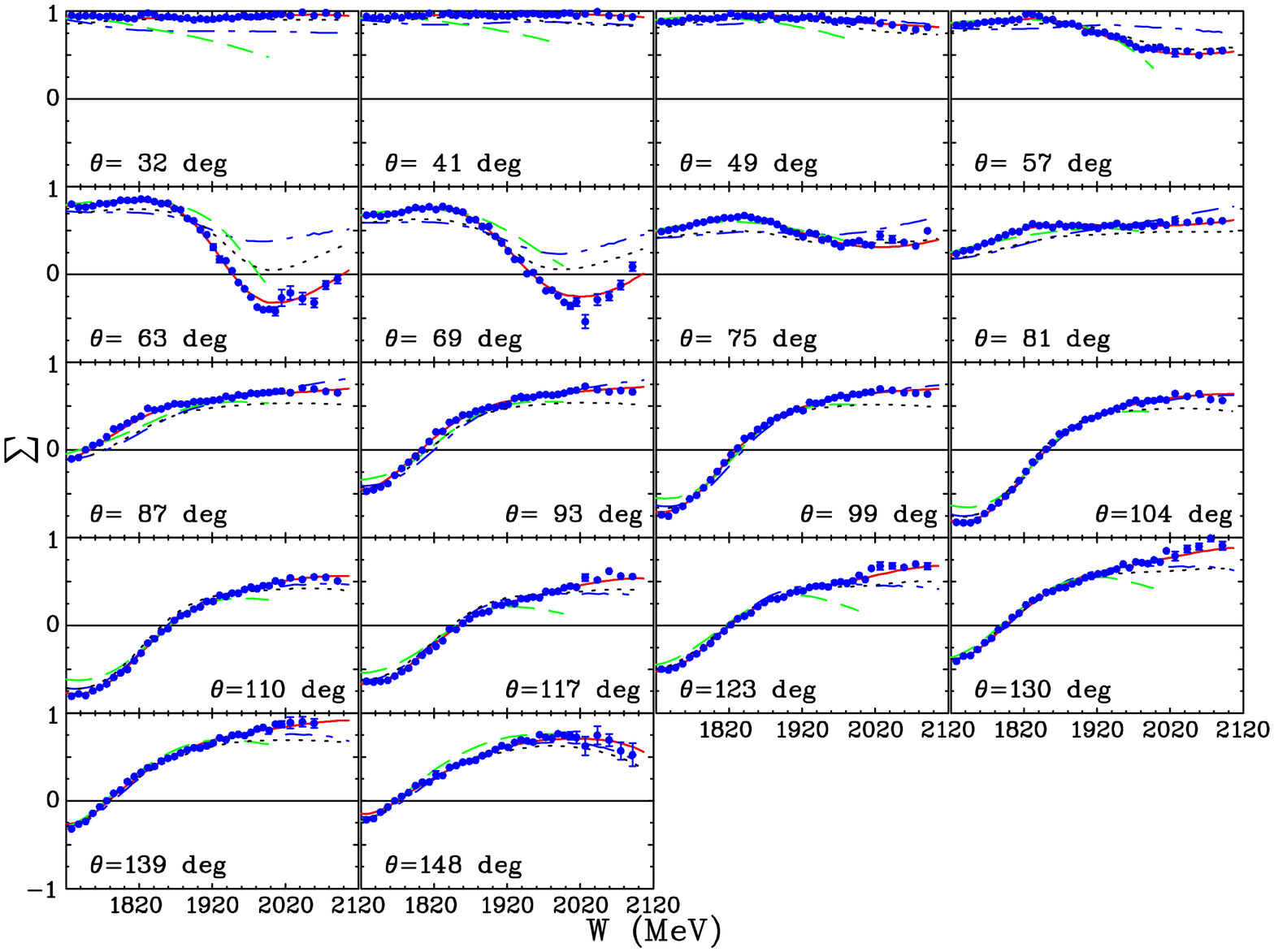}
\caption{(Color online) Fixed angle excitation functions of the
        beam asymmetry $\Sigma$ for $\vec{\gamma} p\to\pi^0p$.
        The pion center-of-mass production angle is shown.
        Notation as in Fig.~\protect\ref{fig:g1}. \label{fig:g4}}
\end{figure*}


\begin{figure*}[th]
\includegraphics[height=1\textwidth, angle=90]{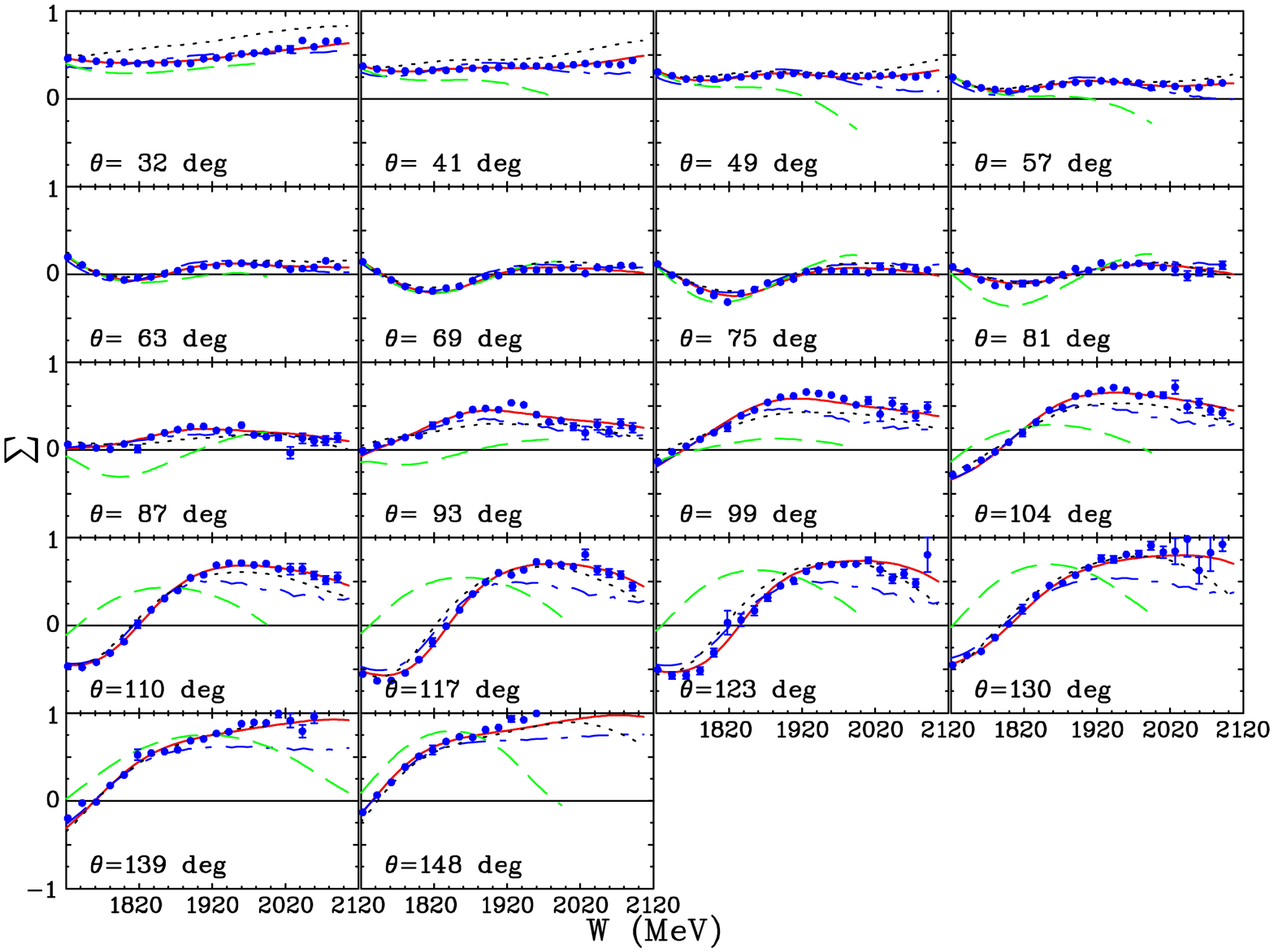}
\caption{(Color online) Fixed angle excitation functions of the
        beam asymmetry $\Sigma$ for $\vec{\gamma} p\to\pi^+n$.
        The pion center-of-mass production angle is shown.
        Notation as in Fig.~\protect\ref{fig:g3}. \label{fig:g5}}
\end{figure*}


\begin{figure*}[th]
\centerline{
\includegraphics[height=0.35\textwidth, angle=90]{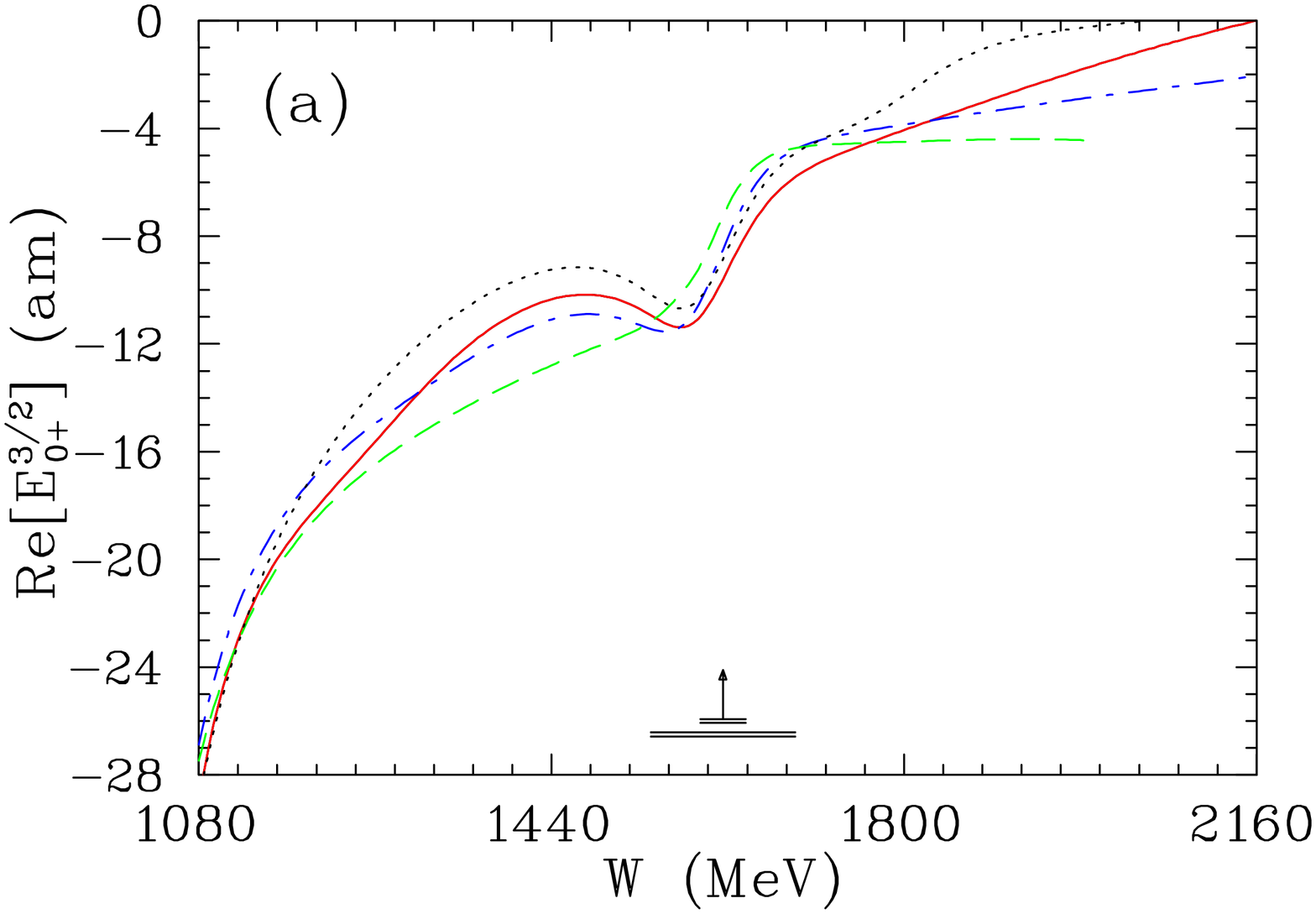}
\includegraphics[height=0.35\textwidth, angle=90]{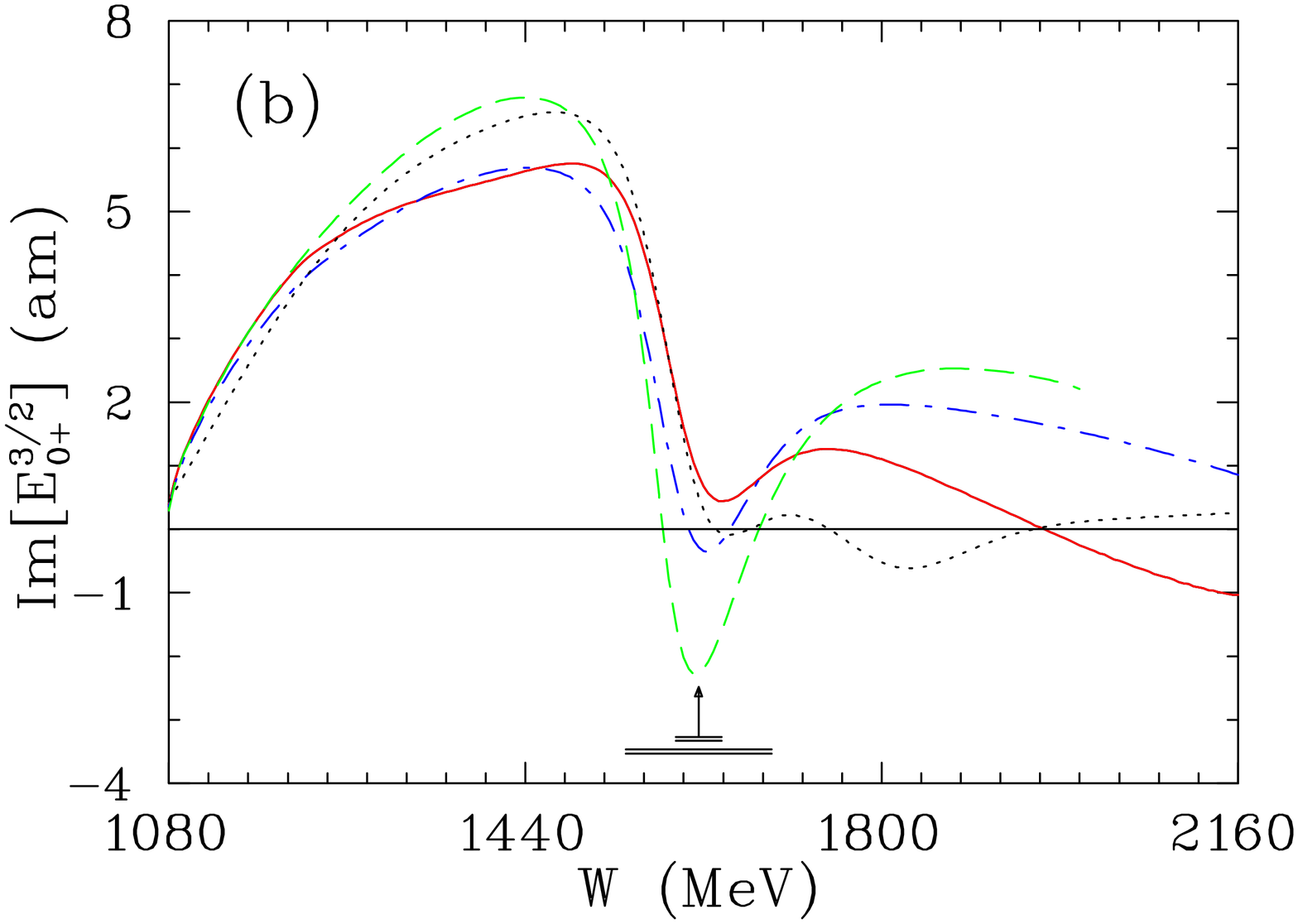}}
\centerline{
\includegraphics[height=0.35\textwidth, angle=90]{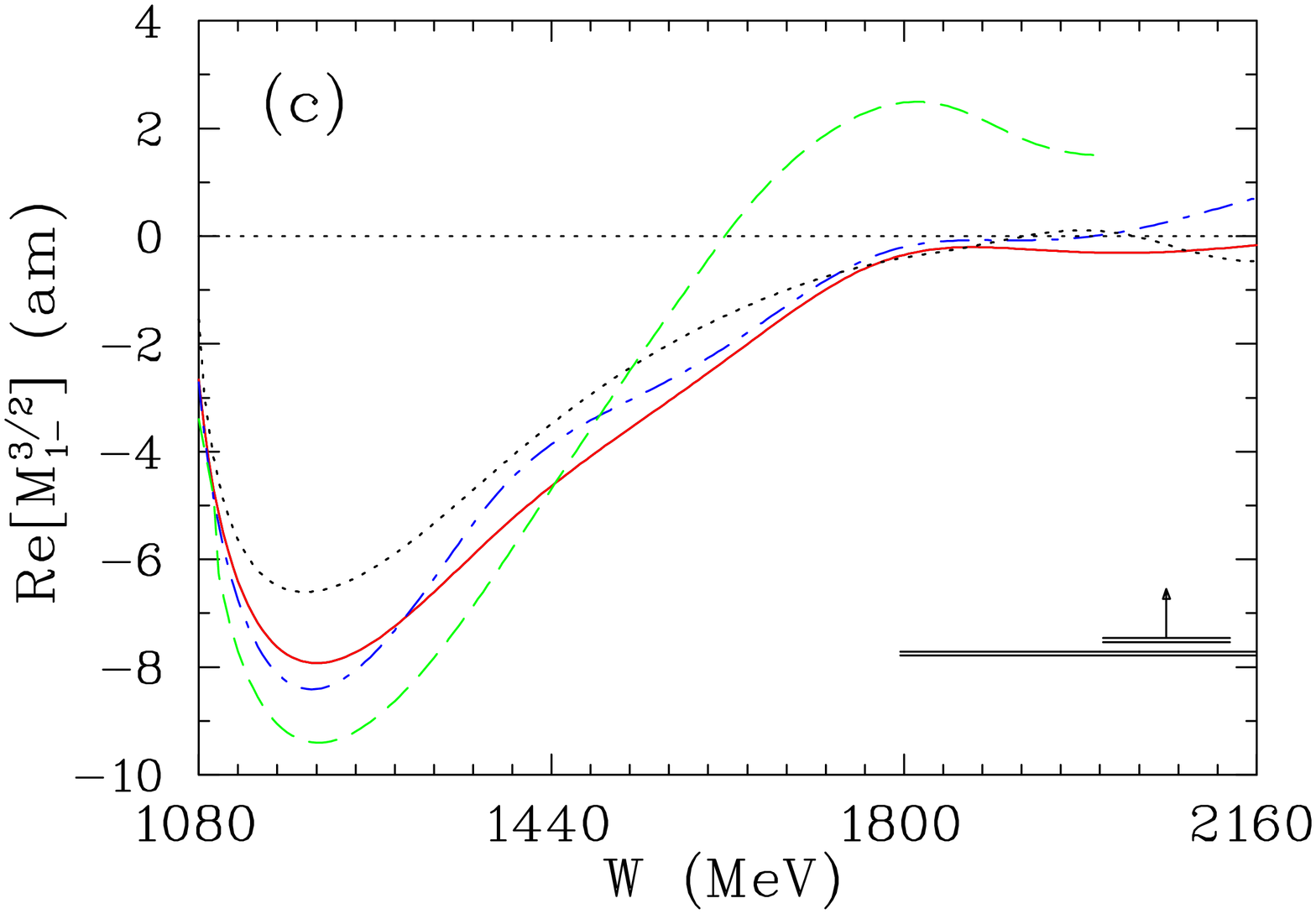}
\includegraphics[height=0.35\textwidth, angle=90]{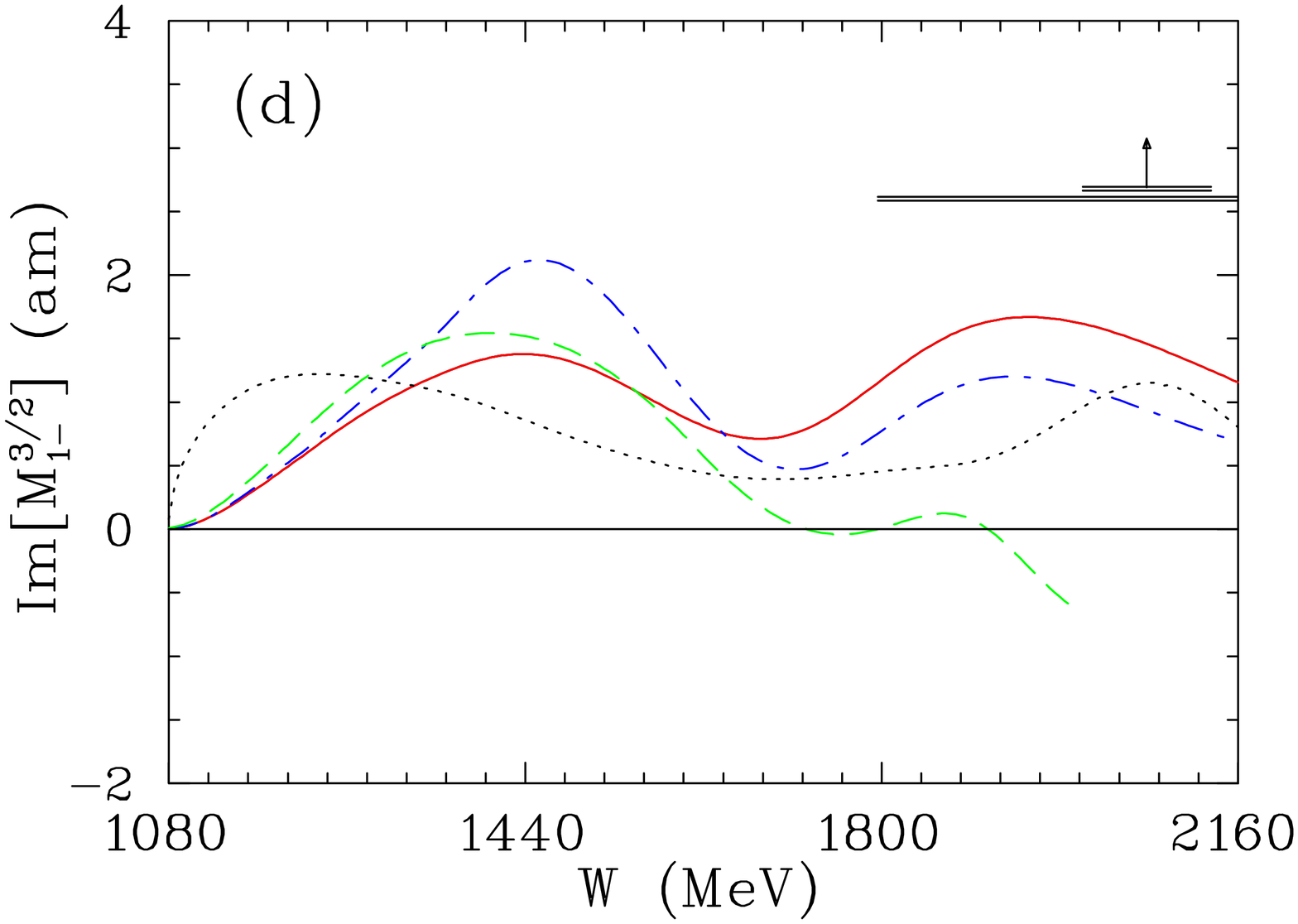}}
\centerline{
\includegraphics[height=0.35\textwidth, angle=90]{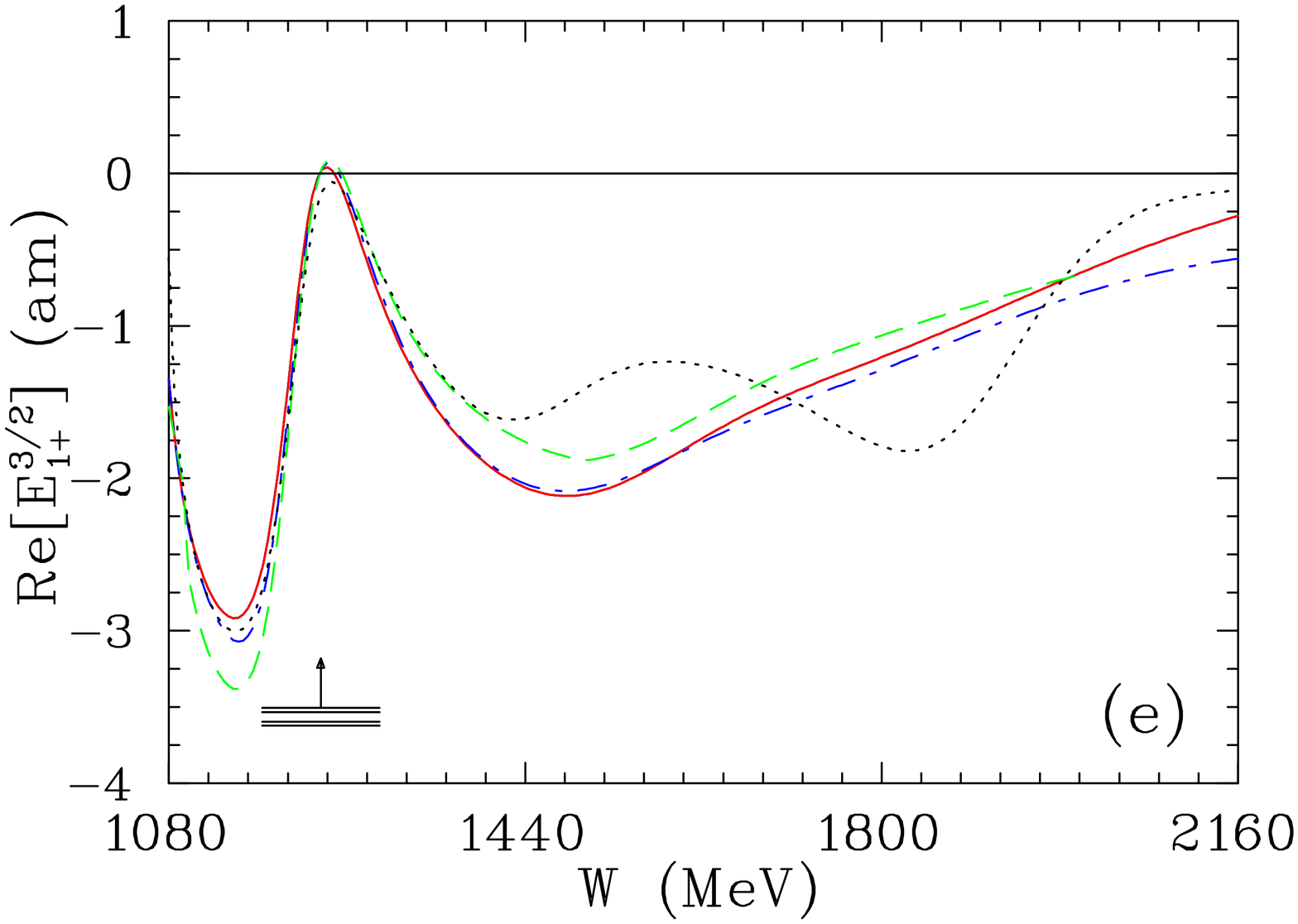}
\includegraphics[height=0.35\textwidth, angle=90]{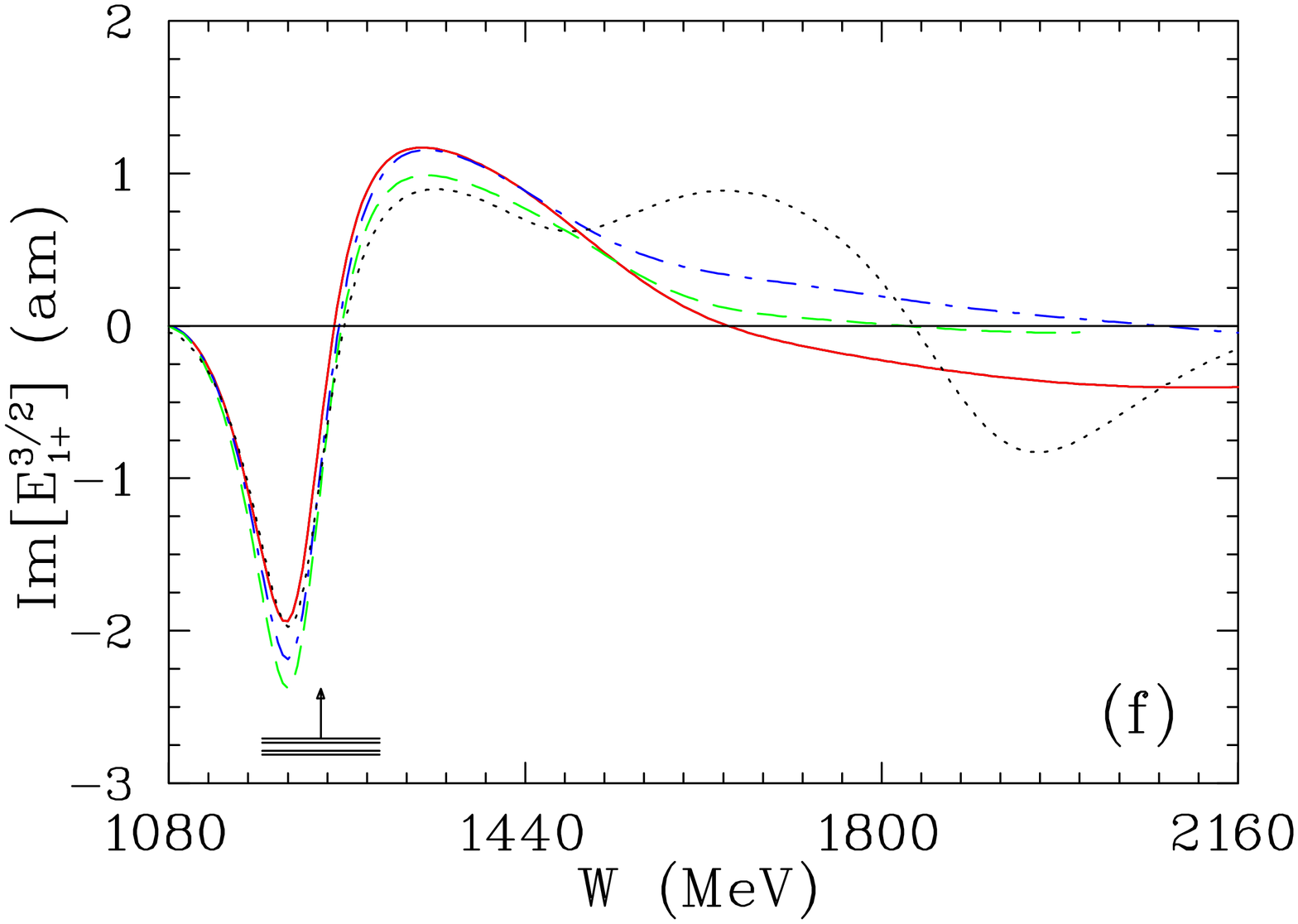}}
\centerline{
\includegraphics[height=0.35\textwidth, angle=90]{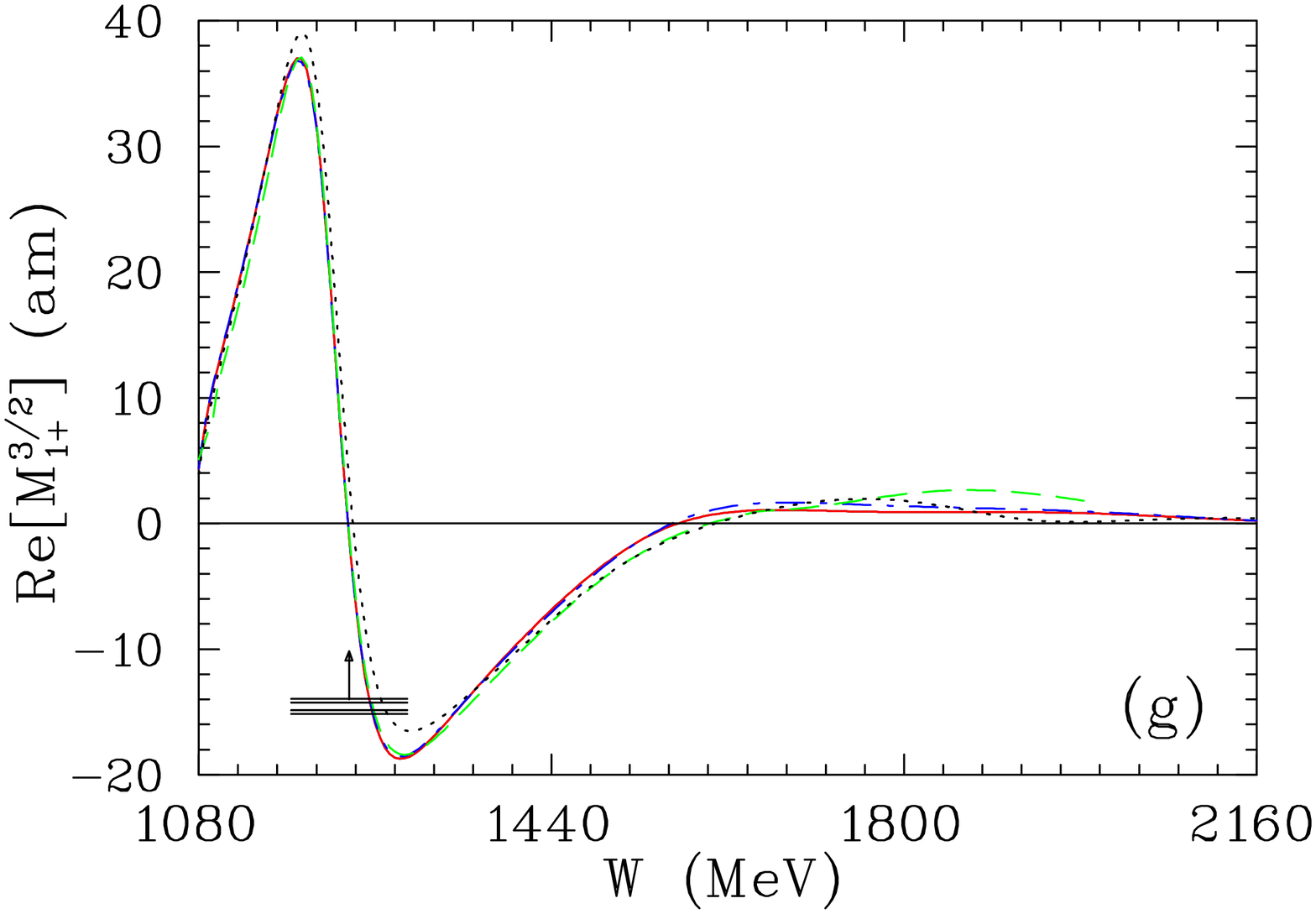}
\includegraphics[height=0.35\textwidth, angle=90]{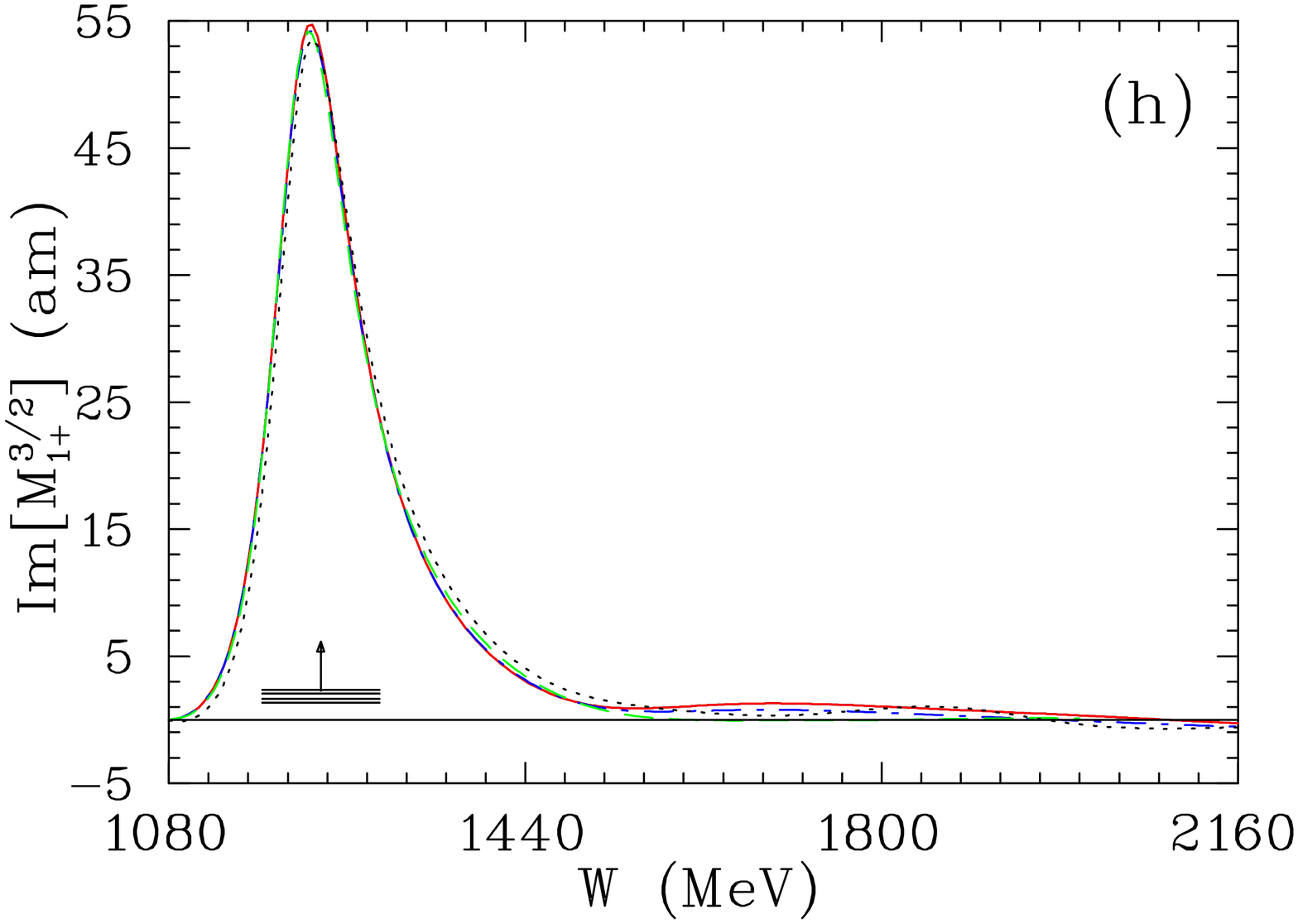}}
\caption{(Color online) Isospin $I$=3/2 multipole amplitudes from
        threshold to $W$ = 2.16~GeV ($E_{\gamma}$ = 2.02~GeV) for 
        $l = 0,1$.
        Solid (dash-dotted) lines correspond to the SAID DU13
        (CM12~\protect\cite{cm12}) solution.  Dashed (short-dashed)
        lines give MAID07~\protect\cite{Maid07}, which terminates
        at $W$=2~GeV (BG2011-02 BnGa solution~\protect\cite{BnGa}).
        Vertical arrows indicate resonance energies $W_R$ and
        horizontal bars show full ($\Gamma$) and partial
        ($\Gamma_{\pi N}$) widths associated with the SAID $\pi N$
        solution SP06~\protect\cite{sp06}. \label{fig:g7}}
\end{figure*}


\begin{figure*}[th]
\centerline{
\includegraphics[height=0.35\textwidth, angle=90]{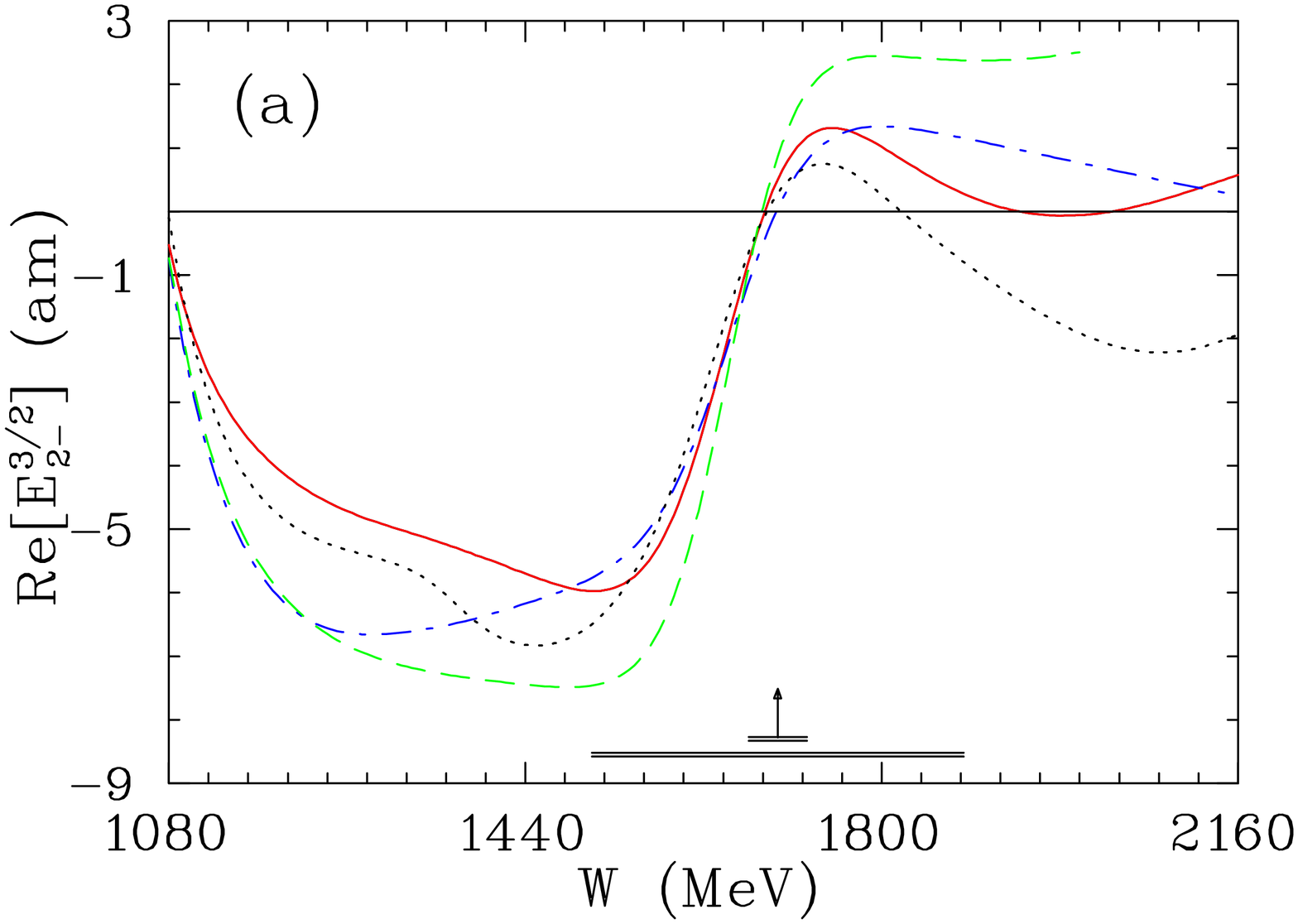}
\includegraphics[height=0.35\textwidth, angle=90]{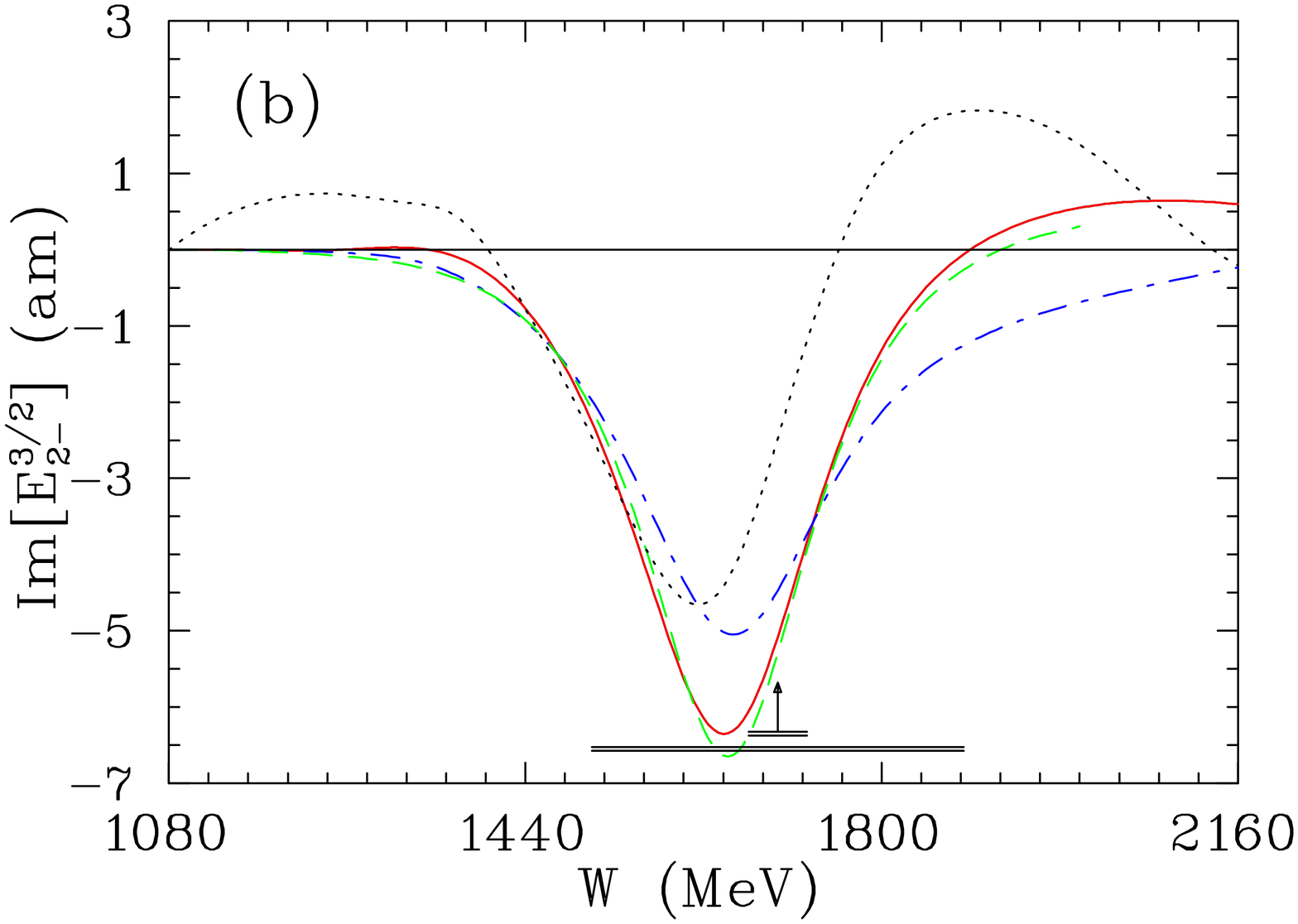}}
\centerline{
\includegraphics[height=0.35\textwidth, angle=90]{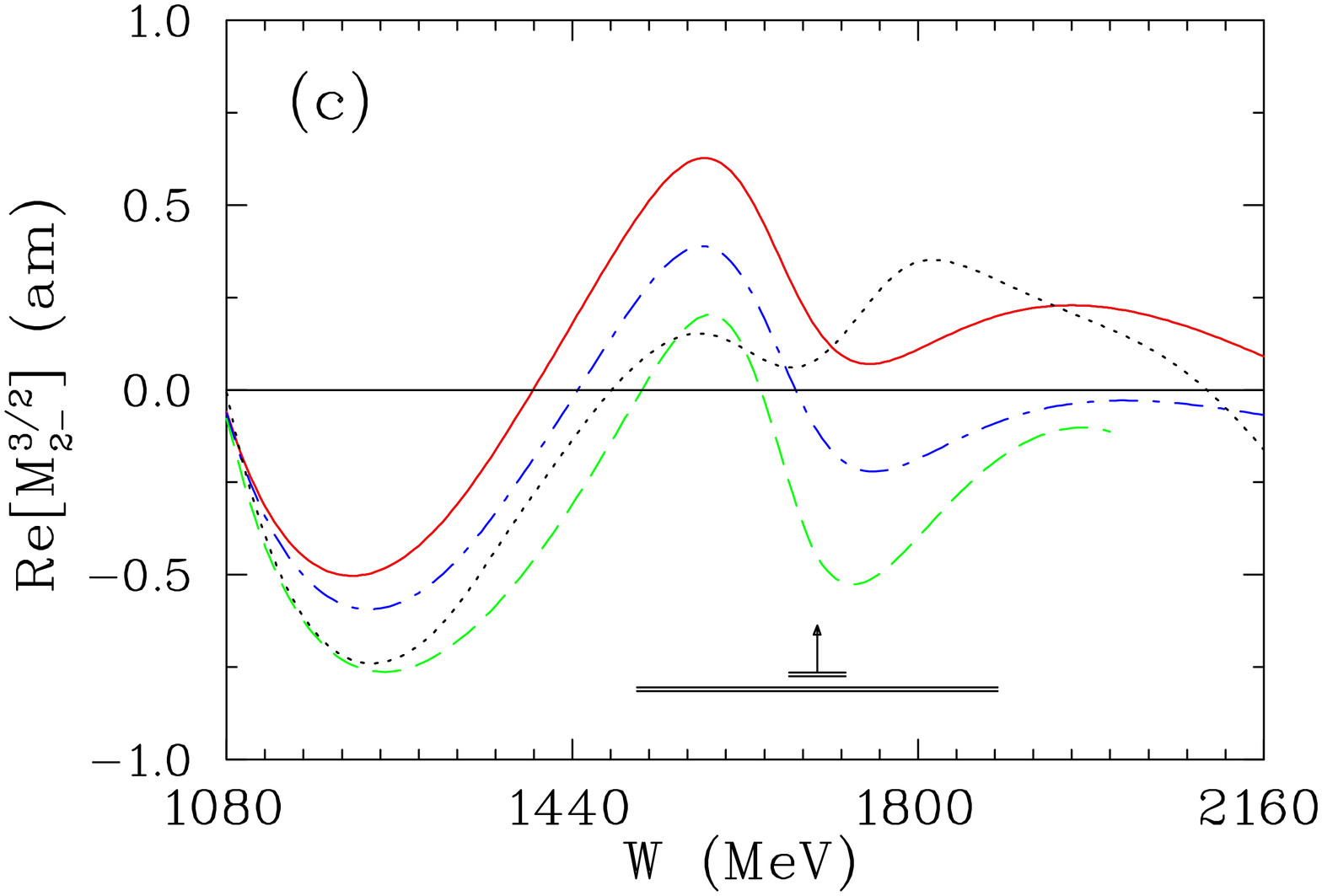}
\includegraphics[height=0.35\textwidth, angle=90]{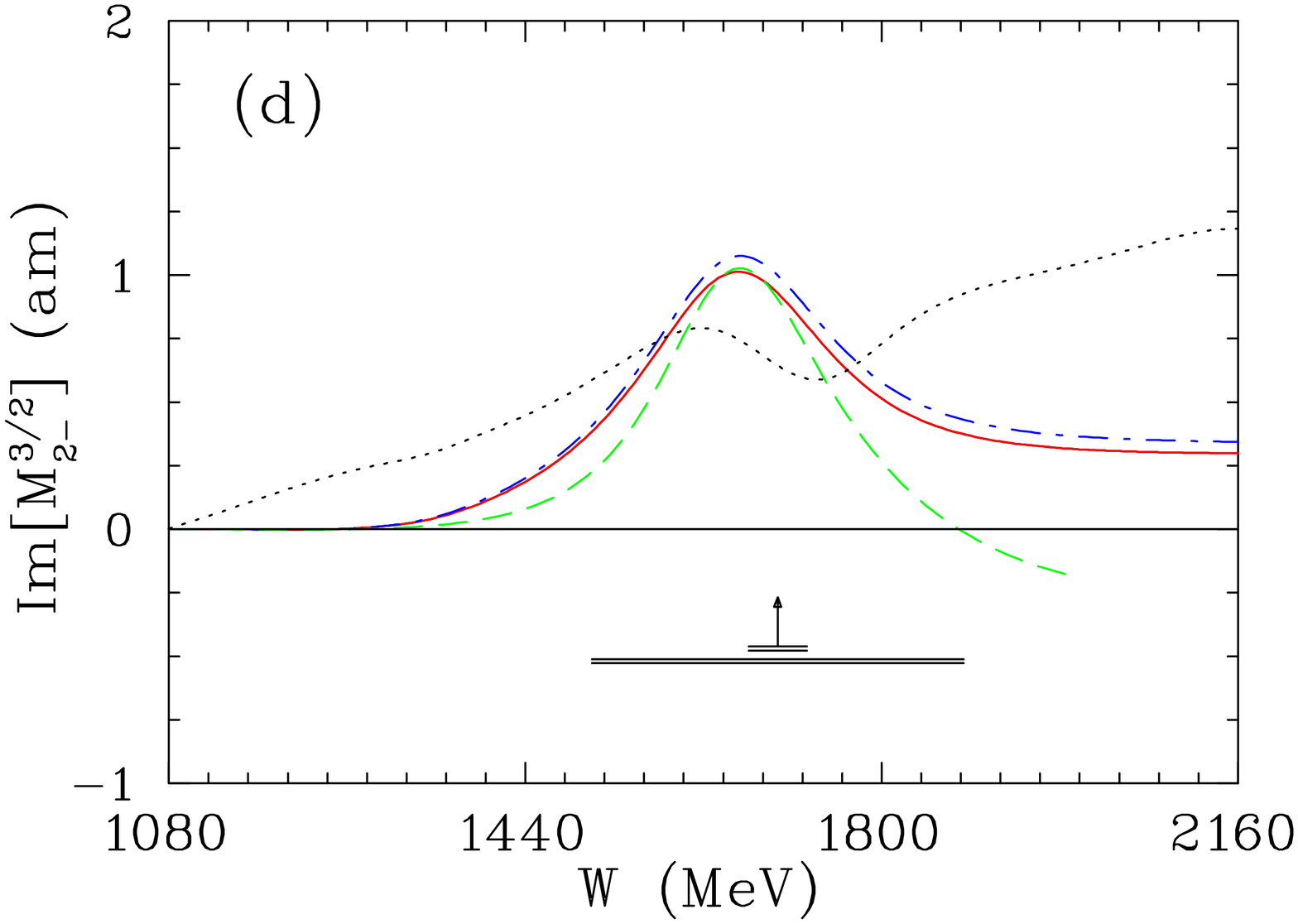}}
\centerline{
\includegraphics[height=0.35\textwidth, angle=90]{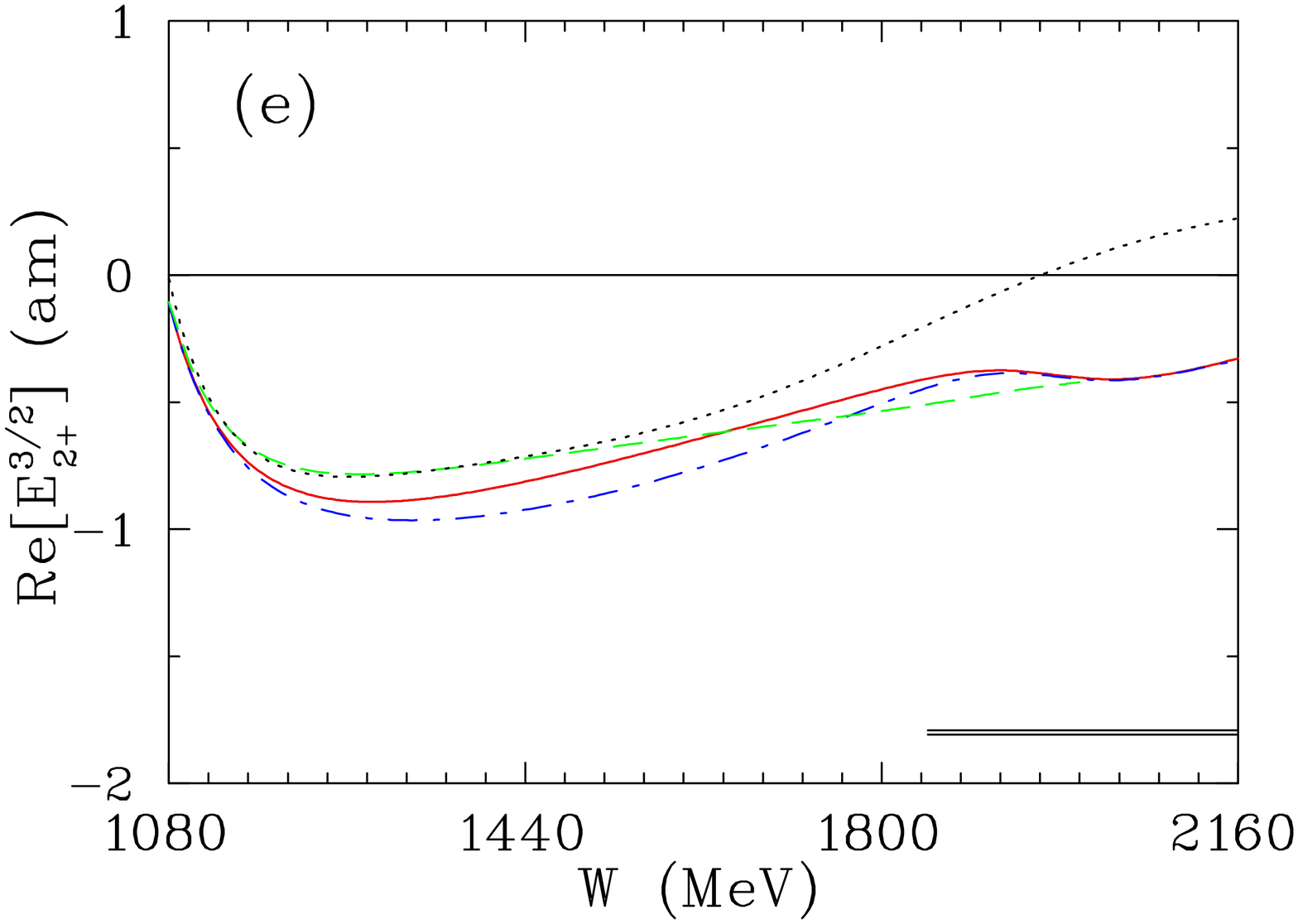}
\includegraphics[height=0.35\textwidth, angle=90]{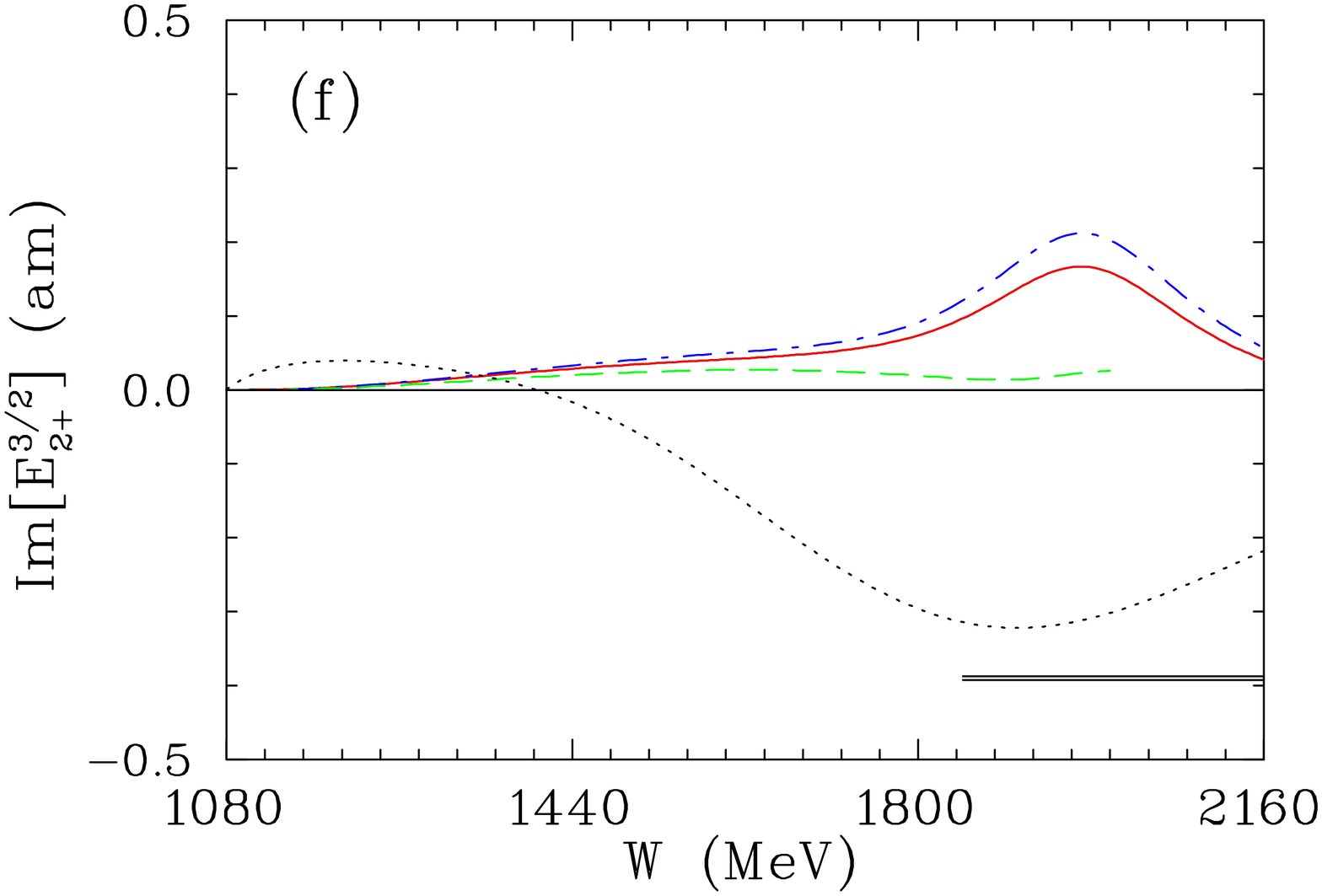}}
\centerline{
\includegraphics[height=0.35\textwidth, angle=90]{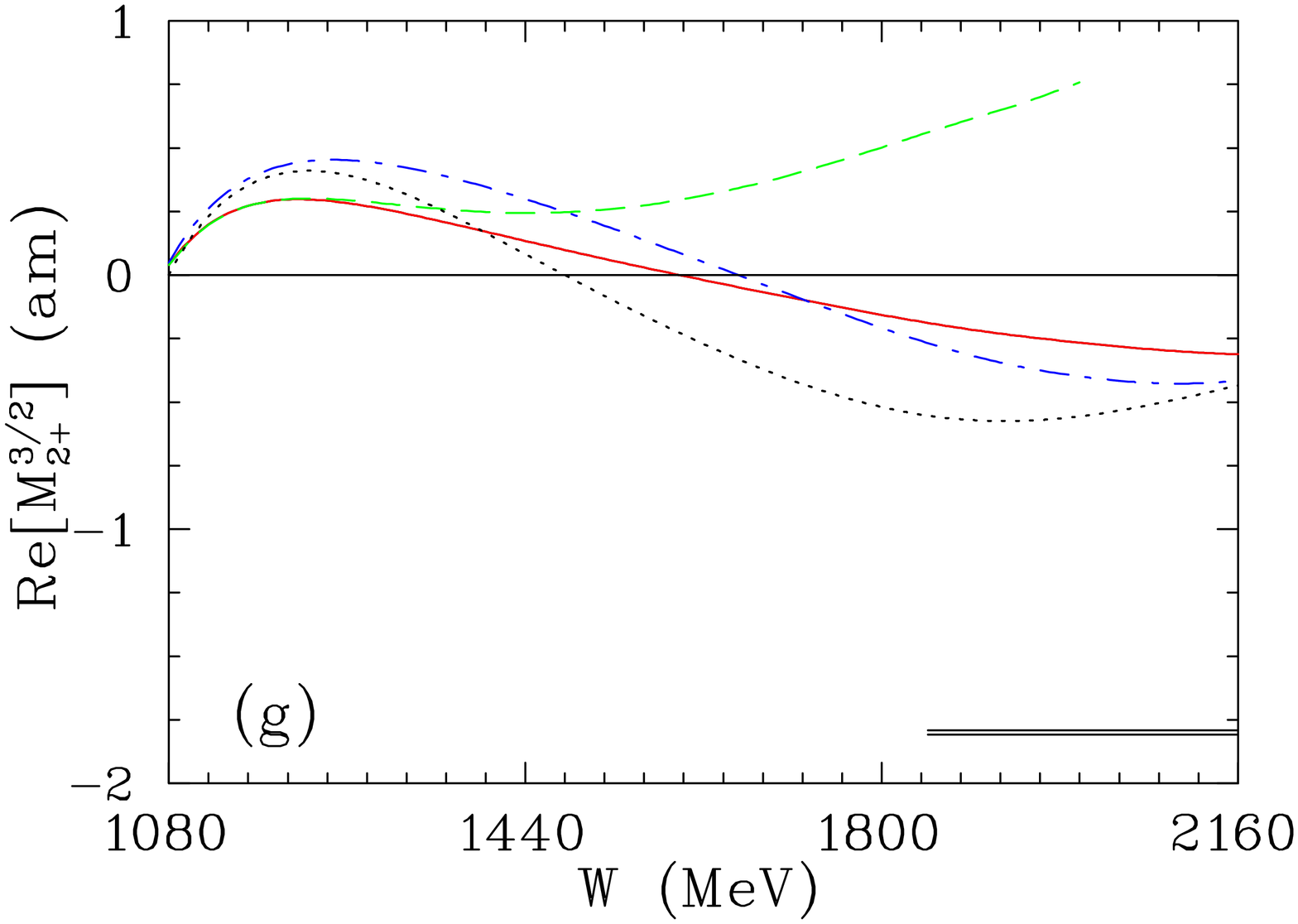}
\includegraphics[height=0.35\textwidth, angle=90]{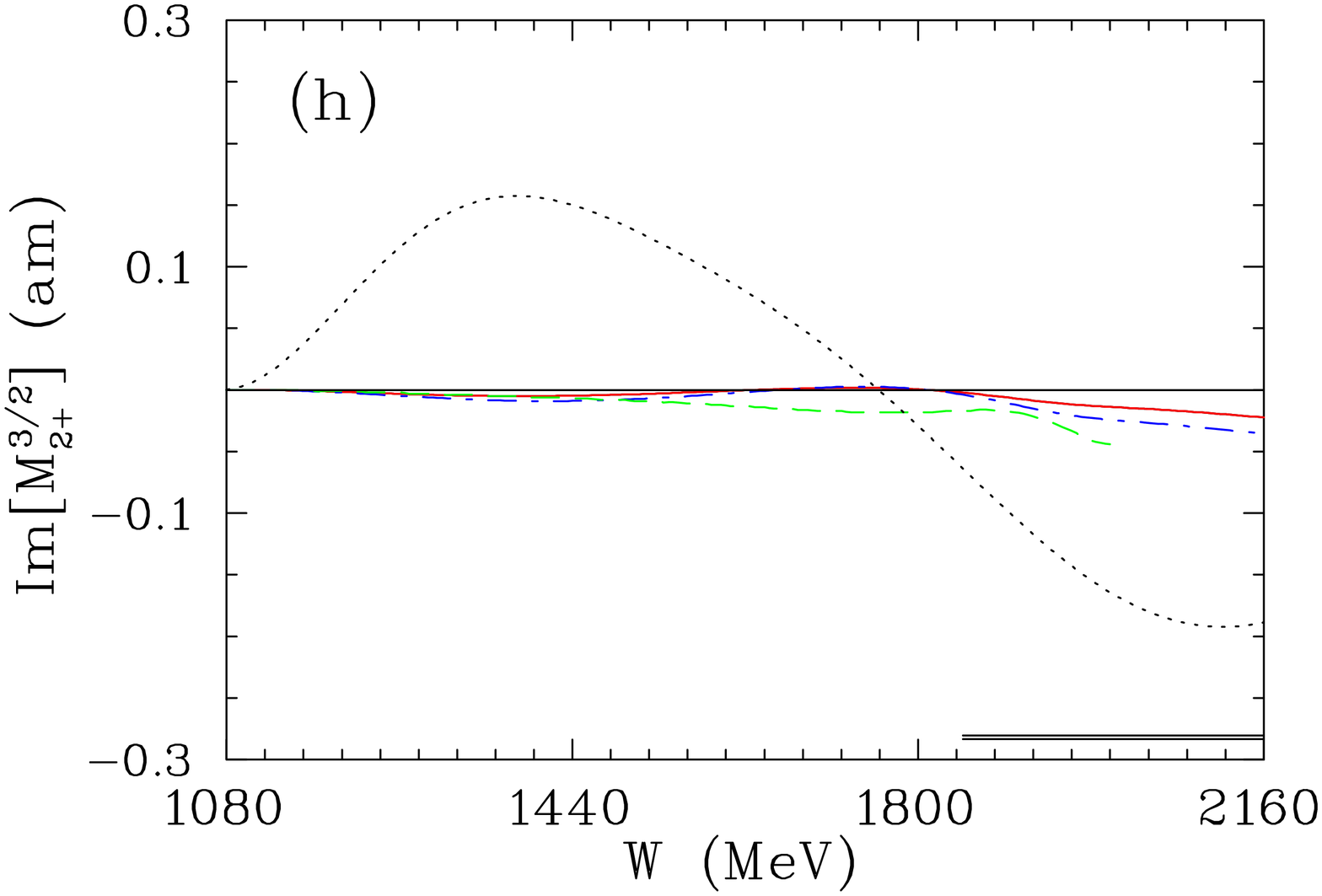}}
\caption{(Color online) Isospin $I$=3/2 multipole amplitudes from
        threshold to $W$ = 2.16~GeV ($E_{\gamma}$ = 2.02~GeV) for 
        $l = 2$.
        Notation as in Fig.~\protect\ref{fig:g7}. \label{fig:g8}}
\end{figure*}


\begin{figure*}[th]
\centerline{
\includegraphics[height=0.35\textwidth, angle=90]{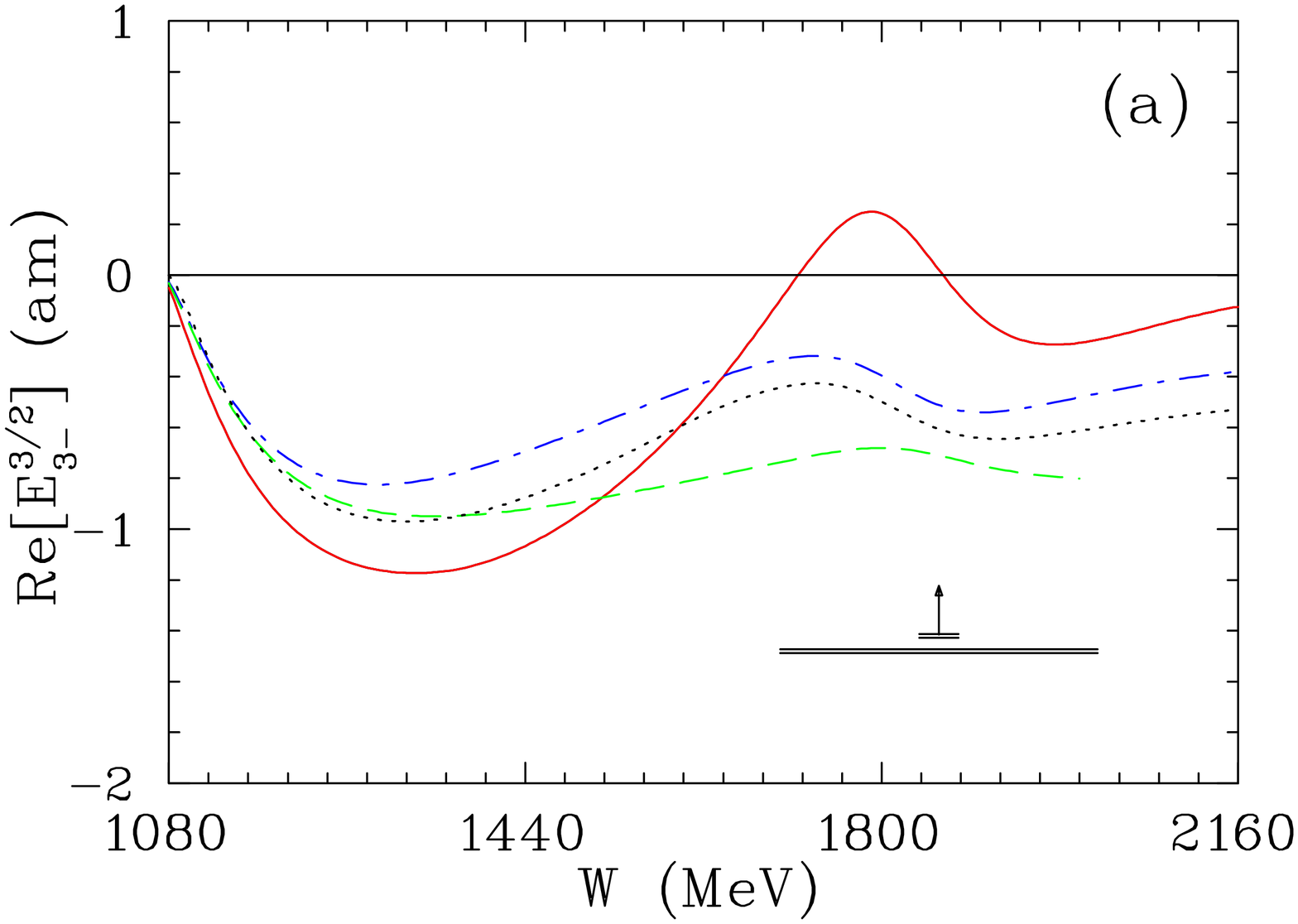}
\includegraphics[height=0.35\textwidth, angle=90]{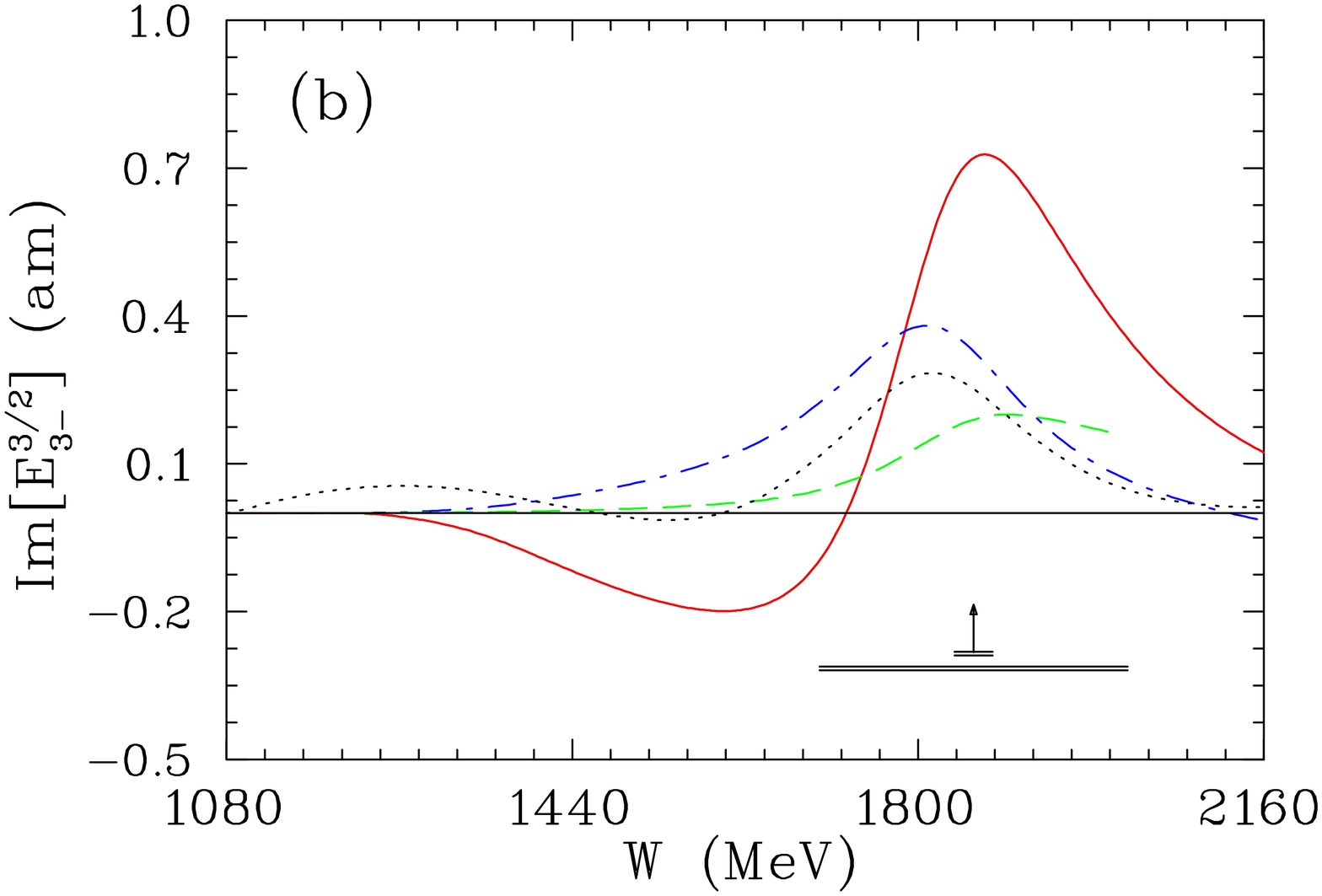}}
\centerline{
\includegraphics[height=0.35\textwidth, angle=90]{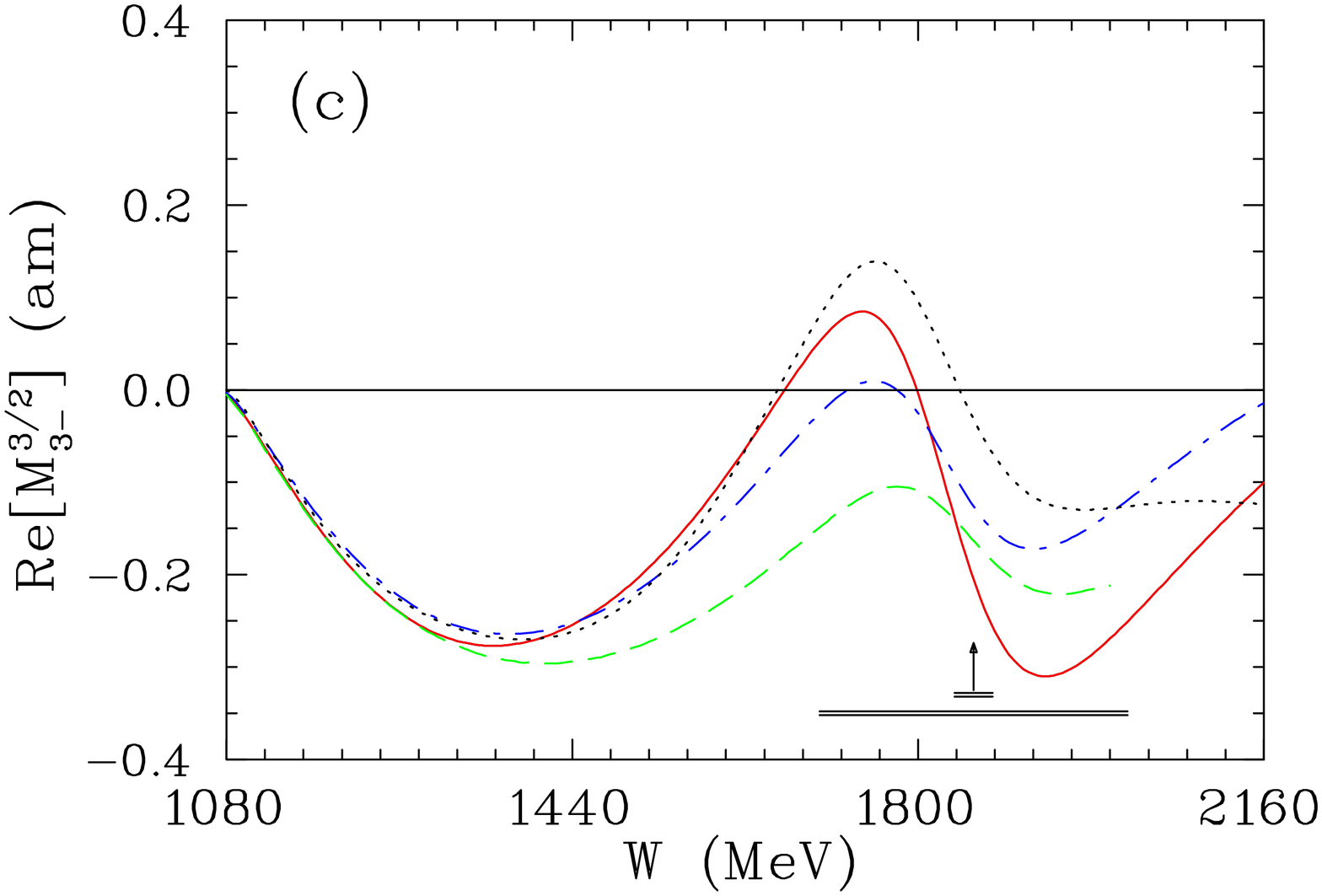}
\includegraphics[height=0.35\textwidth, angle=90]{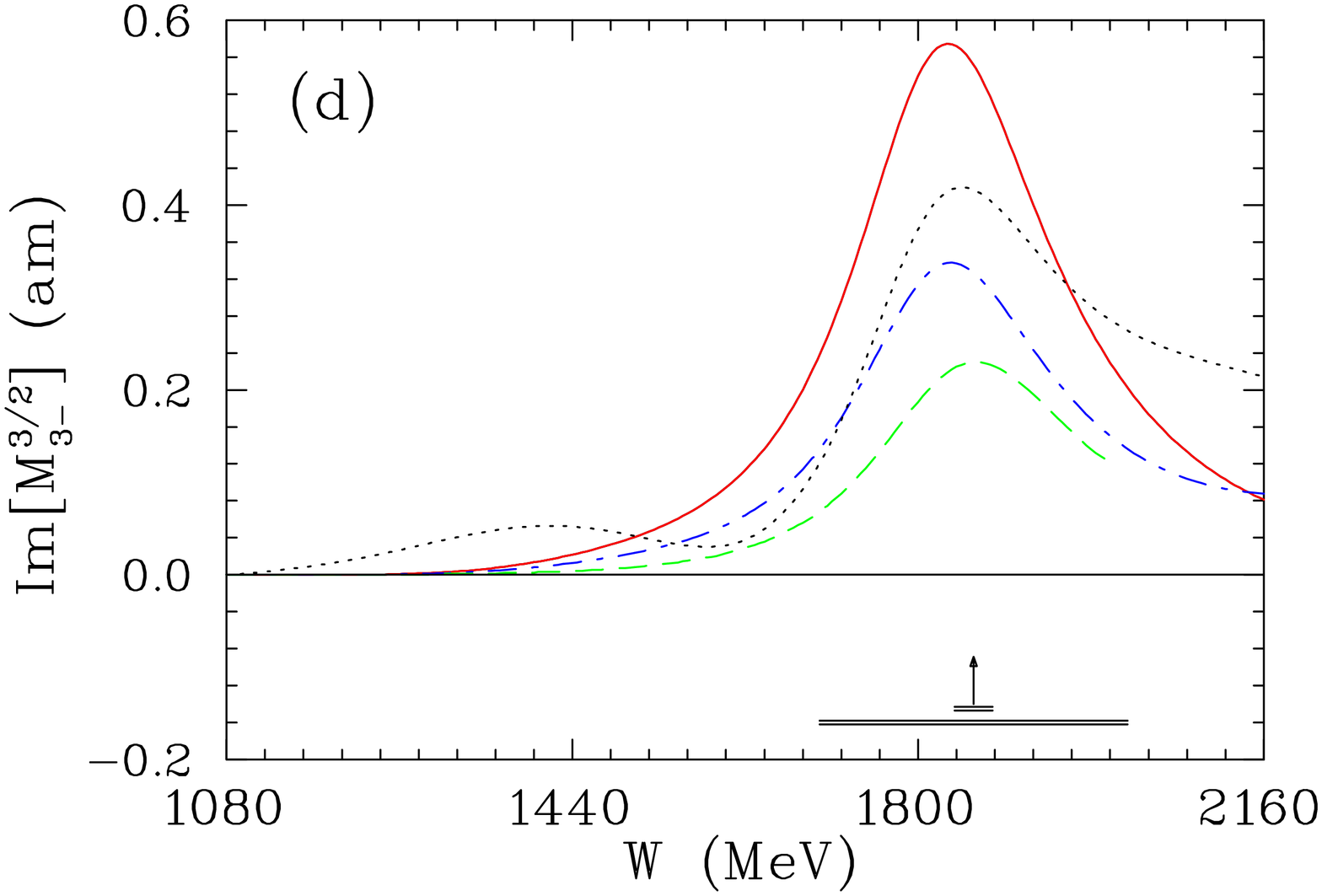}}
\centerline{
\includegraphics[height=0.35\textwidth, angle=90]{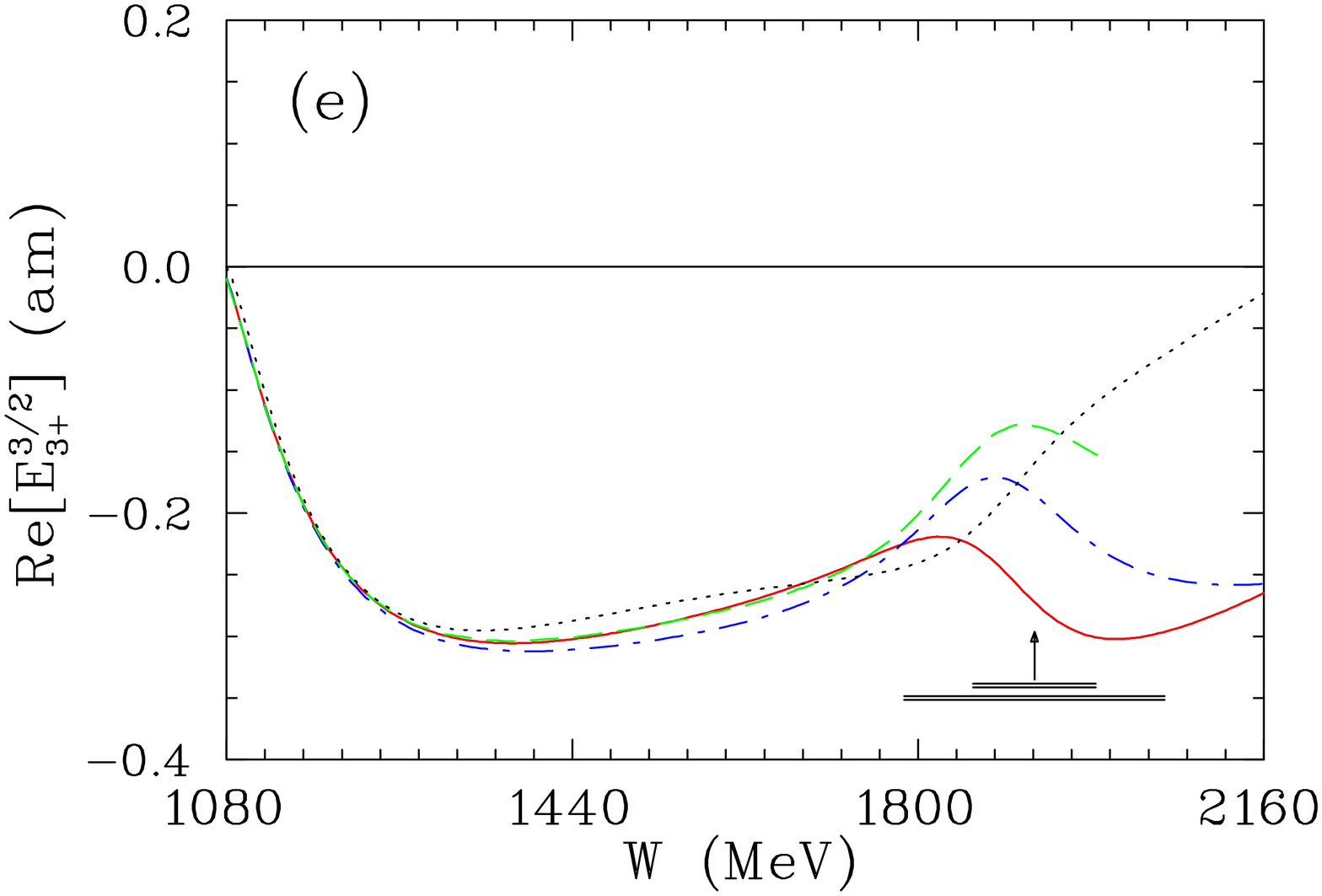}
\includegraphics[height=0.35\textwidth, angle=90]{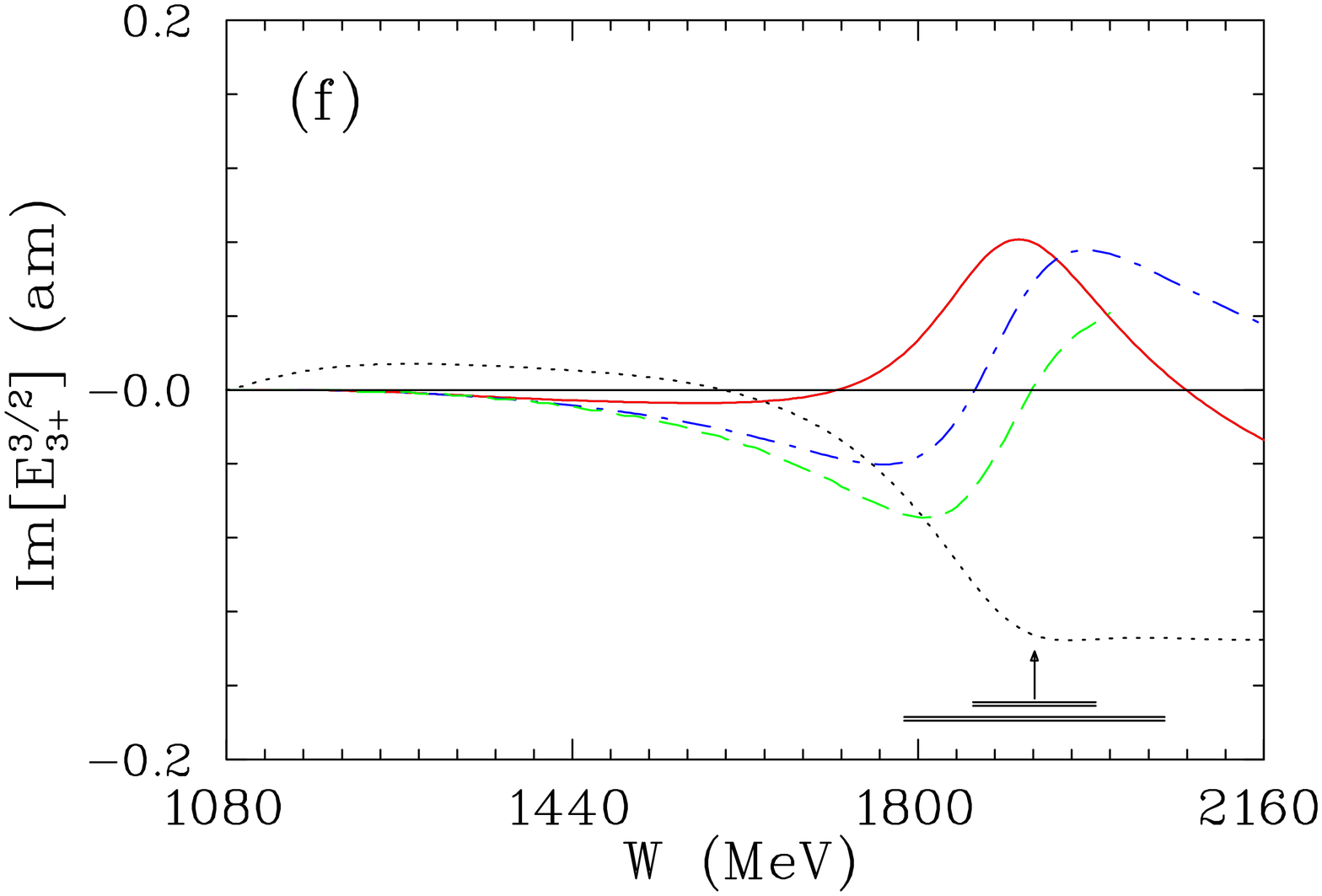}}
\centerline{
\includegraphics[height=0.35\textwidth, angle=90]{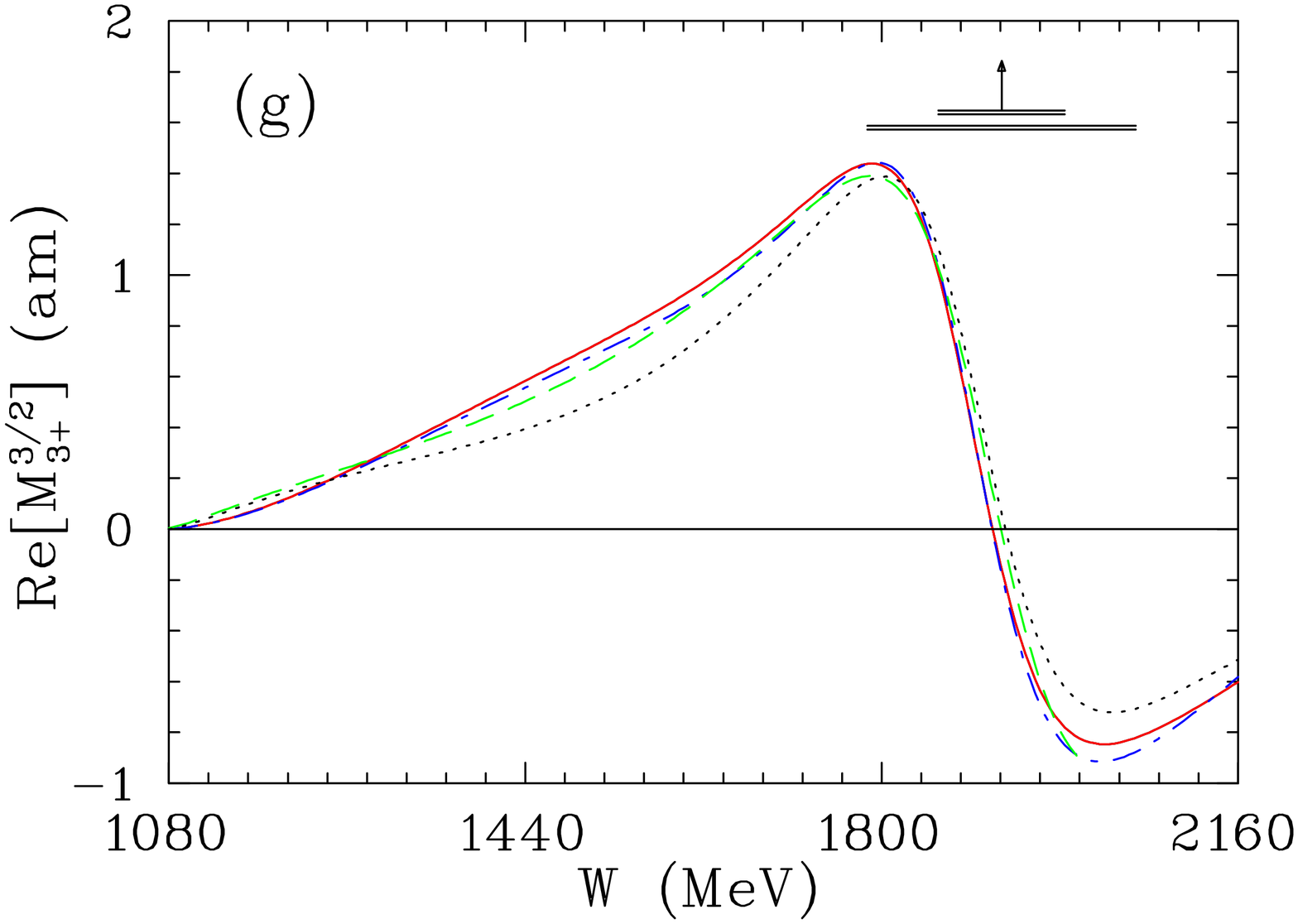}
\includegraphics[height=0.35\textwidth, angle=90]{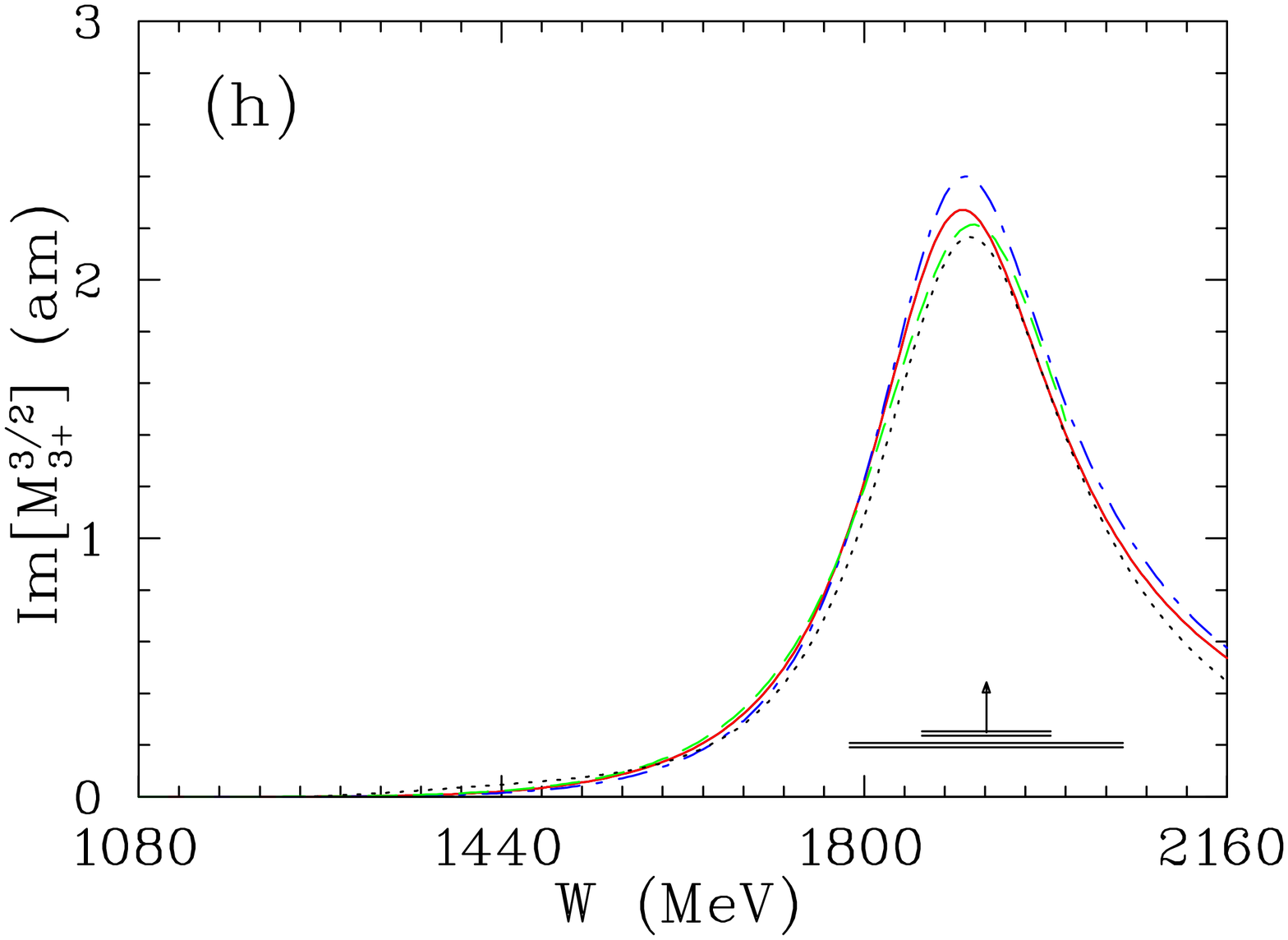}}
\caption{(Color online) Isospin $I$=3/2 multipole amplitudes from
        threshold to $W$ = 2.16~GeV ($E_{\gamma}$ = 2.02~GeV) for 
        $l = 3$.
        Notation as in Fig.~\protect\ref{fig:g7}. \label{fig:g9}}
\end{figure*}


\begin{figure*}[th]
\centerline{
\includegraphics[height=0.35\textwidth, angle=90]{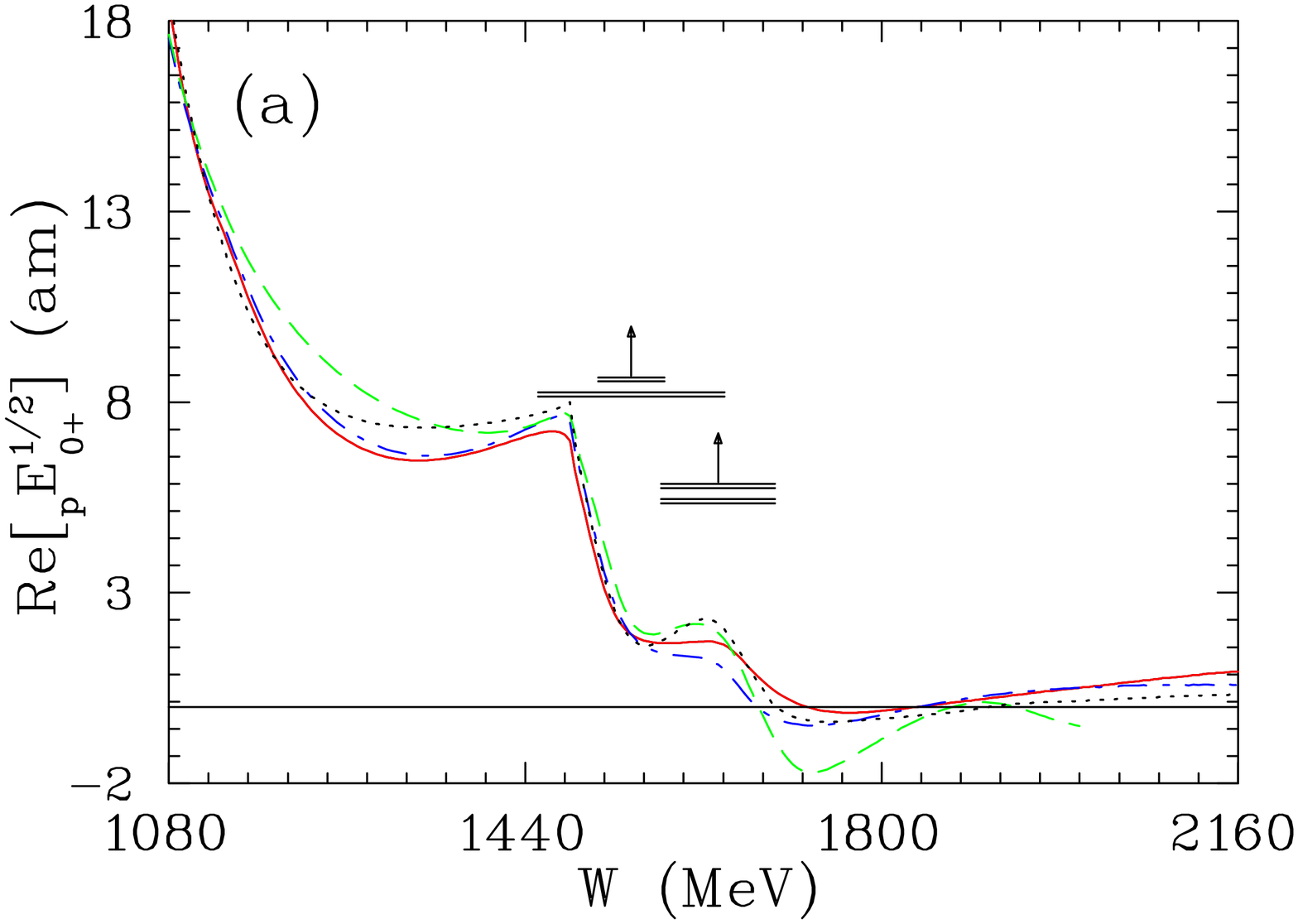}
\includegraphics[height=0.35\textwidth, angle=90]{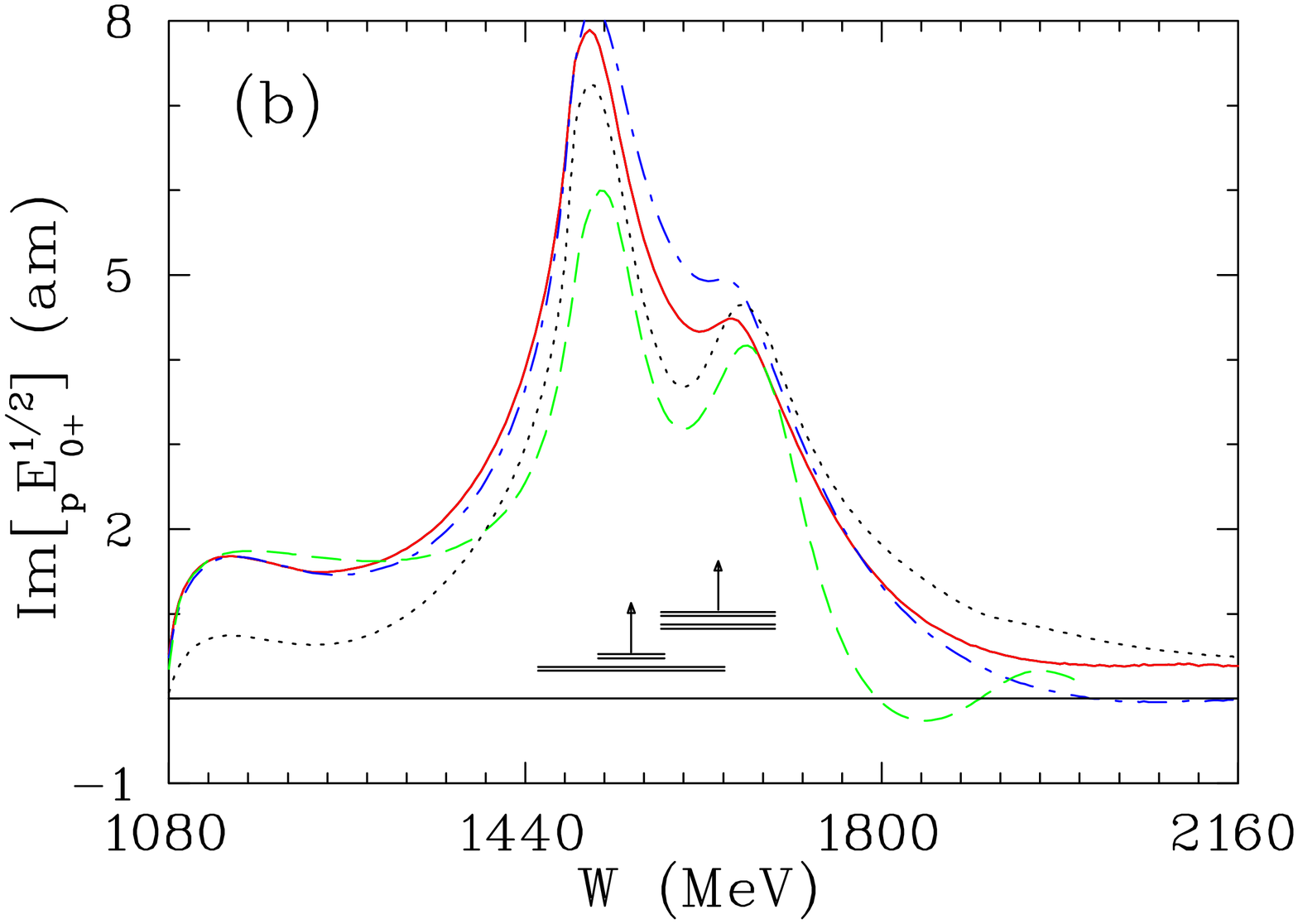}}
\centerline{
\includegraphics[height=0.35\textwidth, angle=90]{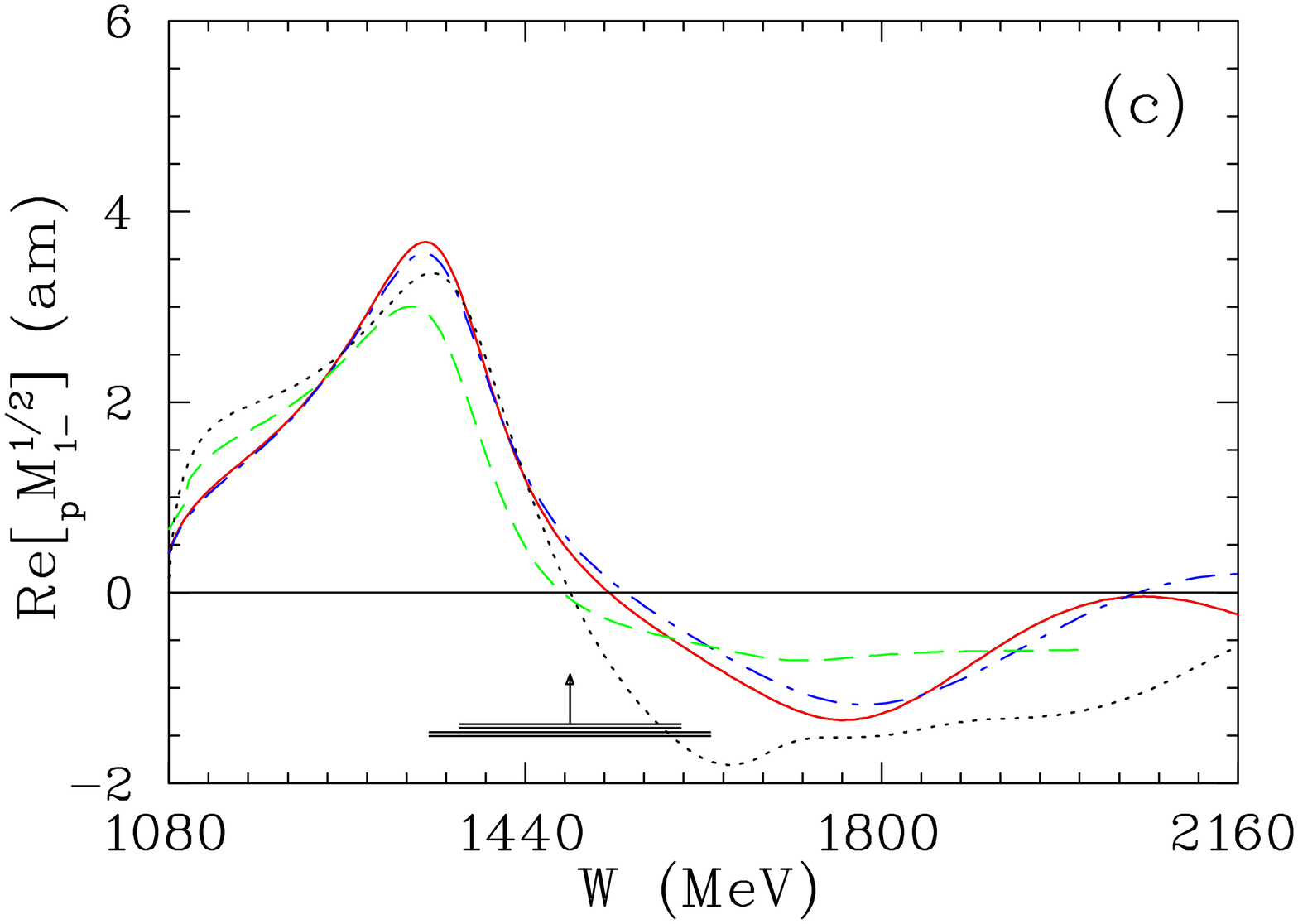}
\includegraphics[height=0.35\textwidth, angle=90]{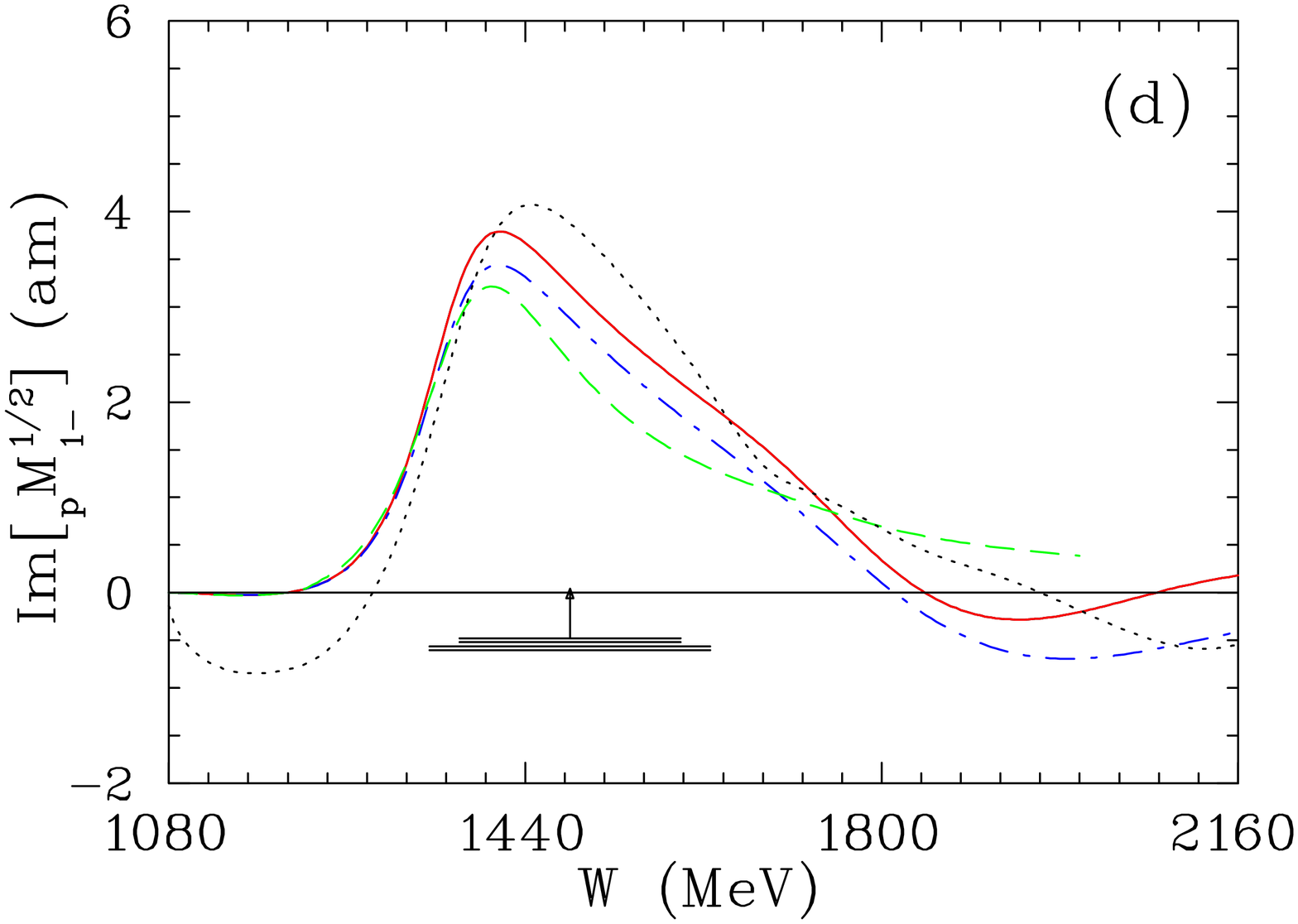}}
\centerline{
\includegraphics[height=0.35\textwidth, angle=90]{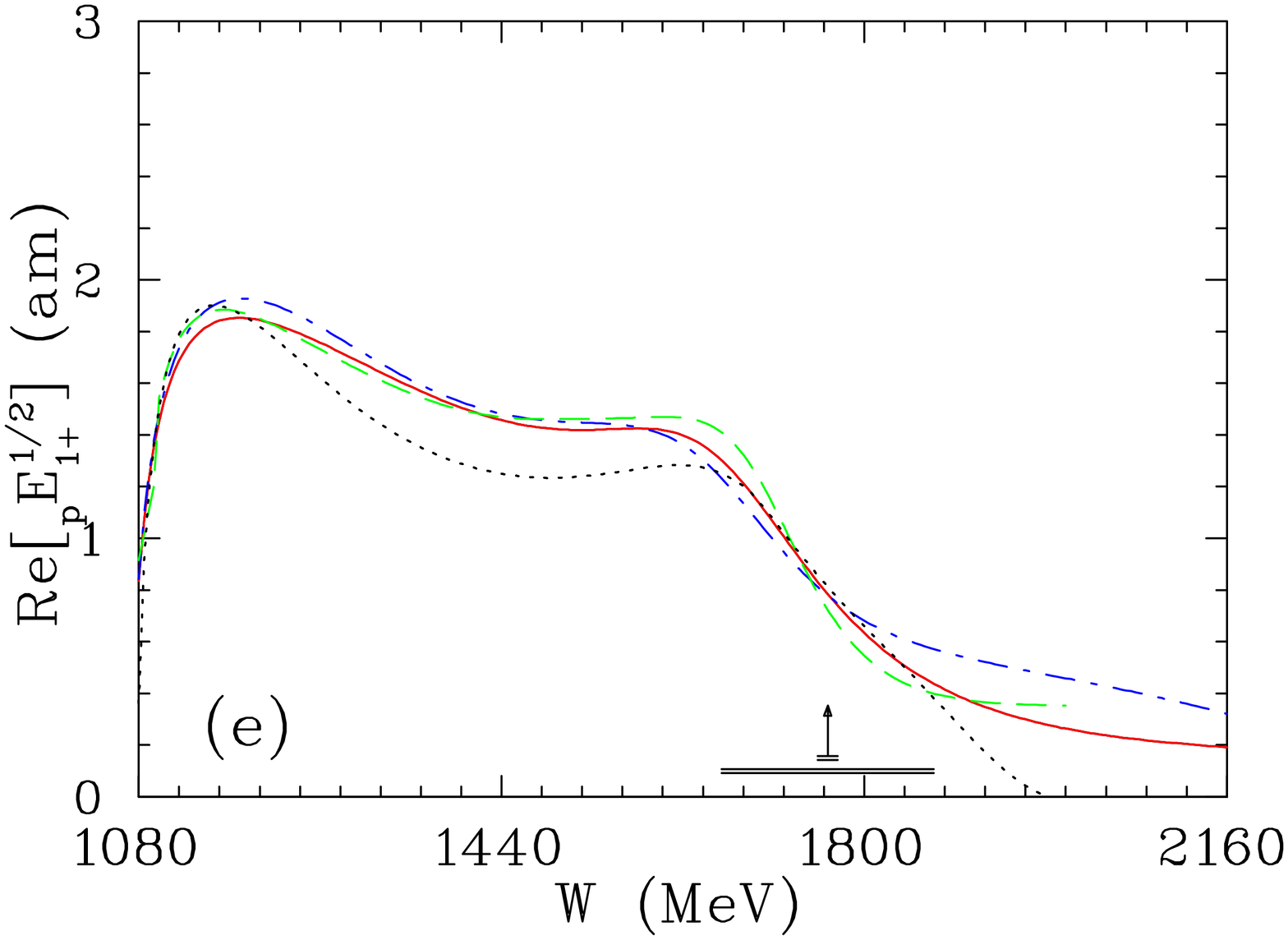}
\includegraphics[height=0.35\textwidth, angle=90]{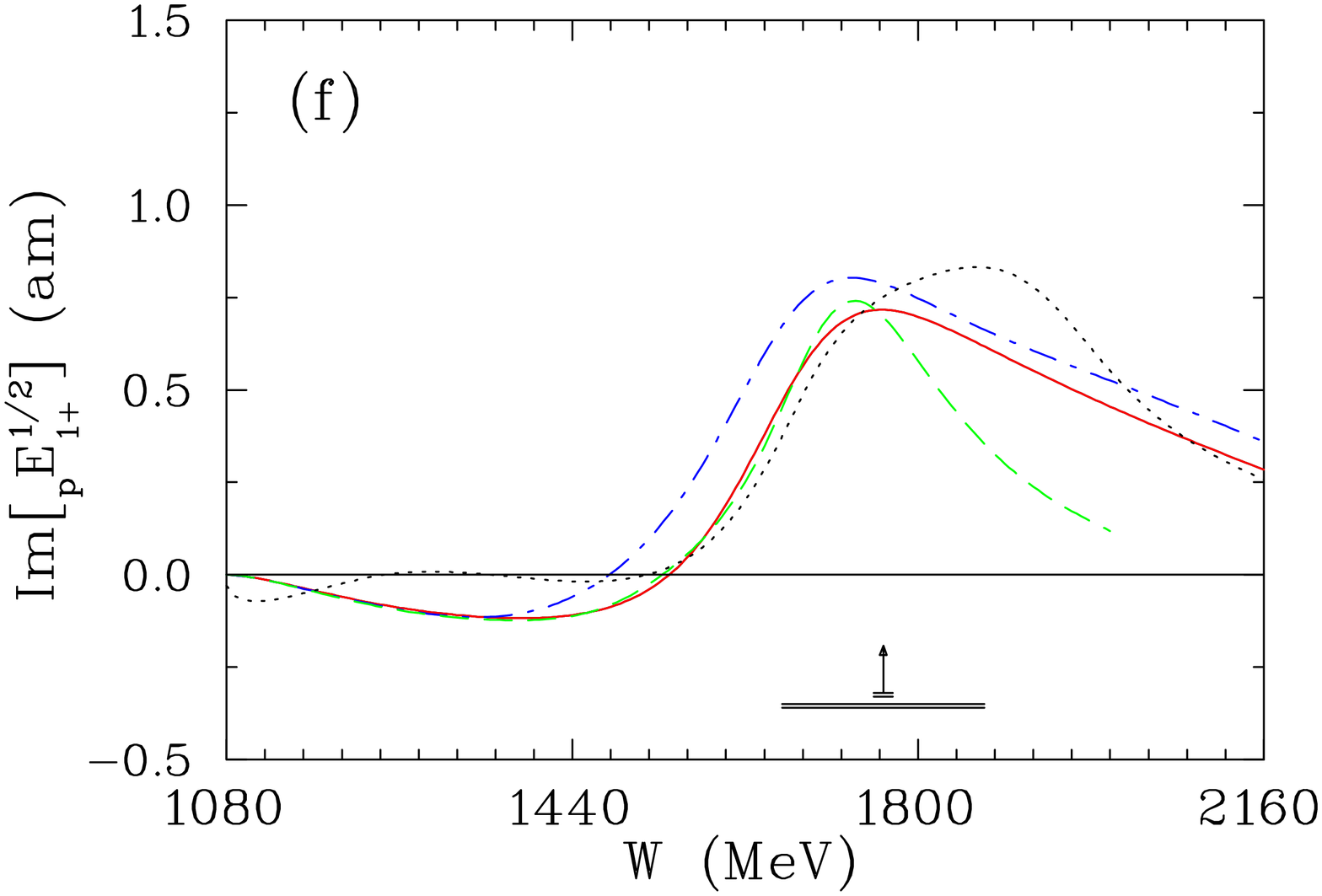}}
\centerline{
\includegraphics[height=0.35\textwidth, angle=90]{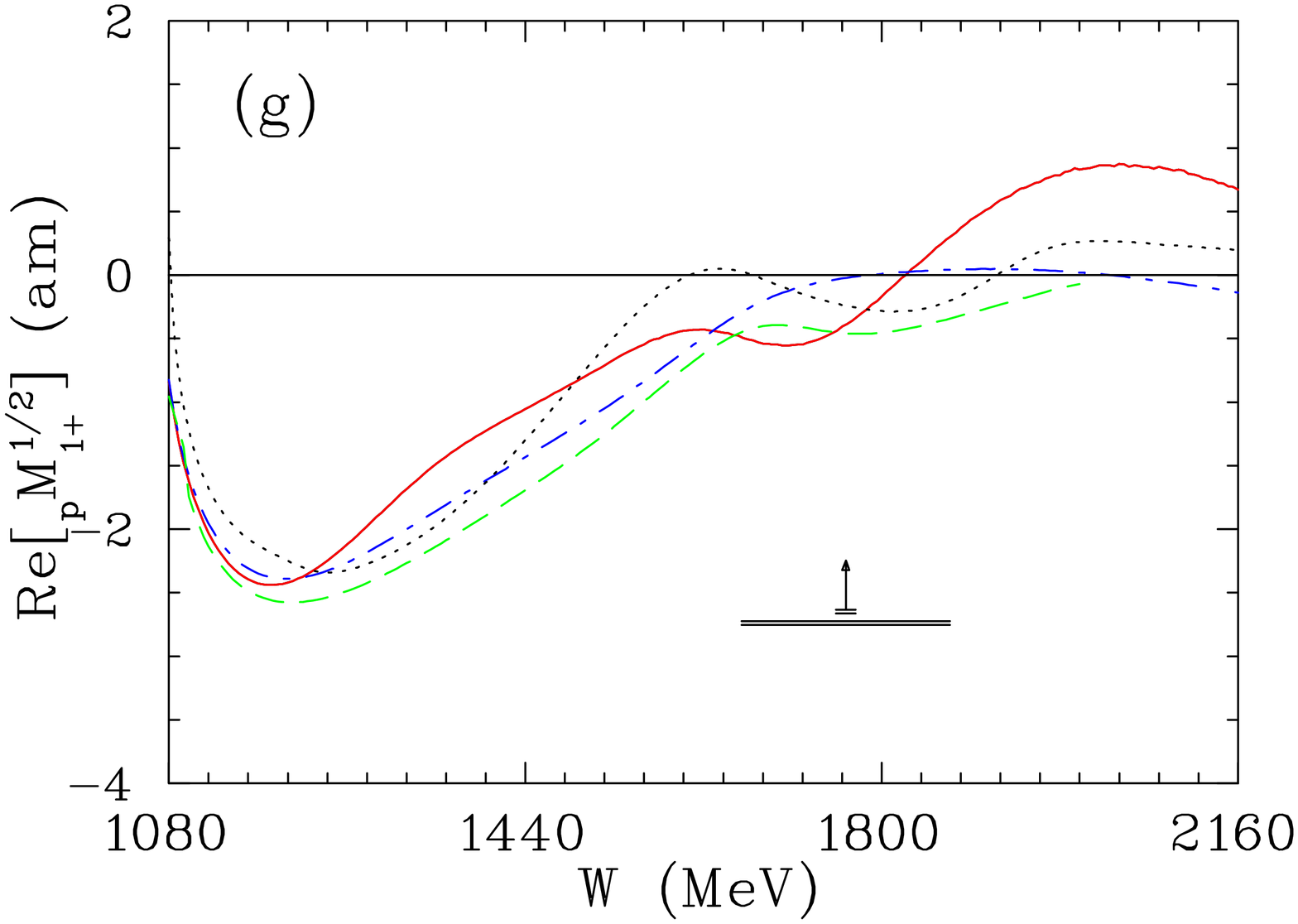}
\includegraphics[height=0.35\textwidth, angle=90]{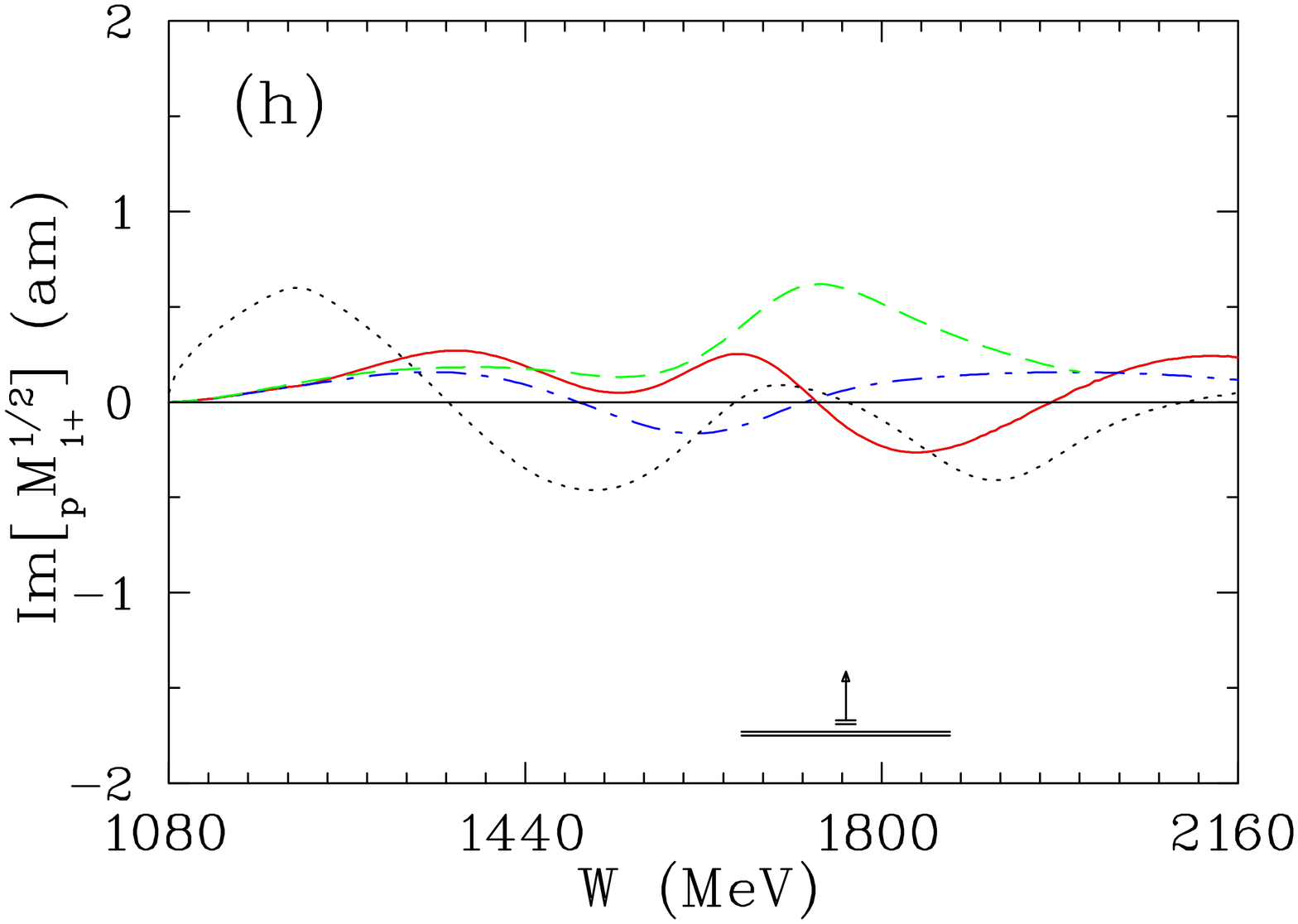}}
\caption{(Color online) Proton multipole $I$=1/2 amplitudes from
        threshold to $W$ = 2.16~GeV ($E_{\gamma}$ = 2.02~GeV) for 
        $l=0,1$.
        Notation as in Fig.~\protect\ref{fig:g7}. \label{fig:g10}}
\end{figure*}


\begin{figure*}[th]
\centerline{
\includegraphics[height=0.35\textwidth, angle=90]{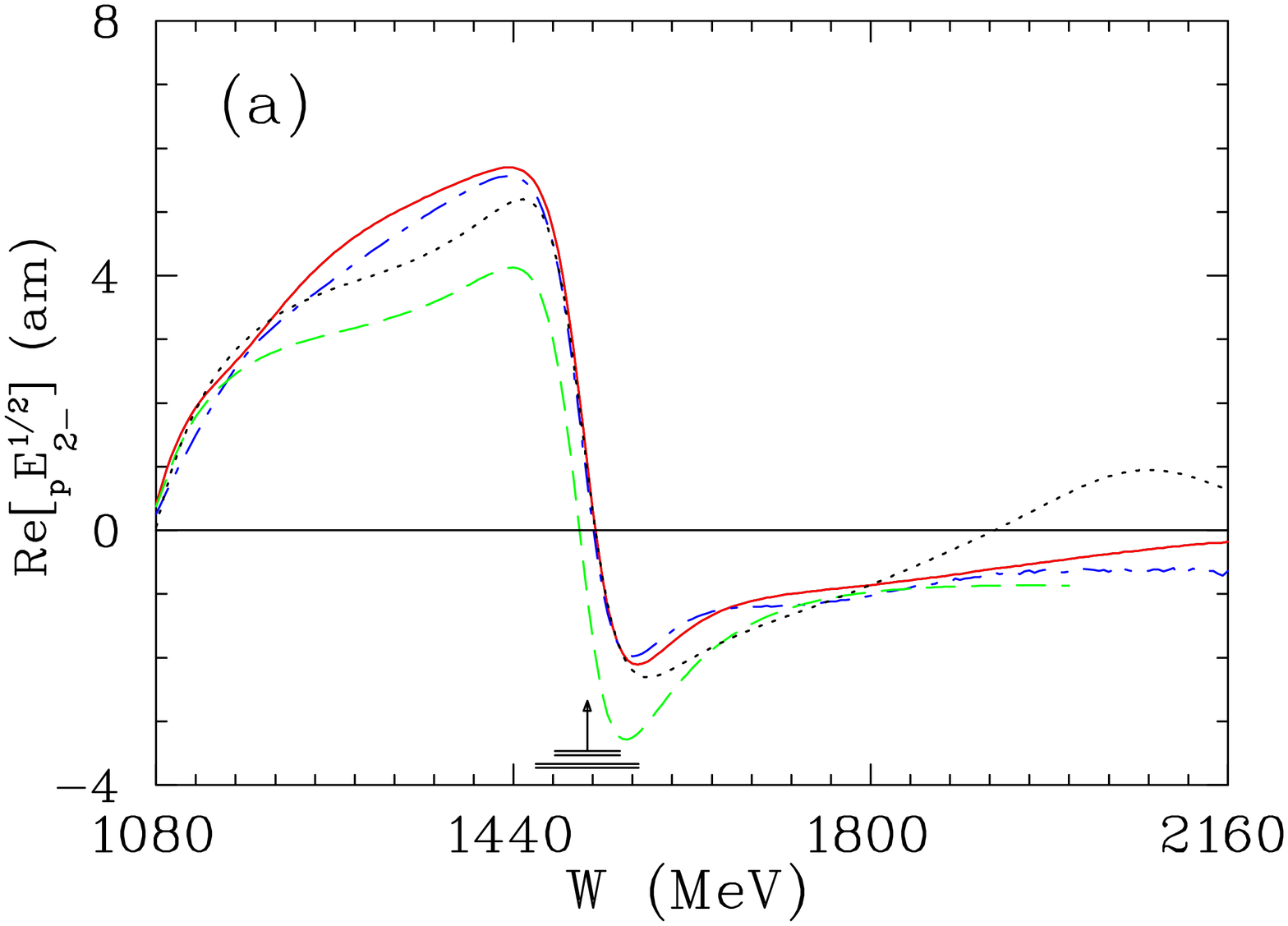}
\includegraphics[height=0.35\textwidth, angle=90]{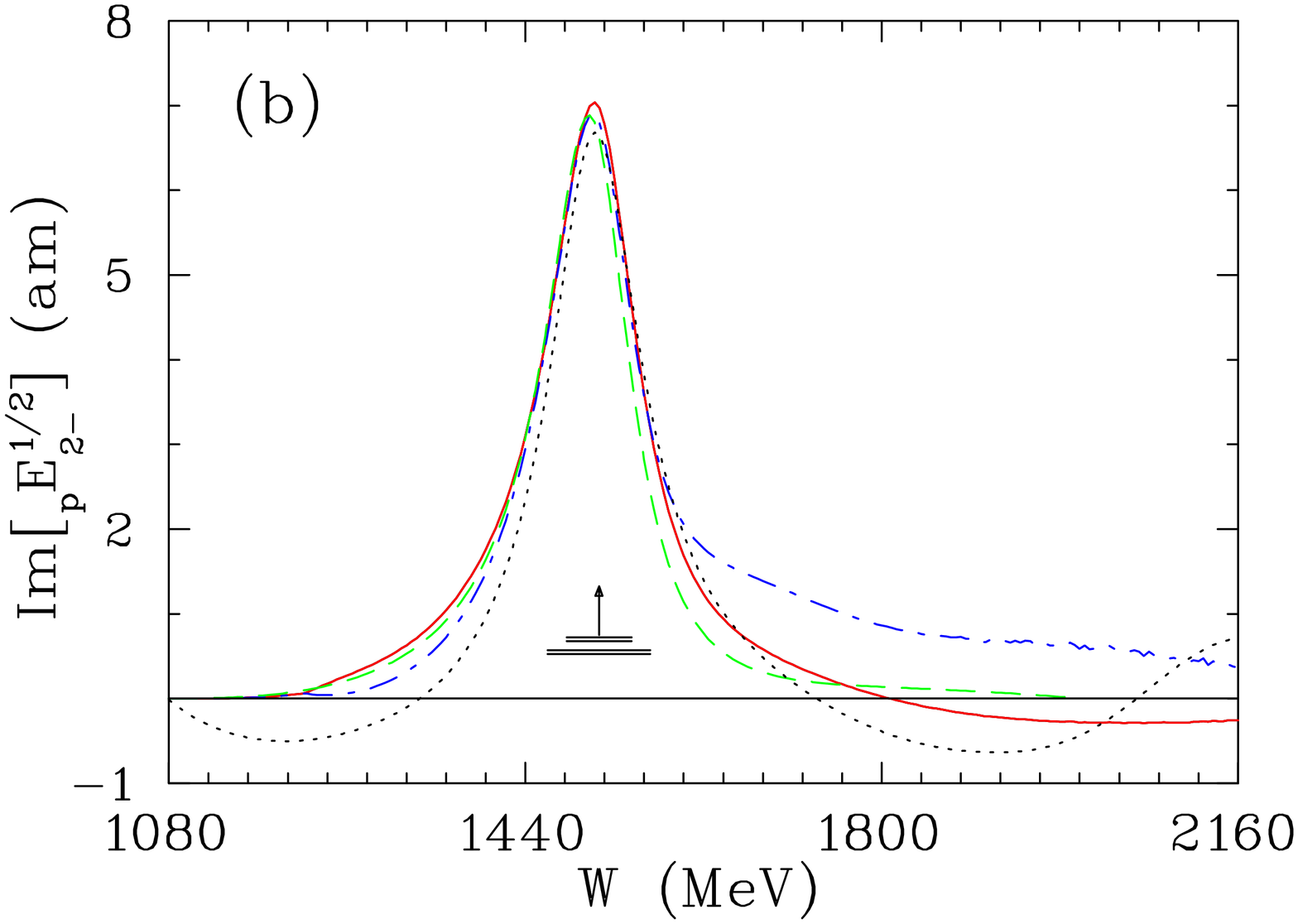}}
\centerline{
\includegraphics[height=0.35\textwidth, angle=90]{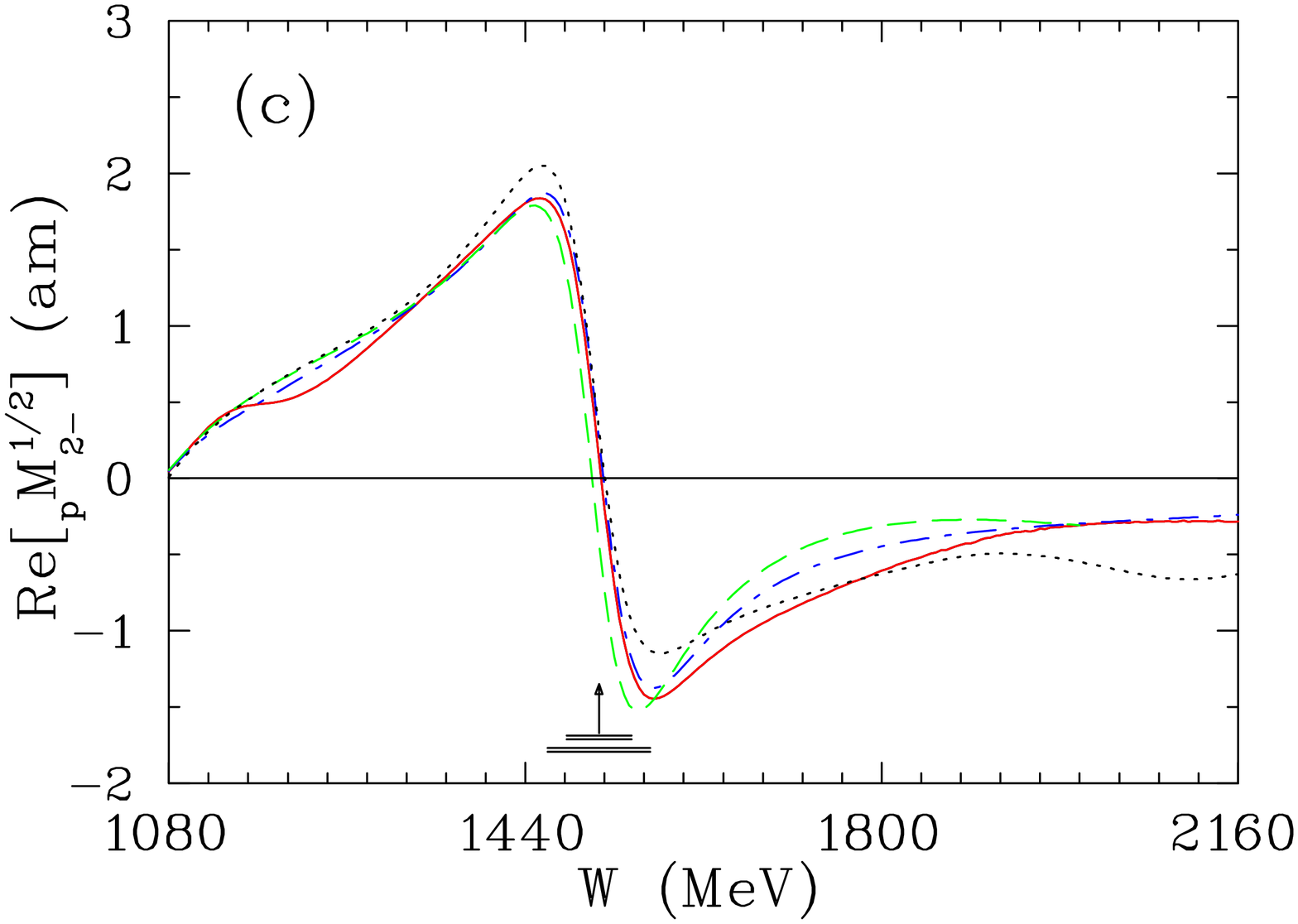}
\includegraphics[height=0.35\textwidth, angle=90]{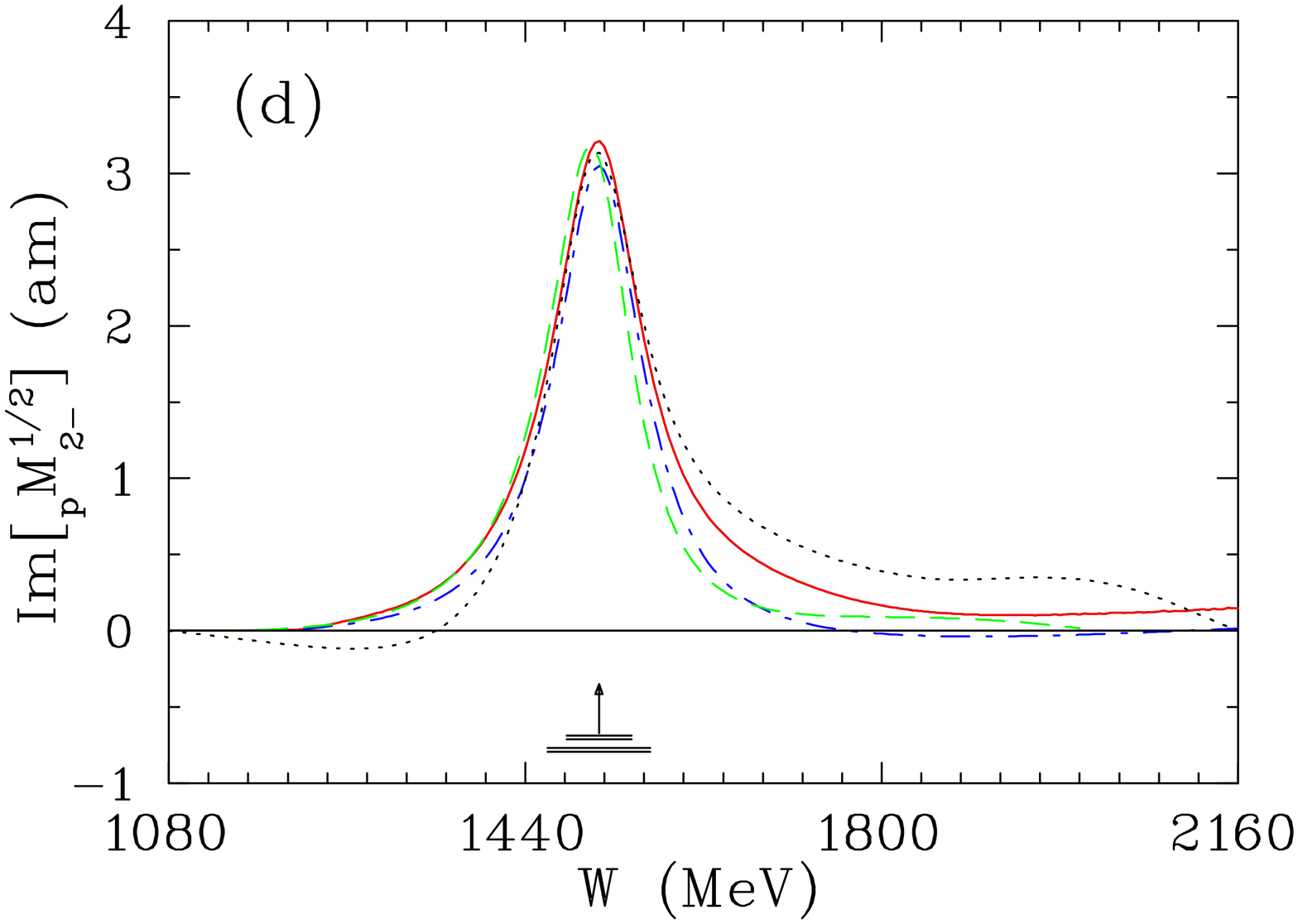}}
\centerline{
\includegraphics[height=0.35\textwidth, angle=90]{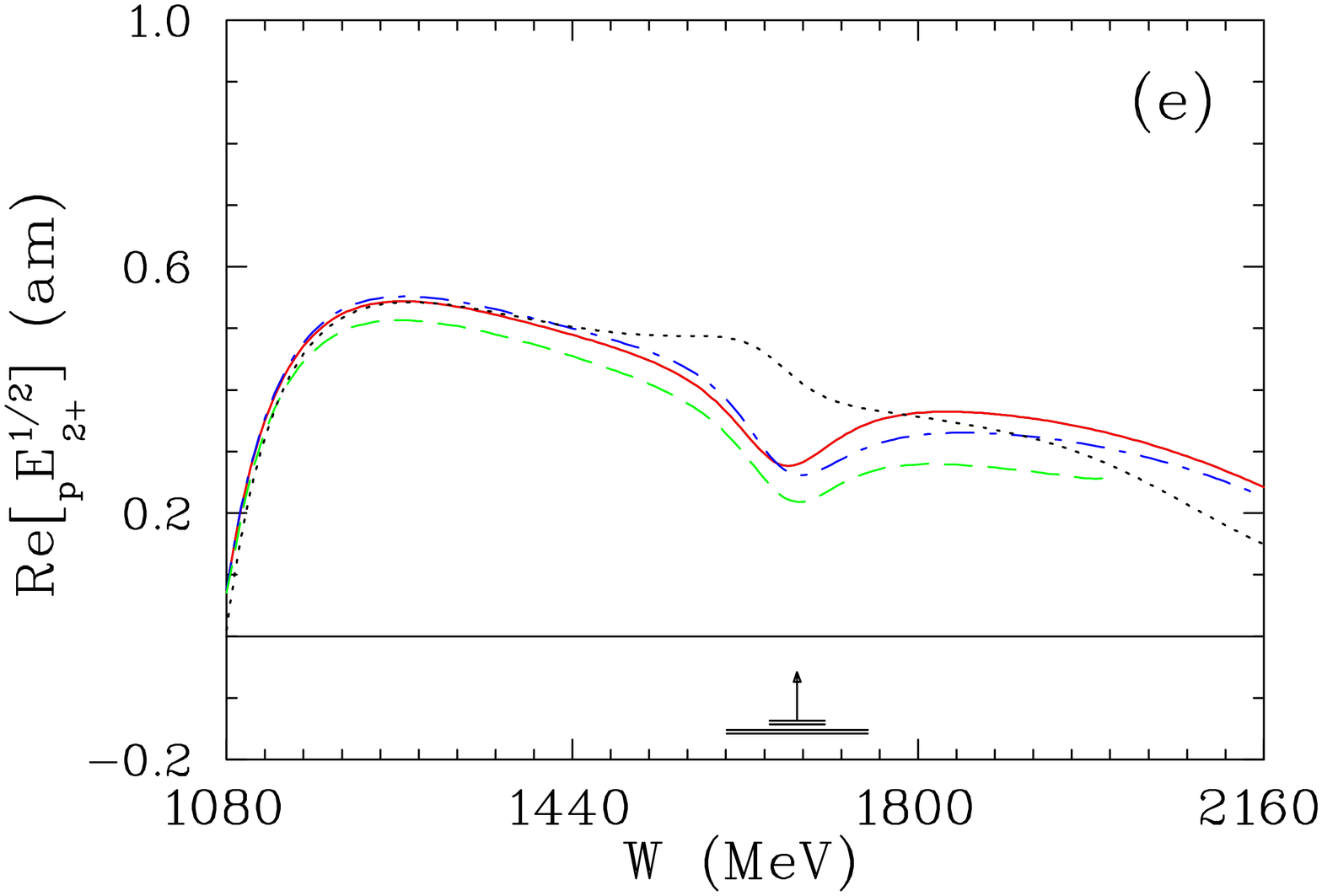}
\includegraphics[height=0.35\textwidth, angle=90]{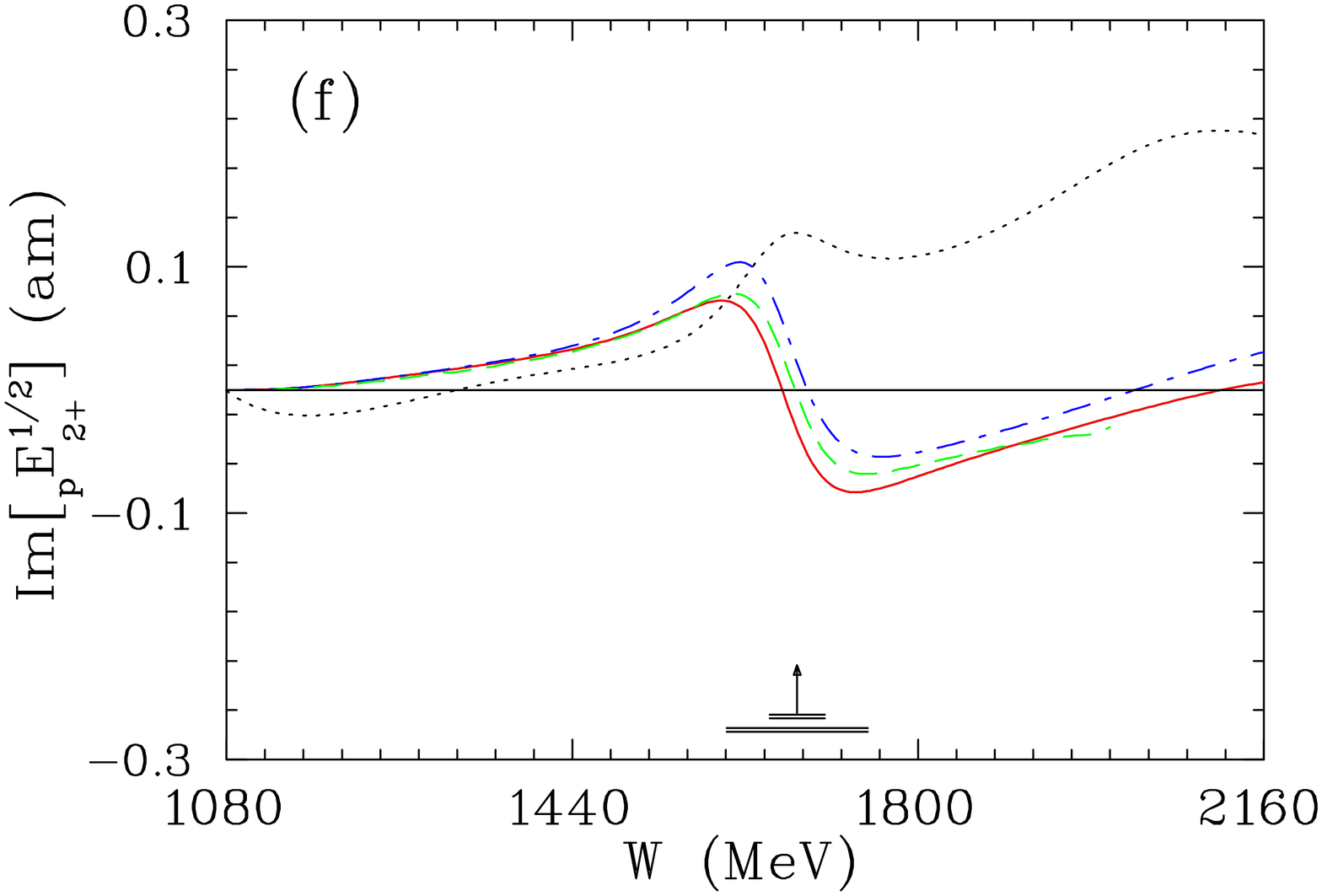}}
\centerline{
\includegraphics[height=0.35\textwidth, angle=90]{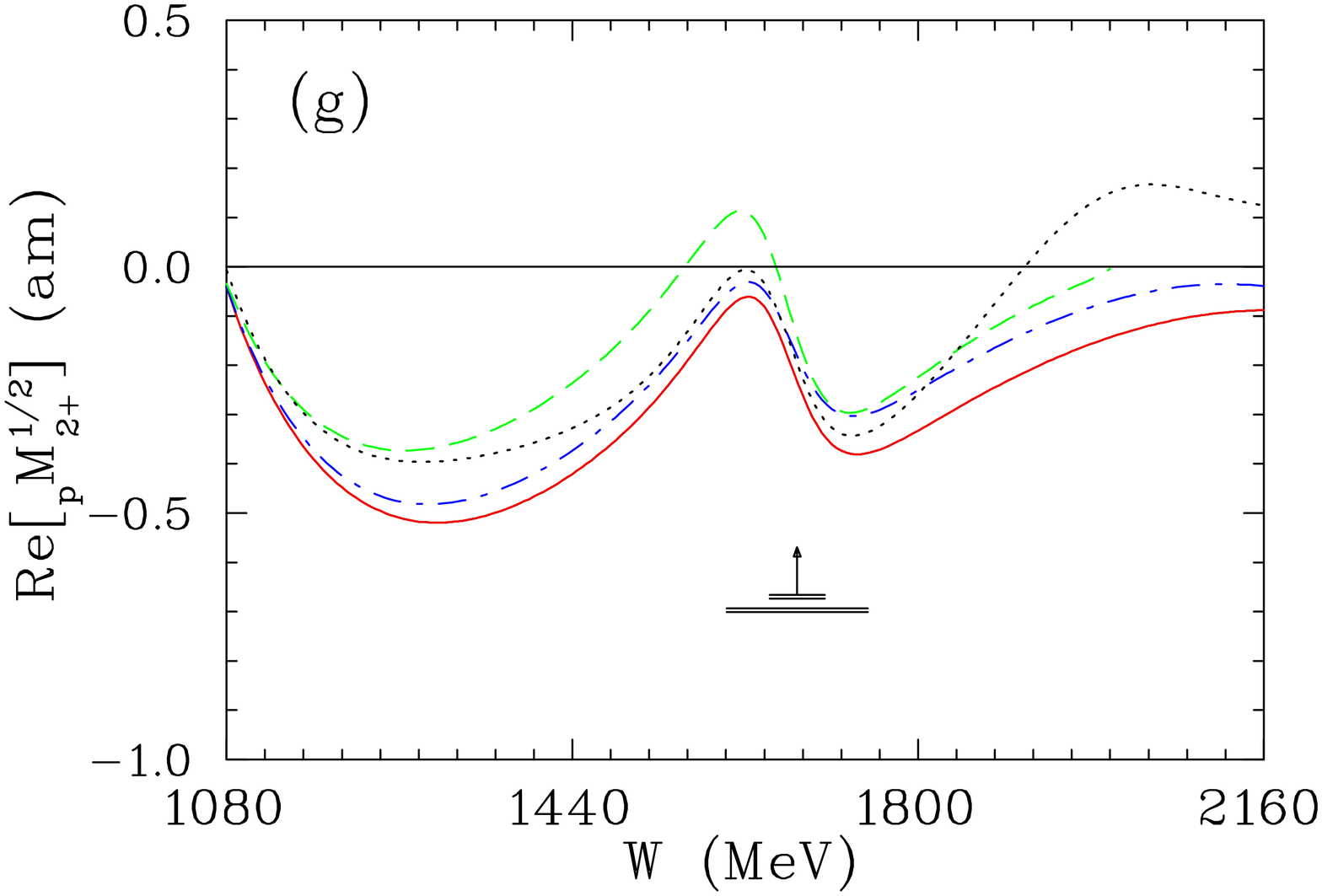}
\includegraphics[height=0.35\textwidth, angle=90]{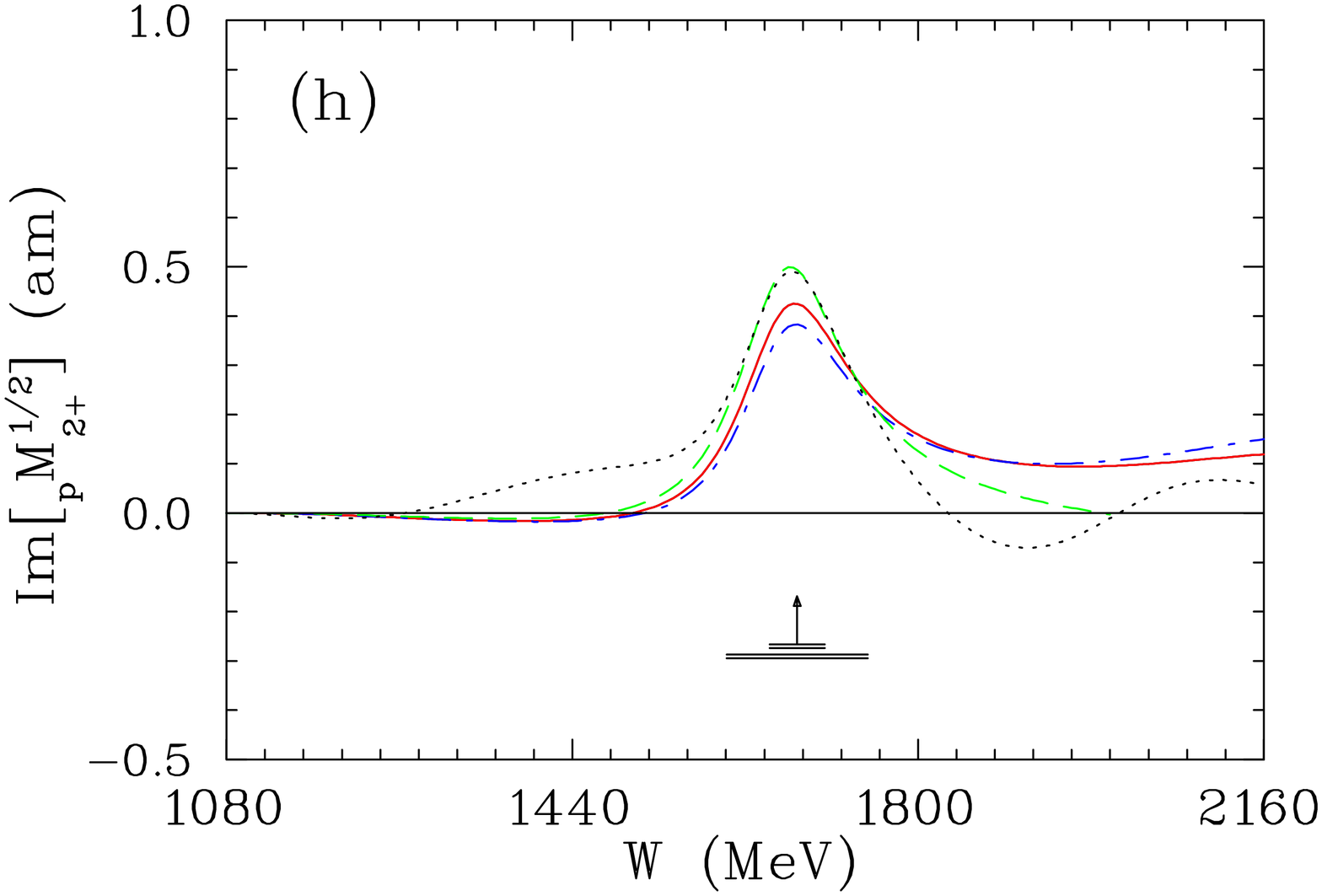}}
\caption{(Color online) Proton multipole $I$=1/2 amplitudes from
        threshold to $W$ = 2.16~GeV ($E_{\gamma}$ = 2.02~GeV) 
        for $l=2$.
        Notation as in Fig.~\protect\ref{fig:g7}. \label{fig:g11}}
\end{figure*}


\begin{figure*}[th]
\centerline{
\includegraphics[height=0.35\textwidth, angle=90]{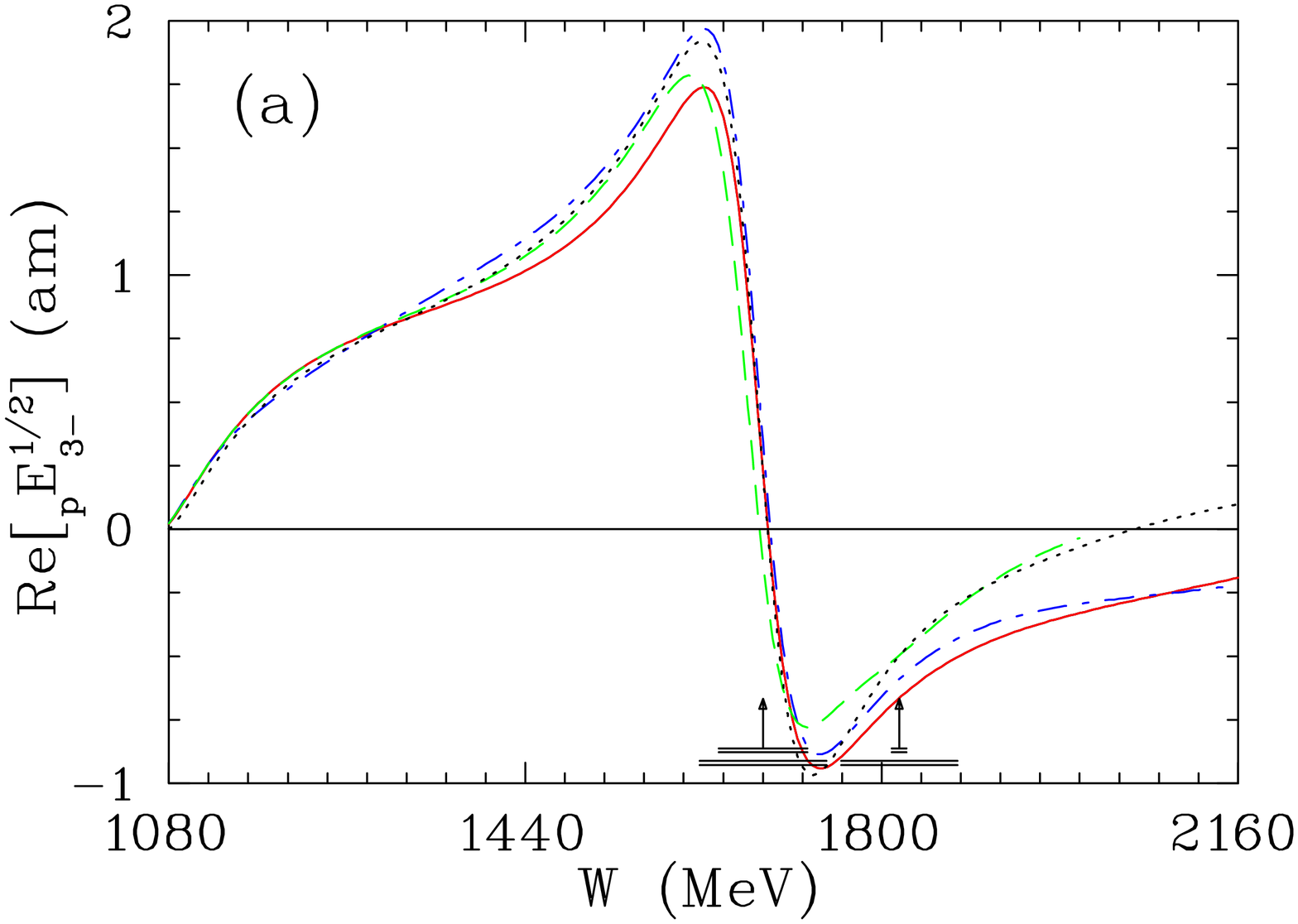}
\includegraphics[height=0.35\textwidth, angle=90]{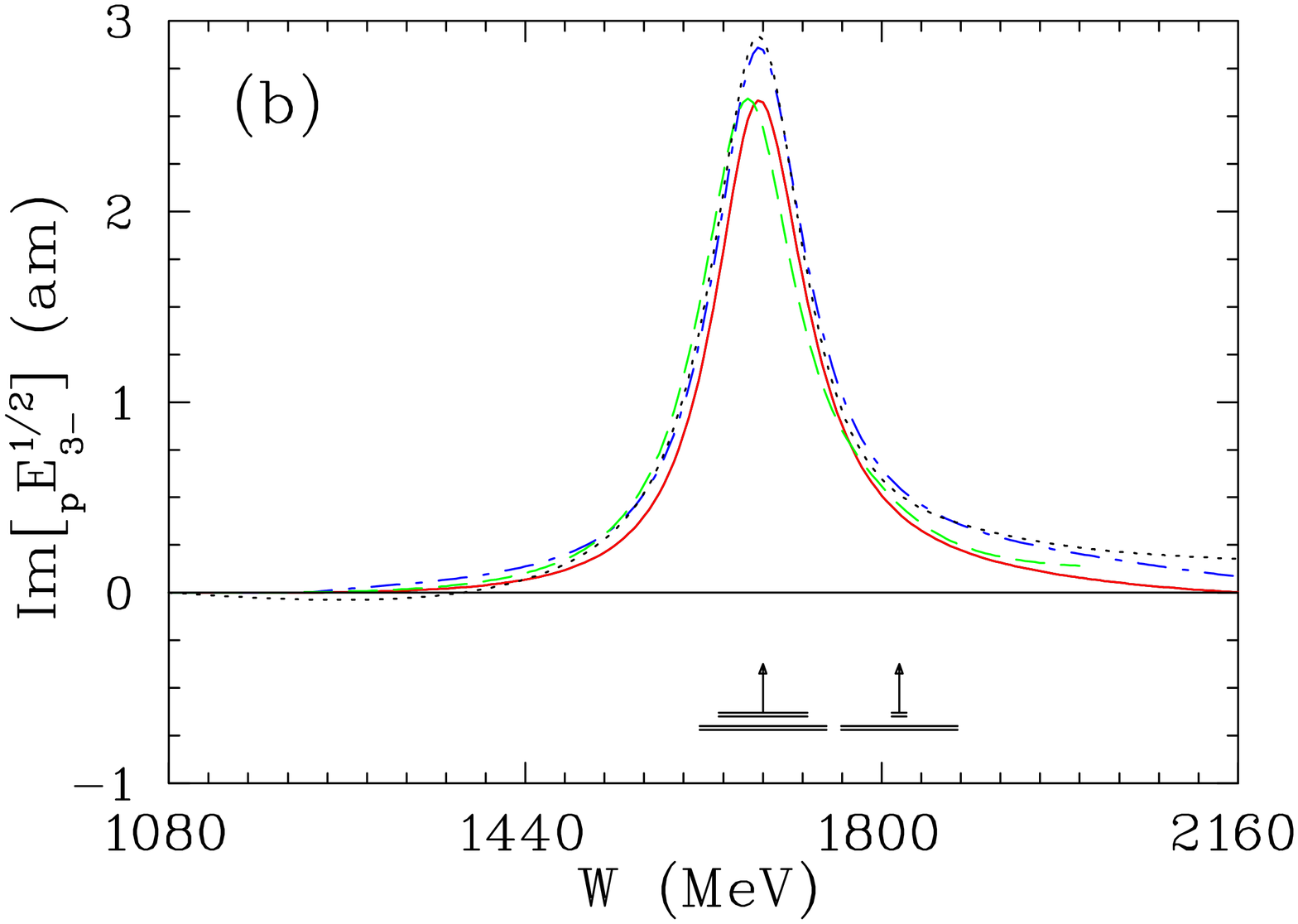}}
\centerline{
\includegraphics[height=0.35\textwidth, angle=90]{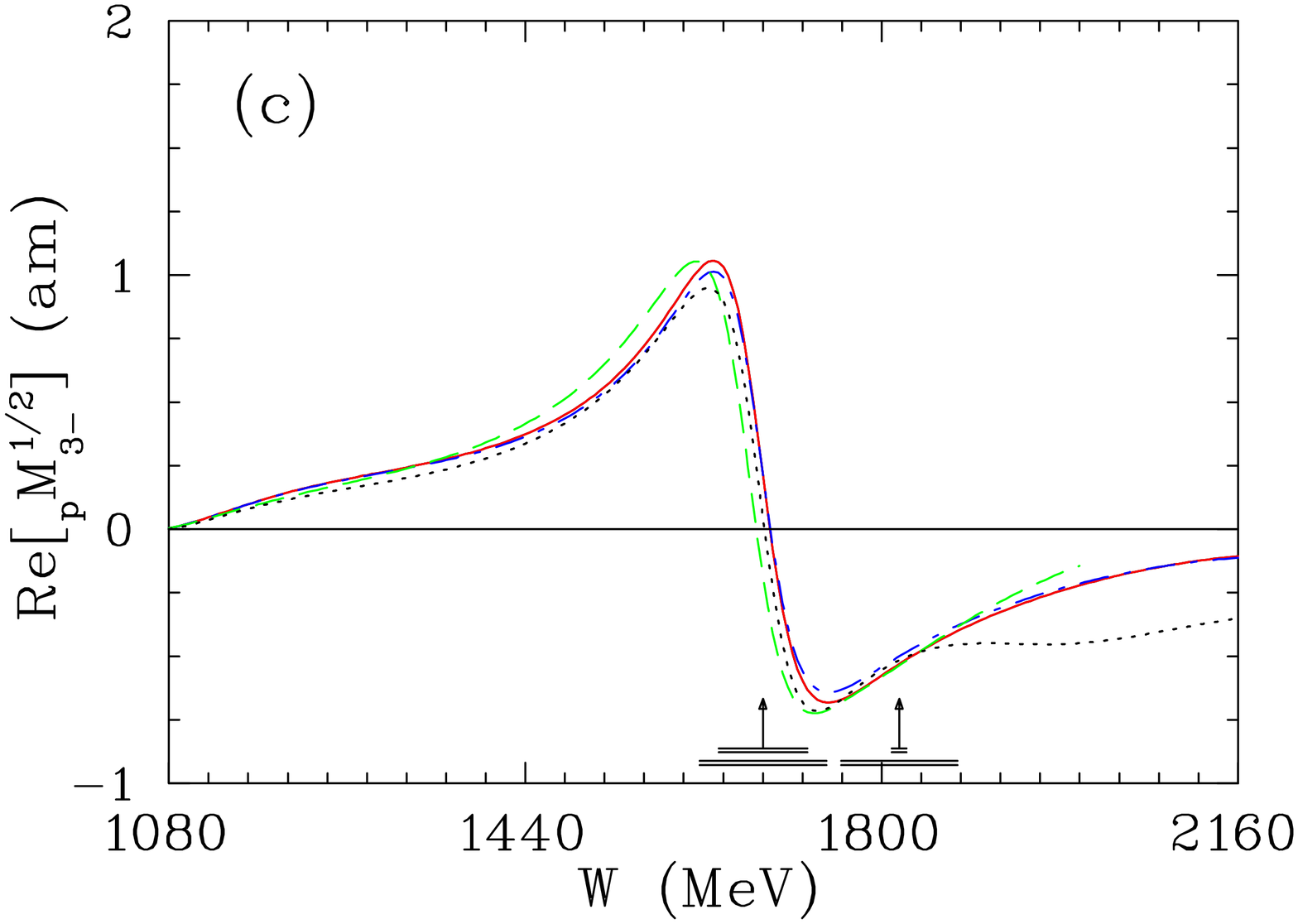}
\includegraphics[height=0.35\textwidth, angle=90]{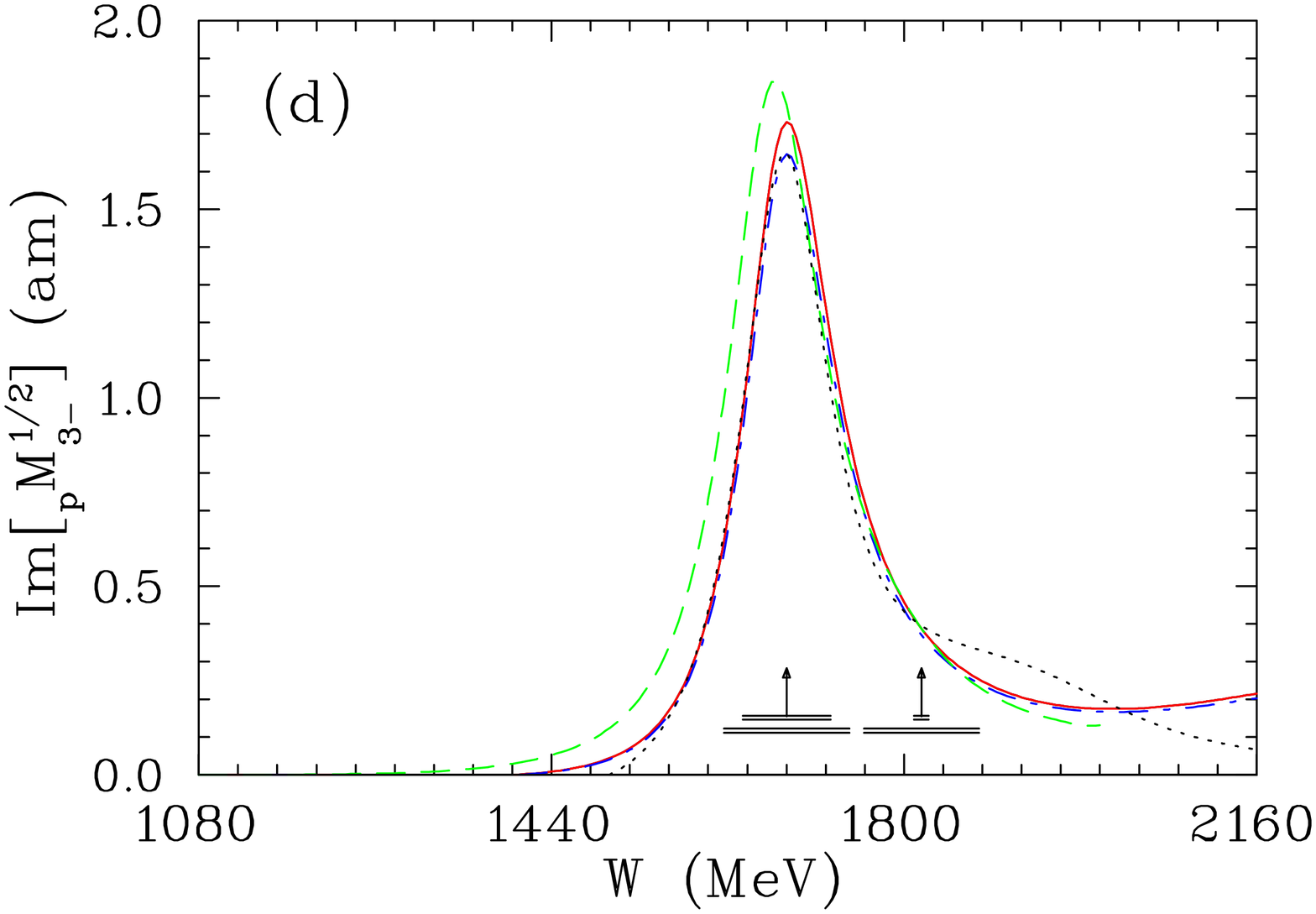}}
\centerline{
\includegraphics[height=0.35\textwidth, angle=90]{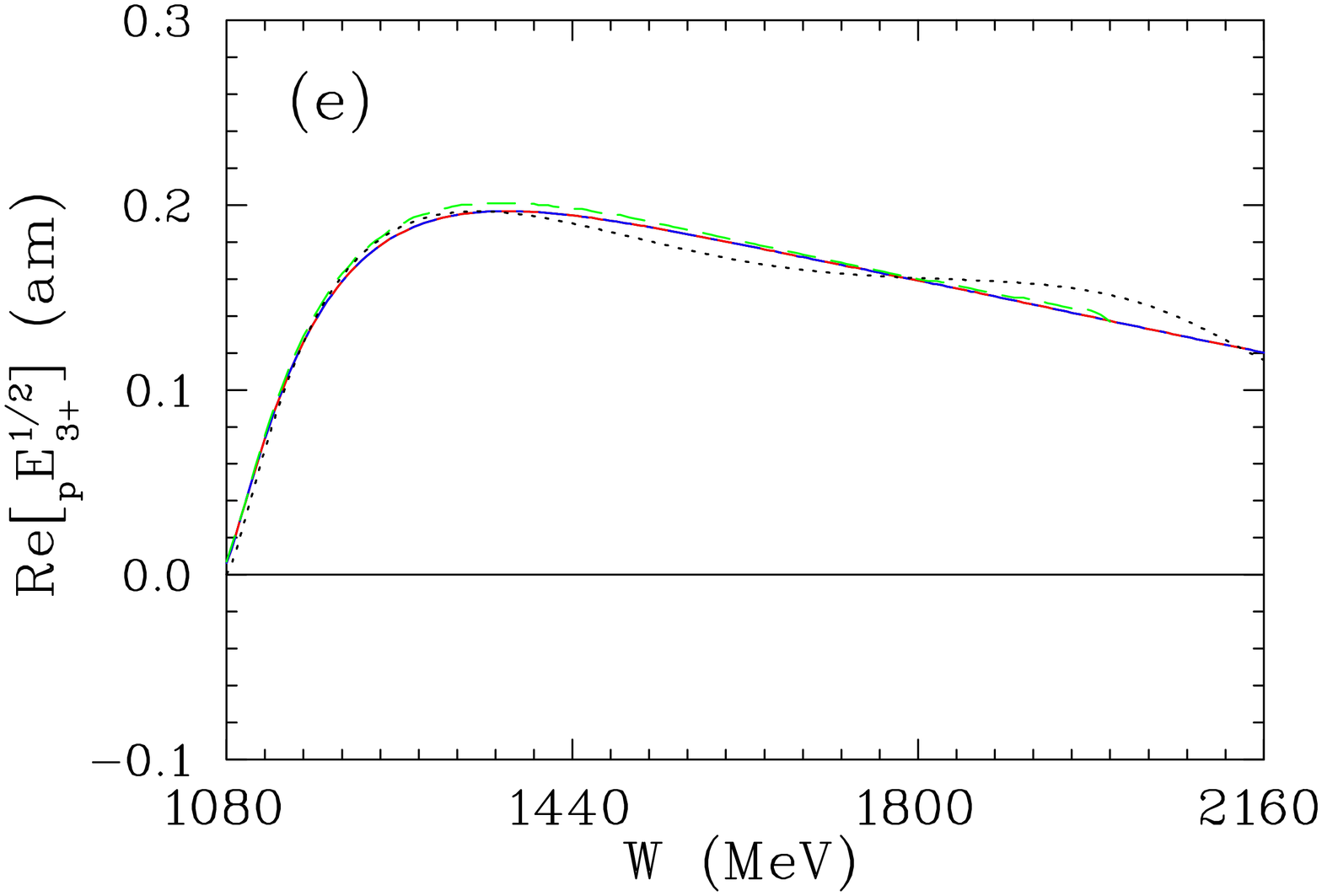}
\includegraphics[height=0.35\textwidth, angle=90]{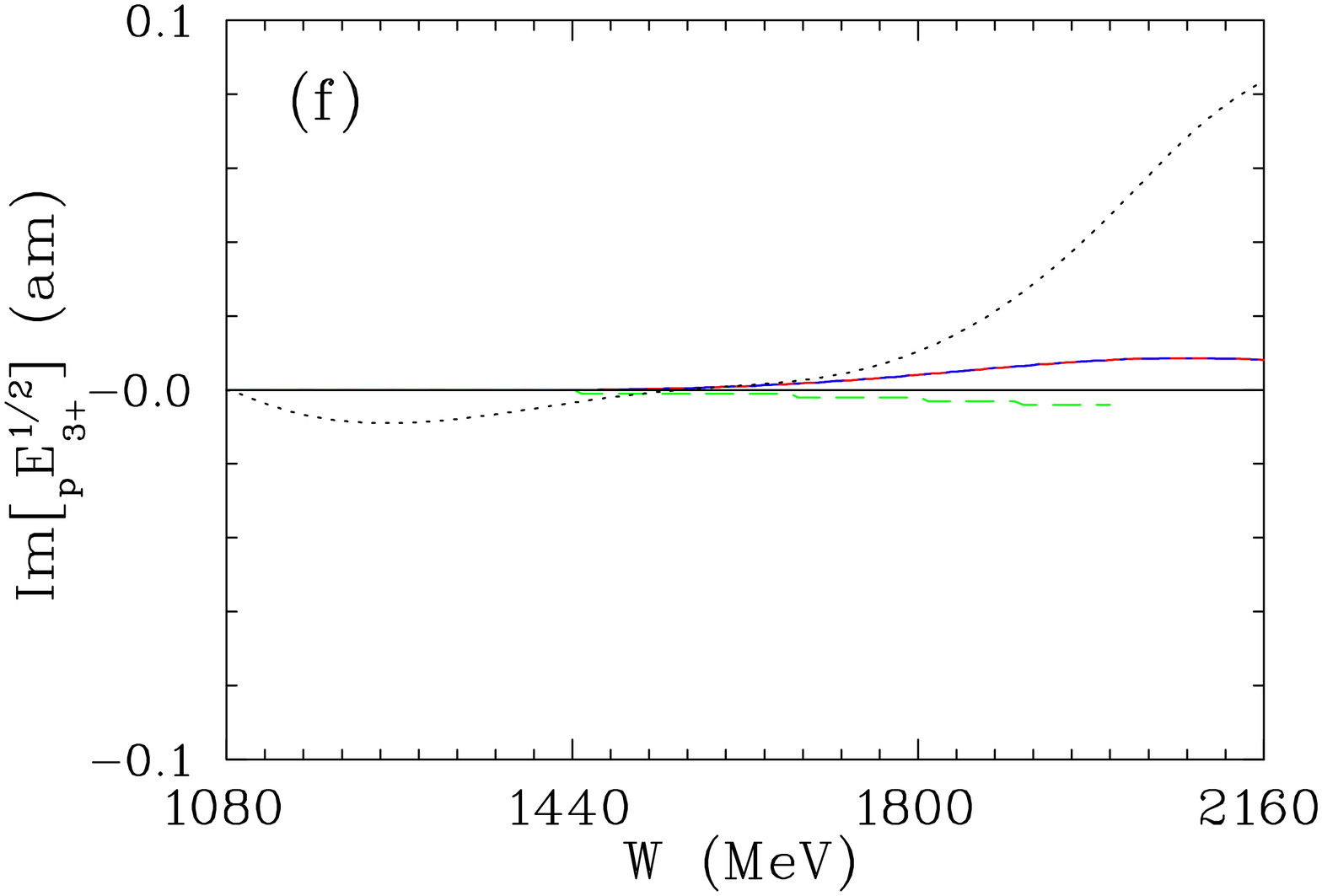}}
\centerline{
\includegraphics[height=0.35\textwidth, angle=90]{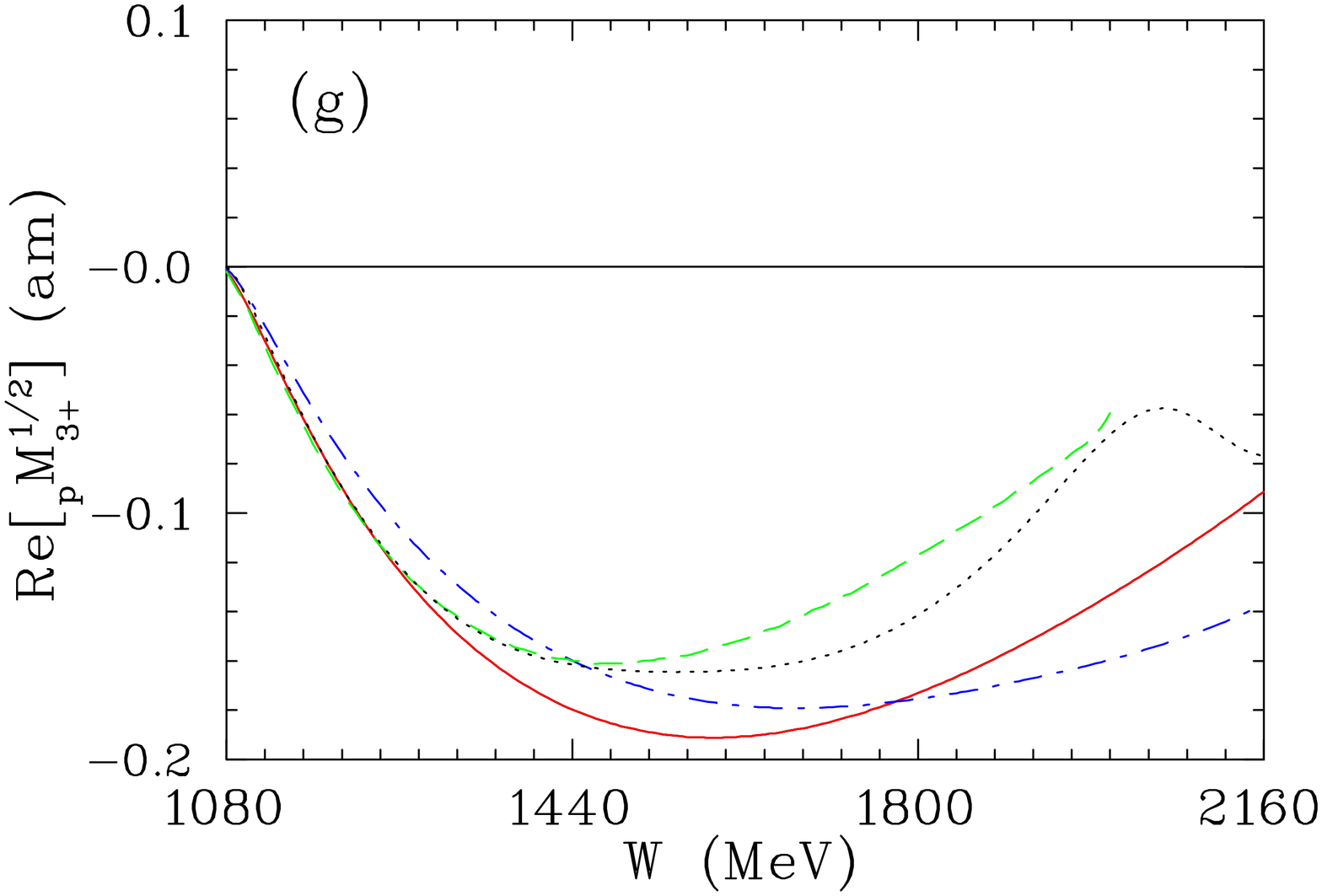}
\includegraphics[height=0.35\textwidth, angle=90]{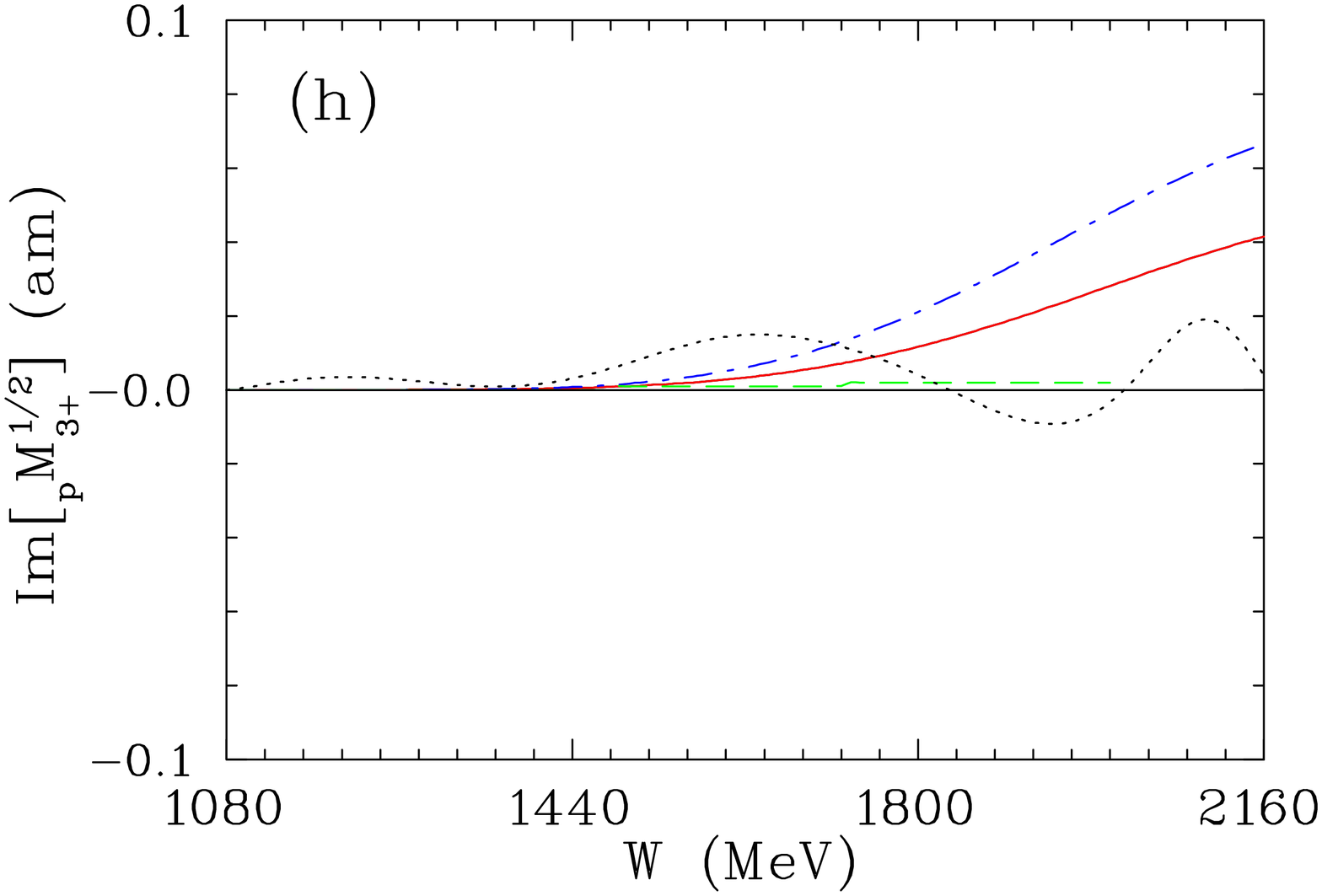}}
\caption{(Color online) Proton multipole $I$=1/2 amplitudes from
        threshold to $W$ = 2.16~GeV ($E_{\gamma}$ = 2.02~GeV) 
        for $l=3$.
        Notation as in Fig.~\protect\ref{fig:g7}. \label{fig:g12}}
\end{figure*}


\begin{figure*}[th]
\centerline{
\includegraphics[height=0.42\textwidth, angle=90]{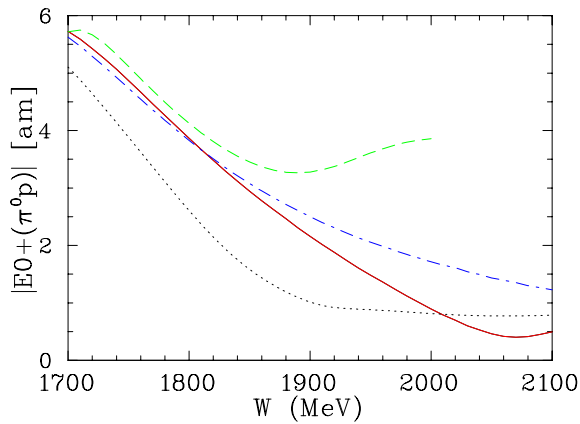}\hfill
\includegraphics[height=0.42\textwidth, angle=90]{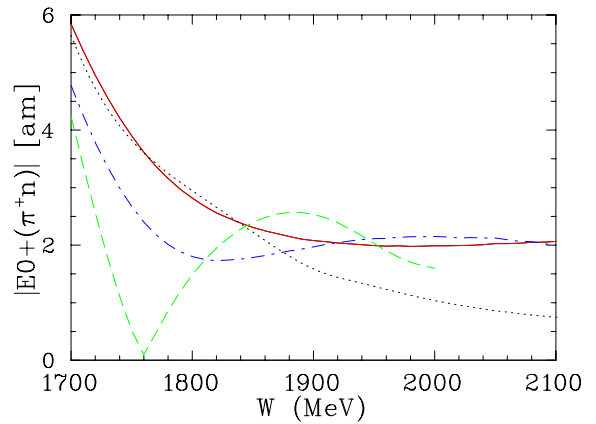}}
\caption{(Color online) Moduli of the $E_{0+}(\pi^0p)$ and
        $E_{0+}(\pi^+n)$ amplitudes.  Notation is the same
        as in Fig.~\protect\ref{fig:g7}. \label{fig:g13}}
\end{figure*}


Obtaining the quantity $(a_0 + \frac{a_4}{2})$ in Eqn~(\ref{eq:hsigma}) 
is straightforward using other moment-$n$ histograms.
In a manner similar to that leading to Eqn~(\ref{eq:hperp2}) and 
(\ref{eq:hpara2}), one obtains for the moment-0 histograms
\begin{eqnarray}
\tilde{Y}^{i,j}_{\bot 0} & = & 
2 \pi a_0 + P_\bot^i \Sigma^{i,j} \pi a_2 \nonumber
\end{eqnarray}
\noindent
and 
\begin{eqnarray}
\tilde{Y}^{i,j}_{\| 0} & = & 2 \pi a_0 - P_\|^i \Sigma^{i,j} \pi a_2 ,\nonumber
\end{eqnarray}
which gives
\begin{equation}
\tilde{Y}^{i,j}_{\bot 0} + \tilde{Y}^{i,j}_{\| 0}\left(\frac{P_\bot^i}{P_\|^i}\right) = 2 \pi a_0 \left(1 + \frac{P_\bot^i}{P_\|^i}\right).
\label{eq:hsum0}
\end{equation}
In a similar fashion, one obtains from the moment-4 histograms
\begin{equation}
\tilde{Y}^{i,j}_{\bot 4} + \tilde{Y}^{i,j}_{\| 4}\left(\frac{P_\bot^i}{P_\|^i}\right) = \pi a_4 \left(1 + \frac{P_\bot^i}{P_\|^i}\right).
\label{eq:hsum4}
\end{equation}
Finally, using the results of Eqs.~(\ref{eq:hsigma}), (\ref{eq:hsum0}), and (\ref{eq:hsum4}),
one obtains
\begin{equation}
\tilde{Y}^{i,j}_\Sigma = \Sigma^{i,j} \left\{ \frac{P_\|^i}{2} ( \tilde{Y}^{i,j}_{\bot 0} + \tilde{Y}^{i,j}_{\bot 4}) +
 \frac{P_\bot^i}{2} ( \tilde{Y}^{i,j}_{\| 0} + \tilde{Y}^{i,j}_{\| 4}) \right\} .
\label{eq:hsigmafinal}
\end{equation}

Combining Eqn~(\ref{eq:hsigmadef}) and (\ref{eq:hsigmafinal}),
the beam asymmetry $\Sigma$ for kinematic bin $i$, $j$ is found to be

\begin{equation}
\Sigma^{i,j} = \frac{ \tilde{Y}^{i,j}_{\bot 2} - \tilde{Y}^{i,j}_{\| 2}}{ \frac{P_\|^i}{2} ( \tilde{Y}^{i,j}_{\bot 0} + \tilde{\
Y}^{i,j}_{\bot 4}) +
 \frac{P_\bot^i}{2} ( \tilde{Y}^{i,j}_{\| 0} + \tilde{Y}^{i,j}_{\| 4}) } .
\label{eq:sigmaresult}
\end{equation}

However, as it stands, the value of $\Sigma^{i,j}$
generated by the ratio in Eqn~(\ref{eq:sigmaresult})
is the beam asymmetry for whatever is in that particular kinematic bin,
which will include not only the particular peak of interest but also
any background within that particular kinematic bin.
The interest here, instead, is the beam asymmetry associated
with the photoproduction of a particular meson,
which appears as a peak in the missing mass spectrum,
and not the associated background beneath that peak.
In practice, then, one extracts from the various histograms
in the numerator and denominator in Eqn~(\ref{eq:sigmaresult})
the yield of the particular meson peak corresponding to the reaction of 
interest.

In order to simplify the notation below, the incident photon energy bin index $i$
and $\cos(\theta)$ bin index $j$ will be suppressed hereafter.
The beam asymmetry is thus written as
\begin{equation}
\Sigma = \frac{ \tilde{Y}_{\bot 2} - \tilde{Y}_{\| 2}  }{ \frac{P_\|}{2} ( \tilde{Y}_{\bot 0} + \tilde{Y}_{\bot 4}) +
 \frac{P_\bot}{2} ( \tilde{Y}_{\| 0} + \tilde{Y}_{\| 4}) }   .
\label{eq:Sigma}
\end{equation}
\noindent

Eqn~(\ref{eq:Sigma}) is the principal result for this method.
With this approach,
rather than partitioning the data
for a given $\Eg$ \ and $\cos(\theta)$ into various $\varphi$ bins,
all the data for a given $\Eg$ \  and $\cos(\theta)$
are used simultaneously to determine the
beam asymmetry $\Sigma$ for the reaction of interest.

\subsection{Statistical uncertainty}
\label{sec:momentMethodStat}

Because the various components of Eqn. (\ref{eq:Sigma}) have
non-vanishing
covariances, the determination of statistical uncertainties,
while straightforward, requires attention.

We begin by defining $w_{m,k,l}$ as the histogram
weighting of the $l^{th}$
Poisson-distributed event, of the $m^{th}$ moment within the $k^{th}$ mass bin of a
moment histogram
$\tilde{Y}_m$. It then follows that the total occupancy $\tilde{Y}_{m,k}$ of
the $k^{\rm{th}}$ bin
within $\tilde{Y}_m$ is
\begin{equation}
\tilde{Y}_{m,k} = \frac{1}{N_{\gamma}}\sum_{l=1}^{Y_{0,k}} w_{m,k,l}  , \nonumber
\end{equation}
where $Y_{0,k}$ is the total number of events in bin $k$.
For $m = 0$ this is simply
\begin{equation}
\tilde{Y}_{0,k} = \frac{1}{N_{\gamma}}\sum_{l=1}^{Y_{0,k}} 1 = \frac{Y_{0,k}}{N_{\gamma}},  \nonumber
\end{equation}
as expected.
For all other moments
\begin{equation}
\tilde{Y}_{m,k} = \frac{1}{N_{\gamma}} \sum_{l=1}^{Y_{0,k}} \cos(m \varphi_l) . \nonumber
\end{equation}
It now follows that the variance $\sigma^2_{\tilde{Y}_{m,k}}$ is given by
\begin{equation}
\sigma^2_{\tilde{Y}_{m,k}} = \frac{1}{(N_{\gamma})^2} \sum_{l=1}^{Y_{0,k}} \cos^2(m \varphi_l) , \nonumber
\end{equation}
which for $m = 0$, reduces to the familiar form for a Poisson distributed
random variable divided by a constant term $N_{\gamma}$,
\begin{equation}
\sigma^2_{\tilde{Y}_{0,k}} = \frac{1}{(N_{\gamma})^2}\sum_{l=1}^{Y_{0,k}} 1 =
\frac{1}{(N_{\gamma})^2} Y_{0,k} = \frac{1}{N_{\gamma}}\tilde{Y}_{0,k} . \nonumber
\end{equation}

It is useful to note that, by way of the double-angle relationship
for the cosine of an angle, the variance of $\tilde{Y}_{m,k}$ can be written as
\begin{equation}
\sigma^2_{\tilde{Y}_{m,k}} = \frac{1}{2 (N_{\gamma})^2} \sum_{l=1}^{Y_{0,k}} \left[ 1 + \cos(2 m \varphi_l) \right] =
\frac{1}{2 N_{\gamma}} \left[
\tilde{Y}_{0,k} + \tilde{Y}_{2 m,k}
\right ] . \nonumber
\end{equation}

The covariance of two variables $\tilde{Y}_{m1,k}$, and $\tilde{Y}_{m2,k}$,
${\rm{Cov}}(\tilde{Y}_{m_1,k},\tilde{Y}_{m_2,k})$, is given by
\begin{eqnarray}
{\rm{Cov}}(\tilde{Y}_{m_1,k},\tilde{Y}_{m_2,k}) & = &
\frac{1}{(N_{\gamma})^2} \sum_{l=1}^{Y_{0,k}}
\cos(m_1 \varphi_{l}) \cos(m_2 \varphi_{l})  \nonumber
\end{eqnarray}

In what follows the identity
\begin{eqnarray}
{\rm{Cov}}(\tilde{Y}_{m,k},\tilde{Y}_{2 m,k})
& = & \frac{1}{(N_{\gamma})^2} \sum_{l=1}^{Y_{0,k}} \cos(m \varphi_{l}) \cos(2 m \varphi_{l})  \nonumber \\
& = & \frac{1}{(N_{\gamma})^2} \sum_{l=1}^{Y_{0,k}} \frac{1}{2} ( \cos(m \varphi_{l}) + \cos(3 m \varphi_{l})   )  \nonumber \\
& = & \frac{1}{2 (N_{\gamma})^2} (\tilde{Y}_{m,k} + \tilde{Y}_{3m,k} ) \nonumber
\end{eqnarray}
will be of use, as well as
\begin{eqnarray}
{\rm{Cov}}(\tilde{Y}_{0,k},\tilde{Y}_{m,k})
& = & \frac{1}{N_{\gamma}} \tilde{Y}_{m,k} . \nonumber
\end{eqnarray}

With these preliminaries,
the statistical
uncertainty for the beam asymmetry $\Sigma$ given by Eqn (\ref{eq:Sigma})
can be determined.

By allowing the following definitions of the numerator and denominator 
of Eqn.  (\ref{eq:Sigma}),
\begin{eqnarray}
n & \equiv & \tilde{Y}_{\bot 2} - \tilde{Y}_{\| 2}   \nonumber \\
d & \equiv &  \frac{P_\|^i}{2} ( \tilde{Y}_{\bot 0} + \tilde{Y}_{\bot 4}) +
 \frac{P_\bot^i}{2} ( \tilde{Y}_{\| 0} + \tilde{Y}_{\| 4}) ,
\label{eqn:nd}
\end{eqnarray}
we can then rewrite the beam asymmetry $\Sigma$ in the form
\begin{eqnarray}
\Sigma = \frac{n}{d} . \nonumber
\end{eqnarray}
The variance of $\Sigma$ is then
\begin{equation}
\sigma^2_{\Sigma} = \Sigma^2 \left\{
\frac{\sigma^2_n}{n^2} + \frac{\sigma^2_d}{d^2} - \frac{2 {\rm{Cov}}(n,d)}{n d} .
\right\}  \nonumber
\end{equation}

We can now determine the variance of $n$, $d$, and the covariance of $n,d$.
The variance of $n$ is
\begin{eqnarray}
\sigma^2_n & = & 
\frac{1}{2 N_{\gamma \bot}}(\tilde{Y}_{\bot 0} + \tilde{Y}_{\bot 4}) +
\frac{1}{2 N_{\gamma \|}}(\tilde{Y}_{\| 0} + \tilde{Y}_{\| 4})  , \nonumber
\end{eqnarray}
where $N_{\gamma \bot}$ ($N_{\gamma \|}$) is the integrated photon flux for perpendicular (parallel)
photon beam orientation.
The variance of $d$ is
\begin{eqnarray}
\sigma^2_d & = & 
\frac{P_\|^2}{4 N_{\gamma \bot}}(\tilde{Y}_{\bot 0} + \frac{1}{2} (\tilde{Y}_{\bot 0} +  \tilde{Y}_{\bot 8}) + 2 \tilde{Y
}_{\bot 4})  \nonumber\\
& + & \frac{P_\bot^2}{4 N_{\gamma \|}}(\tilde{Y}_{\| 0} + \frac{1}{2} (\tilde{Y}_{\| 0} +  \tilde{Y}_{\| 8}) + 2 \tilde{Y}_{\|4})  \nonumber
\end{eqnarray}
and the covariance of $n$, $d$
\begin{eqnarray}
{\rm{Cov}}(n,d) & = & 
\frac{P_\|}{4 N_{\gamma \bot}} ( 3 \tilde{Y}_{\bot2} + \tilde{Y}_{\bot6} ) -
\frac{P_\bot}{4 N_{\gamma \|}} ( 3 \tilde{Y}_{\|2} + \tilde{Y}_{\|6} ) . \nonumber
\nonumber
\end{eqnarray}

All the necessary quantities needed to calculate $\Sigma$
and the associated uncertainty $\sigma_{\Sigma}$ have now been derived.
\section{Yield determination for each kinematic bin}

To determine the $\pi^0$ yields, a technique very similar to the one used
for the {\tt{g1c}} experiment of extracted differential cross sections
for $\pi^0$ photoproduction off the proton \cite{ASUpi0} was employed. 
The {\tt{g1c}} experiment
utilized the same CLAS detector and bremsstrahlung photon tagger as the 
{\tt{g8b}} experiment, but had an 18-cm-long liquid hydrogen target placed at the 
center of CLAS, and only used unpolarized incident photons.

Following the previous discussion, the beam asymmetries were
determined for a particular photon energy and $\cos(\theta)$ bin, 
which we call a ``kinematic bin''.
For each missing mass spectrum within each kinematic bin,
the $\pi^0$ yield was extracted by removing the background under the peak.
It was assumed that the background in the missing mass spectra
arises from two particular types of events:
\newcounter{bean2}
\begin{list}{\arabic{bean2}.}{\usecounter{bean2}\setlength{\rightmargin}{\leftmargin}}
\item Events arising from accidental coincidences between CLAS and the photon tagger.
\item Events arising from two-pion photoproduction via the reaction 
$\gamma p \rightarrow p \pi^+ \pi ^-$.
\end{list}
The spectrum for accidental coincidences can be determined by looking at
events that fell outside the designated trigger window. From experience
with the {\tt{g1c}} experiment, 
the background coming from accidentals within the {\tt{g8b}} data set 
was approximated as being linear in missing mass. Figure~\ref{fig:g1cfits} 
shows an example of the background subtraction from the CLAS published {\tt{g1c}}
pion differential cross sections \cite{ASUpi0}, where the accidental contribution 
was determined by looking at events that fell outside the designated trigger window.
As can be seen in
Fig.~\ref{fig:g1cfits} the assumption that the accidentals are well modeled by a linear
function is reasonable.

To determine the two-pion background,
data for the reaction $\ppipiRxn$ were selected by requiring that each
particle in the final state 
had to be identified through normal particle ID procedures,
that the same incident photon was chosen for each particle,
and that the missing mass was consistent with zero; the criterion for consistency with 
zero mass
was if the mass $m_Y^2$, in the reaction $\ppipiyRxn$
was less than 0.005 GeV$^2$ and greater than -0.01 GeV.
These selected data were used to determine the {\em{shape}}
of the $\pi^+ \pi^-$ component of the background for the $\gamma p \rightarrow p \pi^0$ 
reaction in each kinematic bin.

The background subtraction for the $\pi^0$ was then performed in the following manner:
\newcounter{bean3}
\begin{list}{\arabic{bean3}.}{\usecounter{bean3}\setlength{\rightmargin}{\leftmargin}}
\item The spectrum of missing mass $M_X$ in the $\gamma p \to pX$ reaction 
was fit with a functional form that included 
the linear approximation of the accidentals 
and the shape determined for the charged background noted above.
A total of 3 parameters were varied:
two parameters for the accidental contribution (modelled by a linear function) and 
one parameter for the {\em{magnitude}} of the charged background.
\item The backgrounds determined in the previous step were subtracted from the yield.
\item The background subtracted yield was then fit with a Gaussian and the standard deviation
and centroid of the peak were determined.
\item The region of the histogram resulting from step 3 that 
was within three times the 
standard deviation of the peak centroid was then determined to be the 
$\gamma p \to p \pi^0$ yield in the extracted $\pi^0$ peak.
\end{list}

For the extraction of the yield of the neutron peak from the reaction $\gamma p \rightarrow \pi^+ X$,
it was found that the 2$\pi$ background was negligible, and the only significant background
was from accidentals. For this reason, only a linear approximation of 
the accidentals was included in the background determination for the neutron.

An example of the background subtraction for both neutron and $\pi^0$ extraction
can be seen in Fig.~\ref{fig:g8bfits}. 

\section{Relative Normalization}
\label{sec:RelNorm}

For the measurement of beam asymmetry, knowledge of the 
absolute number of incident photons is not required. Instead, 
only the relative photon normalization
between PARA and PERP running conditions is necessary. In order to obtain the 
relative photon normalization of PARA to PERP,  
a ``rough $\pi^0$'' measurement was used,
where ``rough $\pi^0$''
is defined as any event detected from $\xRxnp$ with missing mass $M_X$ between 0.0 
and 0.25 GeV and,
in this instance,
$0 \le \cos(\theta^X_{c.m.}) \le 1$, where $ \cos(\theta^X_{c.m.})$ is the meson
center-of-mass scattering angle.

For the determination of the relative normalization,
the more conventional approach of binning the rough $\pi^0$
data into azimuthal angle bins was used.
The data were binned in the same $E_\gamma$ bins as
for the moment extraction method, and in addition, 
the data were binned further into 36 azimuthal bins.

Once the yield for $\pi^0$ mesons was determined for each
the running conditions, two quantities
for each \{$E_\gamma$, $\varphi$\} bin were formed,
\begin{eqnarray}
g_{\bot} \equiv \frac{Y_{\bot}}{Y_{a}} \nonumber
\end{eqnarray}
and
\begin{eqnarray}
g_{\|} \equiv \frac{Y_{\|}}{Y_{a}} . \nonumber
\end{eqnarray}
These were fit to
\begin{eqnarray}
g_{\bot} = A_{\bot}[1 + B \cos(2 \varphi)] \nonumber
\end{eqnarray}
and
\begin{eqnarray}
g_{\|} = A_{\|}[1 - B \cos(2 \varphi)], \nonumber
\end{eqnarray}
where $A_{\|} = {N_{\bot}}/{N_{a}}$, $A_{\bot} = {N_{\|}}/{N_{a}}$, $B = P \Sigma$,
and $N_{\bot}$, $N_{\|}$, $N_a$ represent the number of incident photons for 
the PERP, PARA, and amorphous running conditions respectively.

The values of $A$ taken from the parallel polarized beam orientation
were divided by the values derived from the
perpendicular orientation. The fractional values of $A_{\|}/A_{\bot}$ were found 
for each energy
to determine the value of $N_{\| \gamma}/N_{\bot \gamma}$
from the relation $A_{\|}/A_{\bot} = N_{\| \gamma}/N_{\bot \gamma}$.
\section{Uncertainties}
\label{sec:Errs}

The statistical uncertainties for $\Sigma$ were obtained using
the expressions given in subsection \ref{sec:momentMethodStat}.
Systematic uncertainties for $\Sigma$ are dominated by the 
systematics of the polarization and relative normalization since
many of the experimental quantities cancel in the ratio $\Sigma$.

The relative normalization
was primarily dependent upon the total number of $\gamma p \rightarrow p X$ events
having a missing mass (mass $X$) between 0 and 0.25 GeV. The statistics for such
events were quite good and we take the systematic uncertainty of the 
relative normalization as being negligible.

One possible systematic error could come from imperfect knowledge of 
the orientation of beam polarization. To study the orientation
of the beam polarization we took rough $\pi^0$ measurements for
each orientation of the beam polarization (PERP and PARA) and 
normalized each type by the rough $\pi^0$ results from the amorphous
runs. Using the entire set of runs from the 1.3 GeV coherent edge setting, 
the resulting rough $\pi^0$ normalized yields were placed in 90
$\varphi$-bins, and 50 MeV wide photon energy bins. The resulting $\varphi$-distributions
were then fit to the function $A(1 + B \cos(2\varphi + 2C))$, with $A$, $B$, and $C$ being
fit parameters. From the fit we were able to extract the possible 
azimuthal offset by reading out parameter $C$. Figure \ref{fig:phi1275} shows
the resulting fit for both orientations at photon energy of 1275 MeV. (The
figure also clearly shows the six sector structure of CLAS.) We performed the
fitting procedure for five energy bins from the 1.3 GeV data, took the weighted
average, and obtained a possible systematic error in the polarization 
orientation of 0.07 $\pm$ 0.04 
degrees. Since the possible systematic error is so small, we have assumed that such an 
error has a negligible effect on the beam asymmetry measurements.

The overall accuracy of the estimated
photon polarization is difficult to determine.
However, the consistency of the bremsstrahlung calculation could
be checked by comparing predicted and measured polarization ratios
for adjacent coherent edge settings in regions where overlapping
energies exist.
After
consistency corrections were applied \cite{pCor}, the
estimated value for the photon polarization was self-consistent to within 4\%.
Therefore, the estimated systematic uncertainty in the photon 
polarization is taken to be 4\%.

To test the dependence of the Fourier moment method
on the polarization values, 
rough
$\pi^0$ beam asymmetries from the moment method were compared to the 
beam asymmetries obtained using the $\varphi$-bin method (averaged over 
polarization orientations). 
As in Section \ref{sec:RelNorm}, a rough $\pi^0$
azimuthal distribution was extracted for each tagger energy counter (E-counter). This
time, however, the rough $\pi^0$ extraction was performed for the backward 
center-of-mass pion-angles
($-1 \le \cos(\theta^X_{c.m.}) \le 0$), as well as the forward center-of-mass pion-angles
($0 \le \cos(\theta^X_{c.m.}) \le 1$).

For each case (forward and backward angle events), the 
polarized photon data were divided by the corresponding distribution from amorphous data.
As done in Sec. \ref{sec:RelNorm}, the ratios for the azimuthal distributions were then
fit to the expression
\begin{eqnarray}
A \left[ 1 + B \cos(2 \varphi)\right],
\label{eqn:ABfitU}
\end{eqnarray}
where $A$ and $B$ were parameters of the fit. 
The value of beam asymmetry was then determined by $\Sigma = B_{\perp}/P_{\perp}$
($\Sigma = -B_{||}/P_{||}$).

The values of $\Sigma$ determined from the $\varphi$-bin for each polarization 
orientation were averaged to obtain an average $\Sigma$ value.
The average $\Sigma$ value obtained from the $\varphi$-bin
method is compared to the beam asymmetries determined by the moment method, as seen in 
Figure \ref{fig:sigRough}. The top panel of Fig.~\ref{fig:sigRough} shows the rough
$\pi^0$ beam asymmetries as a function of energy counter for the forward center-of-mass
angles, and the backward center-of-mass angles are shown on the bottom panel. In each
panel of Fig.~\ref{fig:sigRough} the black points are $\Sigma$ determined by the 
$\varphi$-bin method and
the blue points represent $\Sigma$ determined from the moment method. A visual inspection
of the plots given in Fig. \ref{fig:sigRough} shows that the $\varphi$-bin and
moment methods give very similar results. 

To quantify the level of agreement between
the two methods, the $\Sigma$ results from the moment method were divided by those of
the $\varphi$-bin method on an E-counter by E-counter basis. A frequency plot of the 
resulting $\Sigma$-fractions ($\Sigma$ from moment method divided by $\Sigma$ from 
$\varphi$-bin method) was created for forward and backward center-of-mass angles of the 
$\pi^0$. In the top panel of Fig. \ref{fig:sigFrac} the frequency of $\Sigma$-fractions
for forward angles is shown, while the bottom panel is the frequency
plot for backward angles. A Gaussian was fit to each distribution of Fig. \ref{fig:sigFrac}
with the results shown in Table \ref{tbl:sigfreq}. 

\begin{table}
\caption{
Gaussian parameters of the fit to the ratios of the results for $\Sigma$ using the 
moment method 
to $\Sigma$ determined by the $\varphi$-bin method on an E-counter by E-counter basis.}
\vspace{2mm}
\centering
\begin{tabular}{|c|c|c|}
\hline
Center-of-mass angles & Center & $\sigma$    \\
\hline
\hline
Forward & 0.9978(3)  & 0.0043(4) \\
\hline
Backward & 1.003(2)  & 0.015(2) \\
\hline
\end{tabular}
\label{tbl:sigfreq}
\end{table}

Since the beam asymmetry results from the moment method are well within 1\%
of the beam asymmetry results coming from the average value (parallel and perpendicular 
orientations) 
determined by the $\varphi$-bin method, we can safely say that the systematic uncertainty
of the moment method due to polarization is nearly identical to the systematics one
obtains when simply averaging the beam asymmetry from each polarization orientation.
Thus, the fractional uncertainty of each polarization systematic 
uncertainty (each estimated as 4\%) is added in quadrature to obtain an estimate of 
the systematic uncertainty in the 
beam asymmetry of 6\%.


\section{Results}
\label{sec:results}

The CLAS beam asymmetries obtained here for $\vec{\gamma} p 
\rightarrow p \pi^0$ (700 data points represented as filled 
circles) are compared 
in Figs.~\ref{fig:g1}$-$\ref{fig:g2}
with previous data from
Bonn~\cite{CBELSA1, CBELSA2} (open circles),
Yerevan~\cite{Yerevan1,Yerevan2,Yerevan3,Yerevan4,Yerevan5,Yerevan6} 
(open triangle),
GRAAL~\cite{GRAAL1} (open squares),
CEA~\cite{CEA} (filled squares),
DNPL~\cite{DNPL1,DNPL2} (crosses), and 
LEPS~\cite{LEPS} (asterisks). 
The results for the reaction $\gamma p \rightarrow n \pi^+$ CLAS 
beam asymmetries (386 data points shown as filled circles)
are compared in Fig.~\ref{fig:g3} 
to previous data from GRAAL~\cite{GRAAL2} (open squares),
Yerevan~\cite{Yerevan7} (open triangles),
CEA~\cite{CEA} (filled squares),
and DNPL~\cite{DNPL2} (crosses). 
Only those world data that are within 
$\pm$3~MeV of the CLAS photon energies $E_{\gamma}$ are shown.  In addition to the 
data, phenomenological curves are included in the above mentioned 
figures and will be discussed further below.

For the CLAS $\pi^0$ data obtained here, the Yerevan results agree well 
except for a few points at $E_{\gamma} =$1265, 1301, and 1337~MeV. 
The Bonn data are 
comprised of two separate experiments~\cite{CBELSA1, CBELSA2}, one 
published in 2009~\cite{CBELSA1} and another published in 
2010~\cite{CBELSA2}. Typically, the CLAS results agree within error 
bars of the Bonn data, and where there is disagreement, it is almost 
always with the earlier 2009 results. The data obtained here is in 
very good agreement with DNPL at $E_{\gamma} = 1337$~MeV and tend to 
be within error bars for all other energies except for $E_{\gamma} = 
1301$~MeV, where several DNPL points are systematically larger than the 
CLAS results. In particular, the data obtained here confirm the
magnitude of the sharp structure seen in the DNPL data near $60^{\circ}$ 
for photon energies greater than about $E_{\gamma}=$ 1600 MeV.  The LEPS results
($E_{\gamma}$ = 1551~MeV, backward angles), as well as the GRAAL 
results look systematically smaller when compared to CLAS.

The $\pi^+$ data obtained here tend to agree well with the previous 
data except for a few points.
Out of the 34 points from GRAAL, easily identifiable 
differences between GRAAL \cite{GRAAL2} and CLAS occur for four with $E_{\gamma} = 1148$ MeV
($\theta = 114^{\circ}$, $122^{\circ}$, $145^{\circ}$, and $ 150^{\circ}$), along 
with a single point at $E_{\gamma} = 1400$ ($\theta = 90^{\circ}$).
The single CEA \cite{CEA} point at 
$E_{\gamma} = 1400 $ MeV ($\theta = 90^{\circ}$) is systematicaly low when compared
to CLAS. For the single Yerevan measurment of beam asymmetry \cite{Yerevan7}, the 
agreement is good.
Comparisons between CLAS and DNPL \cite{DNPL2} are mixed. The DNPL results were taken
with two different sets of beam energies. There was a low energy data set from 
DNPL with photon energies ranging from 520 to 1650 MeV, and a high energy data
set with energies between 1650 and 2250 MeV. Because the DNPL energy ranges
overlap for $E_{\gamma} = 1650$ MeV, they report two sets of beam asymmetries
for that energy. The DNPL data from the low energy data set agrees well with CLAS
except for a single point at $E_{\gamma} = 1400$ MeV ($\theta =  75^{\circ}$), while
the agreement between CLAS and the DNPL high energy data set is sometimes poor.
In particular, at $E_{\gamma} = 1649$ MeV, the DNPL points that are systematicaly
high (low) compared to CLAS occuring at $\theta = $ 30$^{\circ}$, 40$^{\circ}$, 
75$^{\circ}$ ($\theta = $ 105$^{\circ}$, 115$^{\circ}$) are all from the DNPL high
energy data set, while the agreement between CLAS and DNPL at $E_{\gamma} = 1649$ MeV 
from the low energy data set is in good agreement.

Briefly, then, the new CLAS measurements generally are in agreement with older 
results within uncertainties, but the results presented here are far 
more precise and provide finer energy resolution.

\section{Comparison to Fits and Predictions}
\label{sec:fit}

\subsection{Comparison to phenomenological models}

In Figs.~\ref{fig:g1}$-$\ref{fig:g3}, the $\Sigma$ data are 
shown along with predictions from previous SAID~\cite{cm12}, 
MAID~\cite{Maid07} (up to its stated applicability limit at a
center-of-mass energy $W=$ 2~GeV, corresponding to $E_\gamma$ = 
1.66~GeV), and the Bonn-Gatchina (BnGa,~\cite{BnGa}) multipole 
analyses. Also shown are the results of an updated SAID fit (DU13) 
which includes the new data reported here.  In order to increase 
the influence of these new precise data, the CLAS data reported 
here were weighted by an arbitrary factor of 4 in the fit.  
Figs.~\ref{fig:g4} and \ref{fig:g5} show fixed angle excitation 
functions for $\vec{\gamma} p\to\pi^0p$ and $\vec{\gamma} 
p\to\pi^+n$.

For energies below that of the data presented in this paper,
the neutral-pion production data are well represented by
predictions from the multipole analyses up to a center-of-mass
energy of about 1500~MeV. Above this energy, large differences
are seen at very forward angles. The data appear to favor
the SAID and BnGa predictions, with large differences between
the SAID and BnGa values mainly at angles more forward than
are reached in the present experiment. Pronounced dips seen
in Figs.~\ref{fig:g1} and \ref{fig:g2} for the reaction  
$\gamma p\to\pi^0p$, are qualitatively predicted by the three 
multipole analyses. These dips develop at angles
slightly above 60$^\circ$ 
and slightly below 120$^\circ$ (note that these angles are 
related by the space reflection transformation 
$\theta\to\pi -\theta$). 
Our data confirm this feature suggested by earlier 
measurements, however those previous data were not precise enough to 
establish the sharpness of the dips. The revised SAID fit 
(DU13) now has these sharp structures.  Below we shall 
discuss in more detail a possible source of the dip structure
seen in the data.

For the charged-pion reaction, the MAID predictions are 
surprisingly far from the data over most of the measured 
energy range, and particularly at more backward angles. 
Over much of this range the SAID, BnGa, and revised SAID 
curves are nearly overlapping. 

The fit $\chi^2$ per data point $\chi^2/N_p$ for DU13 is 
significantly improved over that from the CM12 SAID prediction 
~\protect\cite{cm12}.  The comparison given in Table \ref{tbl:2} 
shows that, for the new DU13 fit, $\chi^2/N_p$ for the $\pi^0p$ 
channel is 2.77 and $\chi^2/N_p$ for the $\pi^+n$ channel is 2.77, 
an improvement by over an order of magnitude for that $\chi^2/N_p$ 
statistic when compared with the CM12 prediction. While the fit 
$\chi^2$ per datum is 2.77 when solely compared to the new CLAS 
data reported here, Table \ref{tbl:t1} also indicates that the 
fit to the previously published $\Sigma$ data is actually improved 
slightly in DU13 versus CM12, decreasing to 3.67 from 3.99. This 
is due to the added weighting of the $\Sigma$ data reported here 
in the fit, and also provides additional statistical confirmation 
of the consistency of the overall present and prior measurements, 
despite the differences noted above. 
\begin{table}
\caption{Comparison of $\chi^2$ per data point for the beam 
asymmetry $\Sigma$ for the $\pi^0p$
and $\pi^+n$ channels using the predictions of the 
CM12 SAID solution ~\protect\cite{cm12}
and the results of a fit (DU13) to the CLAS data reported here. 
Comparisons are provided for the CLAS data,
previously published data, 
and for a dataset containing both the CLAS data  
(weighted by a factor of 4) and the previously published data.
The number of data points used in each comparison is 
indicated by $N_p$.\label{tbl:t1}}
\vspace{2mm}
\begin{tabular}{|c|c|c|c|c|}
\colrule
Data & Solution & $\Sigma(\pi^0p)$ & $\Sigma(\pi^+n)$ \\
     &          & $\chi^2/N_p$     & $\chi^2/N_p$    \\
\colrule
New CLAS     &   DU13     &   1940/700 = 2.77      &   1070/386 = 2.77 \\
data only     &   CM12     &   53346/700 = 76.2     &  11795/386 = 30.6 \\
\colrule
Previous &   DU13     &   1531/654 = 2.34      &    738/201 = 3.67 \\
data only    &   CM12     &   1704/654 = 2.61      &    801/201 = 3.99 \\
\colrule
CLAS and &   DU13     &   3471/1354 = 2.56     &   1808/587 = 3.08 \\
previous data &   CM12     &   55050/1354 = 40.7    &  12596/587 = 21.5 \\
\colrule
\end{tabular} \label{tbl:2}
\end{table}

In Figs.~\ref{fig:g7} $-$ \ref{fig:g12}, we compare the dominant 
multipole contributions from SAID (CM12 and DU13), MAID, and BnGa. 
While the CM12 and DU13 solutions differ over the energy range of 
this experiment, the resonance couplings are fairly stable. The 
largest change is found for the $\Delta(1700)3/2^-$ and 
$\Delta(1905)5/2^+$ states (Table~\ref{tbl:t2}), for which the 
various analyses disagree significantly in terms of photo-decay 
amplitudes.
 
\begin{table}
\caption{$\Delta(1700)3/2^-$ and $\Delta(1905)5/2^+$
	state Breit-Wigner parameters from SAID (DU13 and 
	CM12~\protect\cite{cm12}), MAID~\protect\cite{Maid07}, 
	BnGa~\protect\cite{BnGa}, and PDG12~\protect\cite{PDG}. 
	\label{tbl:t2}}
\vspace{2mm}
\begin{tabular}{|c|c|c|c|}
\colrule
$\Delta^\ast$ & Solution & A$_{1/2}$ & A$_{3/2}$ \\
              &          & (GeV$^{1/2}\times 10^{-3}$) & (GeV$^{1/2}\times 10^{-3}$) \\
\colrule
$\Delta(1700)3/2^-$ & CM12     & 105$\pm$ 5 &   92$\pm$ 4 \\
                    & DU13     & 132$\pm$ 5 &  108$\pm$ 5 \\
                    & BnGa     & 160$\pm$20 &  165$\pm$25 \\
                    & MD07     & 226        &  210        \\
                    & PDG12    & 104$\pm$15 &   85$\pm$22 \\
\colrule
$\Delta(1905)5/2^+$ & CM12     &  19$\pm$ 2 &$-$38$\pm$ 4 \\
                    & DU13     &  20$\pm$ 2 &$-$49$\pm$ 5 \\
                    & BnGa     &  25$\pm$ 5 &$-$49$\pm$ 4 \\
                    & MD07     &  18        &$-$28 \\
                    & PDG12    &  26$\pm$11 &$-$45$\pm$20 \\
\colrule
\end{tabular}
\end{table}

The reason that MAID better describes the neutral-pion data but 
misses the charged-pion data appears to be tied partly to the 
$E_{0+}^{1/2}$ and $E_{0+}^{3/2}$ multipoles. As can be seen in 
Figs.~\ref{fig:g7} $-$ \ref{fig:g12}, both MAID multipoles differ 
significantly from the SAID values. In Fig.~\ref{fig:g13}, we plot 
for comparison the moduli of those linear combinations of isospin 
amplitudes producing the $E_{0+}^{\pi^0 p}$ and $E_{0+}^{\pi^+ n}$ 
amplitudes. 

\subsection{Associated Legendre function expansion}


\begin{figure*}
\includegraphics[height=1.0\textwidth, angle=90]{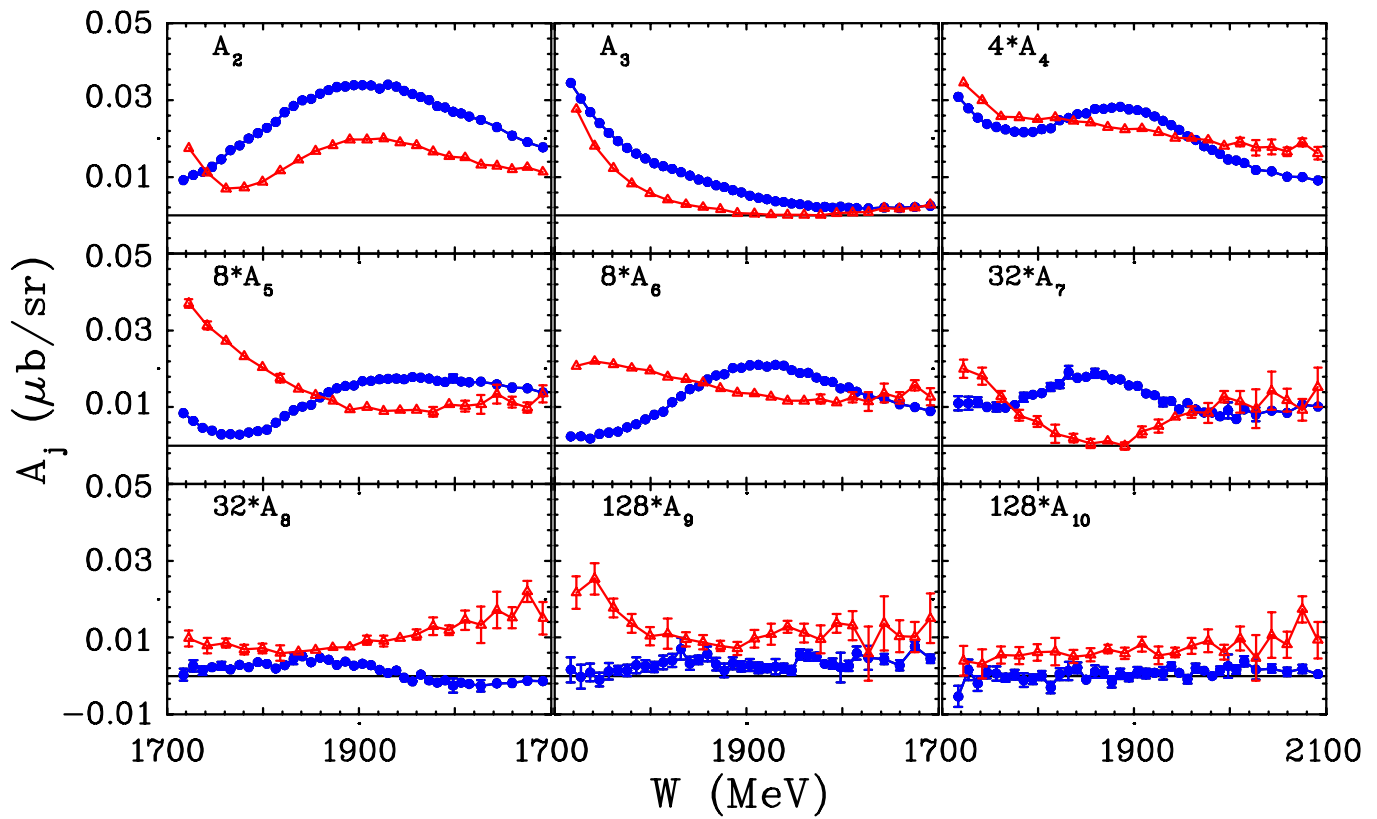}
\caption{(Color online) Coefficients for
associated Legendre functions of the second order for $\pi^0 p$ (solid
circles) and $\pi^+ n$ (solid triangles). Solid lines are plotted to help
guide the eye. \label{fig:gx}}
\end{figure*}

The photoproduction of
a pseudoscalar meson is described by four independent
helicity amplitudes which may be decomposed
over Wigner harmonics $d^j_{\lambda\mu}(\theta)$.~\cite{JW59}
After Barker \textit{et al.}~\cite{BDS74,BDS75},
those amplitudes are commonly denoted $N$, $S_1$, $S_2$, and $D$,
where $\mu = 1/2$ and $\lambda =+1/2,                                                      
-1/2, +3/2, -3/2$, respectively.
The amplitude $N$ is the non-flip helicity amplitude,
the $S$ amplitudes correspond to the single-flip helicity amplitudes,
and the $D$ amplitude corresponds to the double-flip helicity amplitude.
The beam asymmetry $\Sigma$ is related to these helicity amplitudes by the
relation ~\cite{BDS75}
\begin{equation}
[\Sigma~d\sigma/d\Omega] \sim 2 Re[S_1^\ast~S_2 - N ~D^\ast].
\label{ampSig}
\end{equation}
The first summand of this relation contains terms with products $d^{j_{1}}_{3/2,1/2}$
$d^{j_{2}}_{-1/2,1/2}$, while the second contains products $d^{j_{1}}_{1/2,1/2}$
$d^{j_{2}}_{-3/2,1/2}$.
These products yield Clebsch-Gordan series over the associated second-order
Legendre functions $P^2_j(\theta)$,
with the degree $j$ given by $|j_1 - j_2| \leq j \leq j_1 + j_2$.~\cite{JW59}
The beam asymmetry as a whole, then, may be represented by an infinite series
over these second-order associated Legendre functions of degree $j$, with the degree $j$
running from $j=$2 to infinity,
after recalling that $j$ should not be less than 2.

We have used such a series to fit the data on the beam asymmetry $\Sigma$ reported here,
supplemented by the fact that $\Sigma(0)$ = $\Sigma(\pi)$ = 0. The small statistical
uncertainties of the data obtained here allow
a correspondingly robust determination of the second-order associated
Legendre function coefficients $A_j$; these coefficients were
very difficult to determine unambiguously with previously published data of lower
statistical accuracy.
The results of our fits yield unprecedented detail on the energy dependence of the
Legendre coefficients $A_j$, and should prove
very useful in disentangling the helicity amplitudes associated
with pion photoproduction for the present energy range.

As expected for such a fit using orthogonal polynomials,
the Legendre coefficients $A_j$ decrease markedly for large $j$. At our energies and
precision, a maximum value of $j$ = 10 was found to be sufficient to describe
the data. Thus, we truncate the infinite series accordingly, using the relation
\begin{equation}
        [\Sigma ~d\sigma/d\Omega](\cos\theta) = \sum\limits_{j=2}^{10} (2j + 1)~
        A_j~P^2_j(\cos\theta) ,
\nonumber
\end{equation}
where the degree $j$ runs from 2 to 10.

In Fig.~\ref{fig:gx}, we illustrate Legendre coefficients $A_2 - A_{10}$
as a function of center-of-mass energy $W$ from the best fit of
the product of the experimental CLAS $\Sigma$ data provided by this work and DU13
predictions for
$d\sigma/d\Omega$.
None of the coefficients show a narrow structure in the
energy dependence. However, wide structures are clearly seen in the
range W = 1.8 - 2.0 GeV, most likely attributable to contributions from
one or more nucleon resonances known in this energy region with spins
up to 7/2.~\cite{PDG}  It is interesting that the coefficients $A_3$ for both
final states have no energy structures at all; they are smooth functions
throughout this energy region, with no evidence of the structures seen
for the other coefficients.

For the $\pi^+n$ final state, the behavior of the $A_j$ is noticeably different
for most of the coefficients than
the behavior observed for the $\pi^0p$ final state.
The energy dependence of the $A_2$ term for the $\pi^+n$ final state
has a similar, though smaller, bump as seen in the neutral pion data.
Likewise, the $A_3$ coefficients for
both the $\pi^+ n$ and $\pi^0 p$ final states show similar energy
behavior.
The energy dependences of the $A_4$ - $A_8$ coefficients for
the $\pi^+ n$ final state are seen to lack the narrow structures seen for
the $\pi^0 p$ final state. Moreover, the $A_8$ coefficient for
the neutral pion changes sign near W = 1950 MeV, while staying
positive for the $\pi^+ n$ case.

These pronounced differences
between charged and neutral pion reactions reveal the essential role
of the interferences between the photoproduction amplitudes for
the two final states
with isospin 1/2 and 3/2. Energy structures are less clear for the coefficients
$A_9$ and $A_{10}$. The $A_{10}$ coefficients, especially for the neutral
pion, are statistically consistent with zero, thus justifying our
truncation of the Legendre series.

The pion production angles 60$^\circ$ and 120$^\circ$ are ``mirror'' angles
which reveal dynamics associated with the interference of several amplitudes having
different angular momenta.  The sharpness of both dips seen in the $\Sigma$ data
indicates that important contributions must come from partial waves with large $j$.

This analysis of the angular dependence of the beam asymmetry data
in terms of associated Legendre functions
reinforces the long-recognized complexity of the nucleon resonance spectrum
in this energy region.
That complexity underscores the point that
an accurate interpretation of beam asymmetry in pion
photoproduction will require a comprehensive account
of the amplitude interference effects
both in terms of angular momentum $j$ and isospin.
The complicated interplay of the contributions from
the different resonances demands further clarification
through measurements of other polarization
observables in order to isolate contributions to particular amplitudes.
For example, the expression in equation \ref{ampSig} above
for the beam asymmetry $\Sigma$ in terms of
$N$, $S_1$, $S_2$, and $D$ from Ref.~\cite{BDS75} may be compared to the expression
from the same reference for the double-polarized observable $G$,
\begin{equation}
[G~d\sigma/d\Omega] \sim 2 \rm{Im}[S_1^\ast~S_2 - N ~D^\ast].
\nonumber
\end{equation}
Thus, the combination of $\Sigma$ and $G$ data greatly facilitate
isolating the individual contributions of each helicity amplitude.
New data on polarization observables have been taken (Ref.~\cite{frostProp})
in Hall-B at Jefferson Laboratory using a polarized target (transverse and longitudinal)
with polarized photon beams (circular and linear) that is currently undergoing
analysis for the observables G, F, T, P. The information from these observables,
coupled with the detailed results obtained here for $\Sigma$,
will permit tremendous progress in deconvoluting
the nucleon resonance spectrum.

\section{Conclusion}
\label{sec:conc}

An extensive and precise dataset (1086 data points) on the beam asymmetry 
$\Sigma$ for $\pi^0$  and $\pi^+$  photoproduction from the proton
has been obtained,
and a Fourier moment technique
for extracting beam asymmetries from experimental data has been described.
The measurements obtained here have been compared
to existing data. The overall agreement is good, while the data provided here 
more than double the world database for both pion reactions, are more precise 
than previous measurements, and cover the reported energies with finer 
resolution.

The present data were found to favor the SAID and Bonn-Gatchina analyses
over the older MAID predictions for both reactions.
The present set of beam asymmetries has been incorporated into
the SAID database, and exploratory fits have been made, resulting in 
a significant improvement in the fit chi-squared, and allowing for a much
improved mapping of the sharp structure near $60^{\circ}$ and less sharp one 
near $120^{\circ}$ at 
photon energies 
greater than about 1600~MeV. Resonance couplings have been extracted and the 
largest change from previous fits was found to occur for the 
$\Delta(1700)3/2^-$ and $\Delta(1905)5/2^+$ states.

Beyond these phenomenological analyses, we performed an analysis of our beam 
asymmetry data
using a series based on associated Legendre functions,
coupled with predictions for the differential cross sections from SAID. 
This fit was made possible
by the high statistical accuracy of the current data set.
The analysis clearly shows
the 
important role of interference contributions coming from
the isospin 1/2 and 3/2 basis states to the
$\pi^0$  and $\pi^+$  photoproduction reactions. 
When combined with future measurements of $G$, these data
should greatly help attempts to disentangle the contributions of various
resonances to the photoproduction process. 

\acknowledgments

The authors gratefully acknowledge the work of the Jefferson Lab 
Accelerator Division staff.  This work was supported by the 
National Science Foundation, the U.S. Department of Energy (DOE), 
the French Centre National de la Recherche Scientifique and 
Commissariat \`a l'Energie Atomique, the Italian Istituto 
Nazionale di Fisica Nucleare, 
the United Kingdom's Science and Technology Facilities Council 
(STFC), and the 
National Research Foundation of Korea.
The Southeastern Universities Research 
Association (SURA) operated Jefferson Lab for DOE under contract 
DE-AC05-84ER40150 during this work.


\end{document}